# Network Codes with Overlapping Chunks over Line Networks: A Case for Linear-Time Codes$^{\dagger}$

Anoosheh Heidarzadeh, *Student Member, IEEE*, and Amir H. Banihashemi, *Senior Member, IEEE*

Department of Systems and Computer Engineering

Carleton University, Ottawa, Canada

Email: {anoosheh,ahashemi}@sce.carleton.ca

### Abstract

In this paper, the problem of designing network codes that are both communicationally and computationally efficient over packet line networks with worst-case schedules is considered. In this context, random linear network codes (dense codes) are asymptotically capacity-achieving, but require highly complex coding operations. To reduce the coding complexity, Maymounkov *et al.* proposed chunked codes (CC). Chunked codes operate by splitting the message (a collection of packets) into non-overlapping chunks and send a randomly chosen chunk at each transmission time by a dense code. The complexity, that is linear in the chunk size, is thus reduced compared to dense codes. In this paper, the existing analysis of CC is revised, and tighter bounds on the performance of CC are derived. As a result, we prove that (i) CC with sufficiently large chunks are asymptotically capacity-achieving, but with a slower speed of convergence compared to dense codes; and (ii) CC with relatively smaller chunks approach the capacity with an arbitrarily small but non-zero constant gap. To improve the speed of convergence of CC, while maintaining their advantage in reducing the computational complexity, we propose and analyze a new CC scheme with overlapping chunks, referred to as overlapped chunked codes (OCC). We prove that for smaller chunks, which are advantageous due to lower computational complexity, OCC with larger overlaps provide a better tradeoff between the speed of convergence and the message or packet error rate. This implies that for smaller chunks, and with the same computational complexity, OCC outperform CC in terms of the speed of approaching the capacity for sufficiently small target error rate. In fact, we design linear-time OCC with very small chunks (constant in the message size) that are both computationally and communicationally efficient, and that outperform linear-time CC. Finite-length simulation results consistent with the asymptotic analytical results are also presented. Both the analytical and simulation results suggest great potential for the application of OCC for multimedia transmission over packet networks.

## I. INTRODUCTION

**T**HERE has recently been a surge of interest in application of network coding for large-scale file sharing over packet networks [1]. Network coding has been shown to generally reduce the expected file downloading time for various probabilistic and deterministic models of the flow transmission schedules [2], [3]. In practice, however, not only the network nodes might be blind to the schedule, but also the accurate modeling of the schedule might be too complex and/or infeasible [4]–[6]. The problem of a practical code design is therefore to achieve the capacity of the network with high probability under any arbitrary schedule unknown at the network nodes.

Random linear network codes (a.k.a. dense codes) are known to achieve the capacity in this setting asymptotically (when the message length tends to infinity), while having linear coding costs (the encoding/decoding algorithms at each node require a number of packet operations per message packet linear in the message length) [7], [8]. The rather high coding cost, however, impedes the application of dense codes for the transmission of large files. One would thus be interested in devising coding schemes with relatively low complexity.

To overcome the computational inefficiency of dense codes, Maymuonkov *et al.* proposed *chunked codes* (CC) in [8]. These codes operate by dividing the original message into non-overlapping chunks. Each node then randomly chooses a chunk at any time instant and transmits it by using a dense code. Thus, CC require less complex coding operations as they apply coding on smaller chunks rather than the original message. (The coding costs of such codes are linear in the size of the chunks.) This advantage of CC, however, comes at the cost of lower speed of convergence or higher message or packet error rate compared to dense codes.

In [8], it has been shown that CC asymptotically achieve the capacity so long as the size of chunks (a.k.a. aperture size) is bounded below. This lower bound has been shown to be an increasing function of the message length. Thus the aperture size cannot be reduced down to a constant in the message length. The coding algorithms, therefore, cannot be performed in linear time (with constant costs in the message length). This may hamper the use of such codes in practical applications with severe computational resource limitations.

The need for coding schemes with smaller coding costs motivated the authors in [8] to also analyze CC with smaller apertures. They showed that a CC with chunks of smaller sizes (down to some constant in the message length) approaches the capacity with an arbitrarily small but non-zero constant gap.

## A. Main Contributions

Targeting the design of codes with better tradeoff between the computational complexity, on one hand, and the speed of convergence and the message or packet error rate, on the other hand, the main contributions of this paper are as follows:

- We provide a more comprehensive analysis of dense codes, compared to that of [8]. This will then serve as the basic framework for the analysis of CC and OCC.

- We revise the analysis of CC in [8], and prove that CC with any aperture size provide a better tradeoff between the computational complexity, on one hand, and the speed of convergence and the message or packet error rate, on the other hand, in comparison with what was previously thought, based on the results of [8].

- We propose chunked codes with overlapping chunks, referred to as *overlapped chunked codes* (OCC), and show that (i) for sufficiently large apertures, OCC (OCC with larger overlaps) achieve the capacity, but with a slower speed of convergence compared to CC (OCC with smaller overlaps), for any given message error rate; (ii) for smaller apertures, OCC (OCC with larger overlaps) approach the capacity with an arbitrarily small but non-zero constant gap with a larger speed of convergence when compared to CC (OCC with smaller overlaps), for sufficiently small given message or packet error rate.

- The result of (ii) leads to the design of linear-time network codes (with very small apertures of constant size with respect to the message length), which perform better than the existing codes in the literature with the same computational complexity over networks with arbitrary schedules.

- As part of the machinery used in the analysis, we generalize a recently proposed conjecture in [9] on the rank property of a special class of random matrices with overlapping bands to two classes of more general banded random matrices.[1]

- We also demonstrate the advantage of finite-length OCC (OCC with larger overlaps) over CC (OCC with smaller overlaps) through extensive simulation results. For example, our results show that when compared to a CC with similar coding costs, the application of OCC can decrease the downloading time of a $1$MB file from a file server $4$ hops away by about $17\% - 30\%$.

---

[1] The validity of our conjecture is supported via simulations, but a formal proof is still unknown.

## B. Related Work

There are a number of variations of chunk-based codes in the literature of efficient network codes as well as efficient erasure-correcting codes over a single binary erasure channel (BEC). To the best of our knowledge, however, none of these codes have been provably shown to perform better than that in [8] over arbitrary schedules unknown at the network nodes.

Focusing on the design of computationally efficient codes over BEC, in [9], Studholme and Blake propose "windowed erasure codes." These codes have a similar structure to the chunked codes of [8], except that every two contiguous chunks overlap in all but one packet. From the results of [9], it can be concluded that, compared to non-overlapping chunks, the application of overlapping chunks provides a significant improvement in the speed of convergence to the capacity and/or in the probability of decoding failure. This advantage arises from the large size of the overlap between the chunks. However, the larger is the overlap size, the larger will be the number of chunks. Having a large number of chunks, regardless of whether the chunks overlap or not, hampers the application of such codes over a longer network of erasure links (as shown in [8]). To remedy this situation, we reduce the overlap size in our version of chunked codes with overlapping chunks. We in fact demonstrate that with the right balance between the overlap size and the number of chunks, OCC can be a viable choice for information transmission over packet networks.

The idea of overlapping chunks has also been proposed by Silva *et al.* in [10], independently. Unlike our contiguous overlapping scheme, the overlapping scheme of [10] has a grid structure. Also, no theoretical result is presented in [10]; and the simulation results are only on the application of such coding schemes over a single BEC, not a (longer) line network. In [11], Li *et al.* propose a randomized overlapping scheme and provide a finite-length analysis of such codes over the BEC. Our preliminary simulations demonstrate that the codes proposed in [11] do not perform well over longer line networks. Moreover, the analysis in [11] does not seem to be generalizable to line networks. In this work, we provide an asymptotic analysis of overlapped chunked codes (with contiguous overlaps) over line networks. The proposed codes are superior to those of [10] and [11] in the underlying setting. The structure of overlapping schemes in [10] and [11] implies that for a low-complexity decoding algorithm, the chunks need to be decoded one at a time. In our work, however, the decoding can be performed on the set of all the chunks together while preserving the low complexity of the decoding algorithm. This thus results in a better tradeoff between the speed of convergence to the capacity and the message or packet error rate. In fact, by performing the decoding algorithm on the set of all the chunks simultaneously, for successful decoding, no chunk needs to be recoverable in isolation. Thus a smaller number of successful packet transmissions is sufficient to ensure successful decoding for a given message/packet error rate. This, however, may come at the cost of increasing the memory requirements.

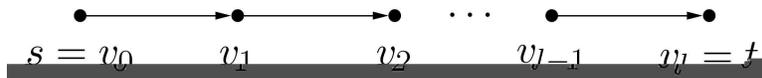

Fig. 1. A line network of length $l$.

The rest of the existing literature on chunk-based codes consider problem settings that are different from ours. In particular, in [12]–[14], some knowledge about the schedule is available at the network nodes. In addition, in [15]–[17], a probabilistic model for the schedule is assumed. Unlike these, in this work, we assume "arbitrary" schedules which are "unknown" at network nodes.

### C. Organization

The remaining of the paper is organized as follows: Section II presents the problem formulation and definitions. In Section III, we study the capacity-achieving codes, i.e., dense codes, CC and OCC with large chunk sizes. Section IV contains the analysis of the capacity-approaching codes, i.e., CC and OCC with small chunks. Section V covers the simulation results, and Section VI concludes the paper.

## II. PROBLEM FORMULATION AND DEFINITIONS

### A. Network Scenario

In this paper, we focus on the information flow problem over line networks. The results however can be generalized to more general network scenarios over arbitrary schedules by a union bound analysis.

*Line Networks:* The collection of $l$ links connecting $l + 1$ nodes $\{v_i : 0 \leq i \leq l\}$ in tandem is called a *line network* of *length* $l$, i.e., for each $0 \leq i < l$, there is a directed link $(u, v)$ between the two nodes $u \triangleq v_i$ and $v \triangleq v_{i+1}$ (see Figure 1). The node $u$ (node $v$) is said to be *transmitting* (*receiving*) over the link $(u, v)$. We consider a *unicast problem* as follows. The node $s \triangleq v_0$, called *source*, generates a vector of messages; the node $t \triangleq v_l$, called *sink* (*receiver*), demands the vector of messages generated at node $s$. The rest of the nodes $\{v_i : 0 < i < l\}$ in the network are called *interior*, and are responsible for relaying the messages from the source to the sink.

*Field and Vector Space:* Suppose that node $s$ is given a set $\mathcal{M}$ of $k$ messages $\{\boldsymbol{x}_i : 1 \leq i \leq k\}$. The messages are each drawn from an $L$-dimensional vector space $\mathcal{F} \ (= \mathbb{F}^L)$ over a finite field $\mathbb{F}$. The index $i$ is called the *label* of $\boldsymbol{x}_i$. We call a vector $\boldsymbol{y} \in \mathcal{F}$ a *packet*. We denote the set of labels of the message packets by $\mathcal{M}_s$.[2]

*Schedule over the Network:* Suppose that following a certain timing schedule, node $u$ transmits a packet over the link $(u, v)$ in any opportunity that it gets. The links may be lossy, i.e., a packet which is sent by a node $u$ may not reach the receiving node $v$. In this case, the packet is called an *erased packet*, otherwise, it is referred to as a *successful packet*. We assume that the erasure rates of different links may be different and may also be time-varying. In addition, the links are assumed to have arbitrary time-varying delays. Thus, the successful packets might be re-ordered at the receiving end of a link. We also assume that the network nodes have infinite memory, i.e., no received packet is discarded over time by any receiving node. Moreover, we consider a scenario where network nodes are blind to the schedule of transmission of successful packets (called *schedule*), i.e., no feedback information is available.

*Graphical Representation of a Schedule:* We use a digraph, called *trellis connectivity graph* (or *trellis* for brevity), to represent a given schedule (see Figure 2). The trellis only represents the successful transmissions through the network. The node set of the trellis includes the nodes $u_\tau$ for each network node $u$ so that a successful packet either departs from or arrives at the node $u$ at time $\tau$, and two extra nodes $s_0$ and $t_\infty$.[3] The edge set of the trellis consists of two groups of directed edges specified as follows; (i) *traffic edges*: every 2-tuple $(u_\tau, v_{\tau'})$, $\tau \leq \tau'$, representing an *in-edge* (*out-edge*) of node $v$ (node $u$), for every pair of nodes $u_\tau$ and $v_{\tau'}$, if there is at least one packet sent by node $u$ at time $\tau$ and received by node $v$ at time $\tau'$, and (ii) *memory edges*: every 2-tuple $(u_\tau, u_{\tau'})$, $\tau \leq \tau'$, if, for all $\tau'' \in (\tau, \tau')$, there is no node $u_{\tau''}$.

Without loss of generality, the traffic edges are assumed to have unit capacity, i.e., only one packet is successfully sent over a traffic edge, as parallel traffic edges are allowed. Also, the memory edges are assumed to have infinite capacity, i.e., at any given time, a network node has access to all its successfully received packets.

---

[2] The finite field $\mathbb{F}$ is defined based on two operations "addition" and "multiplication," respectively represented by symbols "+" and "·". We assume that both operations have the same cost, as one field operation. We also define two types of operations in a vector space $\mathcal{F}$: (i) $\boldsymbol{y} + \boldsymbol{z}$, for $\boldsymbol{y}, \boldsymbol{z} \in \mathcal{F}$, is taken to be symbol-wise with respect to operator $+$ in $\mathbb{F}$, and requires one packet operation, and (ii) $\lambda \boldsymbol{y}$, for $\lambda \in \mathbb{F}$ and $\boldsymbol{y} \in \mathcal{F}$, symbol-wise with respect to operator . in $\mathbb{F}$, and also requires one packet operation.

[3] The node $s_0$ represents the node $s$ at time zero when all the message packets are available, and the node $t_\infty$ represents the node $t$ at a time after which there is no more coded packet to arrive.

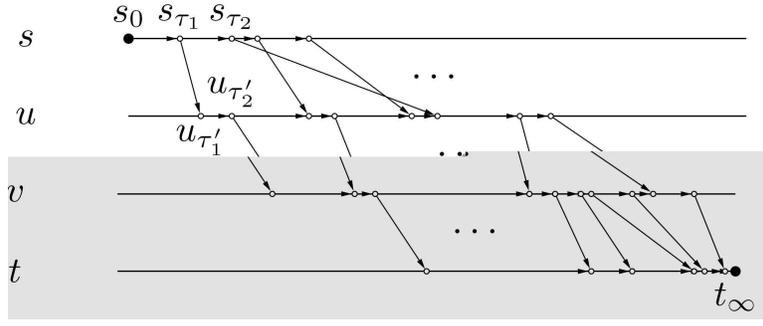

Fig. 2. A trellis for a schedule over a line network of length $l = 3$. The edges $(s_0, s_{\tau_1})$, $(s_{\tau_1}, s_{\tau_2})$, and $(u_{\tau_1'}, u_{\tau_2'})$ are examples of memory edges and the edge $(s_{\tau_1}, u_{\tau_1'})$ is an example of a traffic edge.

*Capacity of Schedule:* The maximum number of message packets that can be successfully sent through a network with a given schedule is called the *capacity of network under the schedule* or simply the *capacity of the schedule*.

Modeling a schedule as a flow network, the capacity of the schedule equals the maximum flow between the nodes $s_0$ and $t_\infty$. By the max-flow min-cut theorem, the capacity of a schedule equals the minimum of the sum of the capacities of the cutset edges among all the cutsets in the trellis. This quantity is called the *min-cut capacity*.

The min-cut capacity is achievable if the network nodes are able to (properly) process their received packets and generate new packets to be sent. When processing is allowed at the nodes, the information flow scheme is called *network coding*, and the min-cut capacity is therefore often referred to as the *network coding capacity*.

When the network nodes are only allowed to generate coded packets by linearly combining their received packets, the network coding scheme is called *linear*. When network codes are restricted to be linear, the maximum number of message packets that can be sent through a network with a given schedule is called the *linear coding capacity*. We focus on linear network codes in this work and for brevity, hereafter, we refer to the linear coding capacity as the *capacity*. It should be noted that the linear coding capacity is equal to the *routing capacity* for the unicast scenario if the schedule is known at the network nodes. This capacity is equal to the number of (traffic) edge-disjoint paths between the source and the sink. In the absence of the knowledge of the schedule at the network nodes, which is the case in this work, however, routing does not achieve this capacity.

In the following, to analyze schedules of a given capacity $n$, we shall focus on schedules in which there are $n$ edge-disjoint paths in the trellis starting from $s_0$ and ending at $t_\infty$.

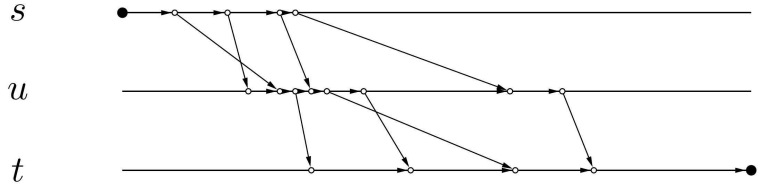

Fig. 3. An example of a network of length 2 under a schedule of capacity 4 with exactly 4 sent and/or received packets at any network node.

Let $\mathcal{I}_{u_\tau}$ ($\mathcal{O}_{u_\tau}$) be the set of in-edges (out-edges) of node $u$ prior to time $\tau$. Clearly $\mathcal{I}_{s_\tau} = \mathcal{O}_{t_\tau} = \emptyset$, for all $\tau \in [0, \infty)$. We label the edges in $\mathcal{I}_{u_\tau}$ ($\mathcal{O}_{u_\tau}$), so that the first in-edge (out-edge) at node $u$ has label 1, the second has label 2, and so forth. With a slight abuse of notation, we also use the notation $\mathcal{I}_{u_\tau}$ ($\mathcal{O}_{u_\tau}$) for the set of labels of edges in $\mathcal{I}_{u_\tau}$ ($\mathcal{O}_{u_\tau}$).

In this work, we are interested in worst-case schedules. We thus assume a given network under a schedule of capacity $n$, so that (i) for every interior node $u$, $|\mathcal{I}_{u_\infty}| = |\mathcal{O}_{u_\infty}| = n$, i.e., precisely $n$ packets are sent and received by node $u$, and (ii) for any time $\tau$, $|\mathcal{O}_{u_\tau}| \leq |\mathcal{I}_{u_\tau}|$, i.e., the number of successful transmissions at each interior node $u$ at any time $\tau$ does not exceed the number of successful receptions at node $u$. A schedule satisfying both conditions is shown in Figure 3.

Condition (i) corresponds to having the minimum number of successful transmissions that can support a schedule of capacity $n$. Condition (ii) ensures that under the constraint of Condition (i), the schedule has $n$ edge-disjoint paths.

*Coding over the Schedule:* In this work, we restrict the codes to be *linear*. A coded packet $\boldsymbol{y}_e$ to be sent over the out-edge $e$ of node $u_\tau$ (or $s_\tau$) is generated by $\sum_{i \in \mathcal{I}_{u_\tau}} \lambda_{e,i} \boldsymbol{y}_i$ (or $\sum_{j \in \mathcal{M}} \mu_{e,j} \boldsymbol{x}_j$), where $\lambda_{e,i}, \mu_{e,j} \in \mathbb{F}$, for all $i \in \mathcal{I}_{u_\tau}$, and all $j \in \mathcal{M}_s$, and $\boldsymbol{y}_i \in \mathcal{F}$ is the packet received along the in-edge $i$ of node $u$, for all $i \in \mathcal{I}_{u_\tau}$. For all $\boldsymbol{x} \in \mathcal{M}$, $\hat{\boldsymbol{x}}$ is estimated by $\sum_{i \in \mathcal{I}_{t_\infty}} \lambda_{\boldsymbol{x},i} \boldsymbol{y}_i$, where $\lambda_{\boldsymbol{x},i} \in \mathbb{F}$, for all $i \in \mathcal{I}_{t_\infty}$, and $\boldsymbol{y}_i \in \mathcal{F}$ is the packet received along the in-edge $i$ of node $t$.

We refer to a method of generating codes as a *coding scheme*, and associated with a coding scheme is a *class of codes* generated by the coding scheme.

We say that a code in a class $\mathcal{C}$ of codes over a vector space $\mathcal{F}$ fails over a given network with a schedule of capacity $n$ if node $t$ fails to recover all the $k$ message packets $\{\boldsymbol{x} \in \mathcal{M}\}$ from the set of $n$ packets $\{\boldsymbol{v}_e : e \in \mathcal{I}_{t_\infty}\}$; otherwise, the code is said to succeed. The ratio $k/n$ is referred to as the *code rate*. The probability that a randomly selected code in $\mathcal{C}$ fails is referred to as the probability of failure of the class $\mathcal{C}$ of codes, and it is denoted by $\epsilon_{k,n}$.

Let $k_n$ be a function of $n$ ($k_n \leq n$), denoting the number of message packets at the source node, and let

$\epsilon_{k_n,n}$ ($0 \leq \epsilon_{k_n,n} \leq 1$) be a function of $n$ (and $k_n$). We say that a coding scheme achieves the capacity or approaches the capacity with gap $\lambda$ over a given network with any schedule of capacity $n$, for $0 < \lambda \leq 1$, if there exists a sequence of codes of rates $\{k_n/n\}$ so that the sequence of failure probabilities $\{\epsilon_{k_n,n}\}$ goes to $\epsilon$, for some $0 \leq \epsilon \leq 1$, arbitrarily small, and the sequence of rates goes to 1 or $1/(1+\lambda)$, as $n$ goes to infinity.

Let $k_{n_{\max}}$ be the largest function of $n$, so that for a given $0 \leq \epsilon \leq 1$, with probability of failure no larger than $\epsilon$, a given coding scheme achieves/approaches the capacity over a network of any schedule of a given capacity $n$. For any given $n$, the larger is the ratio $k_{n_{\max}}/n$, the larger is said to be the speed of convergence of a coding scheme. That is, for a given $n$, with a given probability of failure, a coding scheme with a larger speed of convergence is able to transmit more message packets over any schedule of capacity $n$.

*Complexity of Codes:* The number of packet operations for applying the encoding functions divided by $k \cdot l$ is called the *encoding cost*, (i.e., we add up the number of packet operations needed to generate *all* the coded packets at *all* the non-sink nodes, and normalize it by the number of message packets multiplied by the number of links). The number of packet operations for applying the decoding functions divided by $k$ is called the *decoding cost*.[4]

### B. Problem Formulation

Suppose a network of length $l$ under an arbitrary schedule with capacity $n$. For any given coding scheme, our goal is to derive tight upper bounds on (i) the number of message packets ($k$) drawn from a vector space $\mathcal{F}$ (over $\mathbb{F}_2$)[5] at node $s$, as a function of $n, l$ and/or $\lambda$, in the asymptotic regime (i.e., as $n$ goes to infinity), so that the coding scheme over the vector space $\mathcal{F}$ fails to achieve or approach the capacity with a given gap $0 < \lambda \leq 1$, with probability no larger than $0 \leq \epsilon \leq 1$, and (ii) the encoding and decoding costs.

---

[4]The coding costs exclude field operations that are independent of the size of the packets (i.e., $L$). The reason is that $L$ is usually very large in practice and the computations dealing with packet operations dominate the computational cost.

[5]We restrict the finite field $\mathbb{F}$ to the binary field $\mathbb{F}_2$ (i.e., each message packet is a stream of $L$ bits). The reason for this restriction is two-fold: (i) to have the lowest computational complexity, and (ii) to consider the case with the lowest speed of convergence (the larger is the field, the larger would be the speed of convergence of any coding scheme considered in this paper, but at the cost of increasing the computational complexity).

# III. Capacity-Achieving Codes

We start with the analysis of *random linear codes* (a.k.a. *dense codes*)[6] over line networks of length $l$, with arbitrary schedules of a given capacity $n$, as $n$ goes to infinity.

We address the following issues: (i) a dense code over a vector space $\mathbb{F}_2^L$ (for any integer $L$) is *capacity-achieving*, and (ii) the encoding and decoding costs are each $O(n)$.

## A. Dense Codes

*Encoding:* For every out-edge $e$ of node $u_\tau$, $\boldsymbol{y}_e = \sum_{i \in \mathcal{I}_{u_\infty}} \lambda_{e,i} \boldsymbol{y}_i$ if $u$ is interior, and $\boldsymbol{y}_e = \sum_{j \in \mathcal{M}_s} \mu_{e,j} \boldsymbol{x}_j$ if $u$ is the source, where for all $i \in \mathcal{I}_{u_\infty} \setminus \mathcal{I}_{u_\tau}$, $\lambda_{e,i}$ is 0, and for all $i \in \mathcal{I}_{u_\tau}$, and all $j \in \mathcal{M}_s$, $\lambda_{e,i}$ and $\mu_{e,j}$ are symbols independently and uniformly drawn from $\mathbb{F}_2$.

Since every coded packet is a result of linearly combining message packets, every coded packet $\boldsymbol{y}_e$ over an out-edge $e$ of node $u_\tau$, for an interior node $u$, can be written as

$$
\begin{aligned}
\boldsymbol{y}_e &= \sum_{i \in \mathcal{I}_{u_\infty}} \lambda_{e,i} \sum_{j \in \mathcal{M}_s} \mu_{i,j} \boldsymbol{x}_j \\
&= \sum_{j \in \mathcal{M}_s} \sum_{i \in \mathcal{I}_{u_\infty}} \lambda_{e,i} \mu_{i,j} \boldsymbol{x}_j \\
&= \sum_{j \in \mathcal{M}_s} \mu_{e,j} \boldsymbol{x}_j,
\end{aligned}
$$

where $\mu_{e,j} := \sum_{i \in \mathcal{I}_{u_\infty}} \lambda_{e,i} \mu_{i,j}$, and $\mu_{i,j}$'s can be defined recursively.

The vector $\boldsymbol{\lambda}_e$ of $n$ elements $\{\lambda_{e,i} : i \in \mathcal{I}_{u_\infty}\}$ is called the *local encoding vector* of packet $\boldsymbol{y}_e$. The vector $\boldsymbol{\mu}_e$ of $k$ elements $\{\mu_{e,j} : j \in \mathcal{M}_s\}$ is called the *global encoding vector* of packet $\boldsymbol{y}_e$. The global encoding vector of a packet is sent along with the packet in its *header*.[7]

Let $\mathcal{N}_v$ be a subset of $\mathcal{I}_{v_\infty}$. We say that a collection $\{\boldsymbol{y}_e : e \in \mathcal{N}_v\}$ of packets at a receiving node $v$ is *innovative* if their global encoding vectors are *linearly independent*. We refer to such a collection of packets by its set of in-edges $\mathcal{N}_v$.

*Decoding:* For every $\boldsymbol{x} \in \mathcal{M}$, node $t$ provides an estimate $\hat{\boldsymbol{x}} = \sum_{i \in \mathcal{I}_{t_\infty}} \lambda_{\boldsymbol{x},i} \boldsymbol{y}_i$ by solving the system of linear equations $\{\boldsymbol{y}_e = \sum_{j \in \mathcal{M}_s} \mu_{e,j} \hat{\boldsymbol{x}}_j : e \in \mathcal{I}_{t_\infty}\}$ for $k$ packets $\{\hat{\boldsymbol{x}}_j : j \in \mathcal{M}_s\}$.

---

[6]In a random linear coding scheme as explained later in detail, each packet sent by a node is a random linear combination of *all* previously received packets. Thus the number of non-zero coefficients in linear combinations is rather large, resulting in *dense* linear combinations.

[7]A dense code is set up to never transmit a coded packet with all-zero global encoding vector. Whenever such a packet is generated by a node, it will be discarded and a new packet will be generated till the resulting packet has a global encoding vector with at least one non-zero entry.

This system is uniquely solvable and for all $j \in \mathcal{M}_s$, $\hat{\boldsymbol{x}}_j$ equals $\boldsymbol{x}_j$ if there is an innovative collection $\mathcal{N}_t$ at node $t$ such that $|\mathcal{N}_t| = k$. Thus, a dense code succeeds if node $t$ receives a collection of $k$ packets which form an innovative collection.

Let $Q_v$ be a matrix of size $n \times k$ whose rows are global encoding vectors of packets received by a node $v$, i.e., for all $e \in \mathcal{I}_{v_\infty}$, and for all $j \in \mathcal{M}_s$, $(Q_v)_{e,j} = \mu_{e,j}$. We call $Q_v$ the *decoding matrix* at node $v$; if node $v$ needs to recover the message packets, it has to solve a system of linear equations whose unknowns, known constants and coefficients are the message packets, the received packets at node $v$ and their global encoding vectors, respectively. This can be done, e.g., using Gauss-Jordan elimination [18].

*Analysis [Outline]:* The packets over the first link are random linear combinations of the message packets and their global encoding vectors' entries are all independently uniformly distributed (i.u.d.) Bernoulli random variables. Suppose that the receiving node of the first link in the network is to recover the $k$ message packets after receiving $n$ packets. This can be done so long as there exist $k$ packets with linearly independent global encoding vectors. Now, a question is to find the probability of existence of such a collection of packets at the node. A lower bound on the probability that a set of $n$ vectors of length $k$ whose entries are i.u.d. Bernoulli random variables includes $k$ linearly independent vectors is well-known, see, e.g., [19, Proposition 2].

The packets over any lower link are also each a random linear combination of the packets previously received over the upper link. The entries of the global encoding vectors of these packets are uniformly distributed Bernoulli random variables but not necessarily independent. To the best of our knowledge, there is no non-trivial lower bound on the probability that a set of $n$ such vectors of length $k$ (whose entries are *all* uniform Bernoulli but not *all* independent) includes $k$ linearly independent vectors.

In the following, we show that the entries of the global encoding vectors of any collection of packets at any given node are i.u.d. Bernoulli random variables so long as the packets' local encoding vectors are linearly independent. This implies that the global encoding vector of any packet over any lower link can be written as a linear combination of those global encoding vectors over the upper link whose entries are *all* i.u.d. The main idea in our analysis is therefore to track the distribution of the size of a maximal collection of linearly independent local encoding vectors at each node.

The number of packets with linearly independent local encoding vectors in general depends on the schedule. In this work, however, the schedule is arbitrary. We hence study the two extremal categories of worst-case schedules in which the probability that the local encoding vector of any packet sent by any network node is linearly independent to those of the packets sent earlier by the node is the "largest" or the "smallest" possible, respectively.

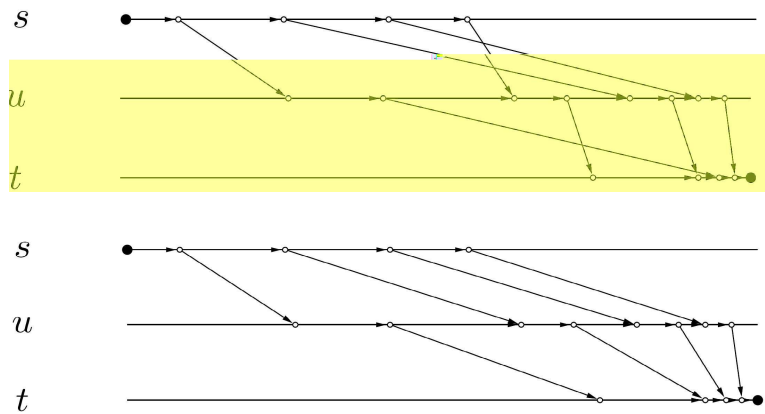

Fig. 4. Two examples of a network of length 2 with one-in-one-out schedules of capacity 4 (with and without re-ordering of received packets).

The first category of worst-case schedules consists of those in which any network node successfully transmits one and only one packet between any two contiguous arrivals. In this case, the smallest number of received packets are available to contribute in generating any packet to be transmitted (see Figure 4); and thus the local encoding vector of any given packet is linearly dependent on those of the rest of the packets with the largest probability. We refer to such schedules as *one-in-one-out*.

The second category of worst-case schedules consists of those in which any network node transmits its first successful packet after receiving all the $n$ packets over the upper link (see Figure 5). We refer to such schedules as *all-in-all-out*.

The local encoding vectors not only depend on the schedule but also are random. The number of linearly independent local encoding vectors over any given link is therefore a random variable. In the following, we derive probabilistic lower bounds on the value of this random variable for each type of worst-case schedules of interest. Taking a union bound over the number of links, we then provide a lower bound on the number of packets with i.u.d. Bernoulli entries over the last link. Finally, by applying a lower bound on the probability of existence of $k$ linearly independent vectors in a set of vectors with i.u.d. Bernoulli entries, we are able to derive a lower bound on the probability that the decoding is successful at the sink node in each case.

*Analysis [Details]:* Every node combines its received packets in a random fashion. Intuitively, the likelihood of linear dependence among the global encoding vectors of the packets increases as they travel down the links from node $s$ to node $t$.

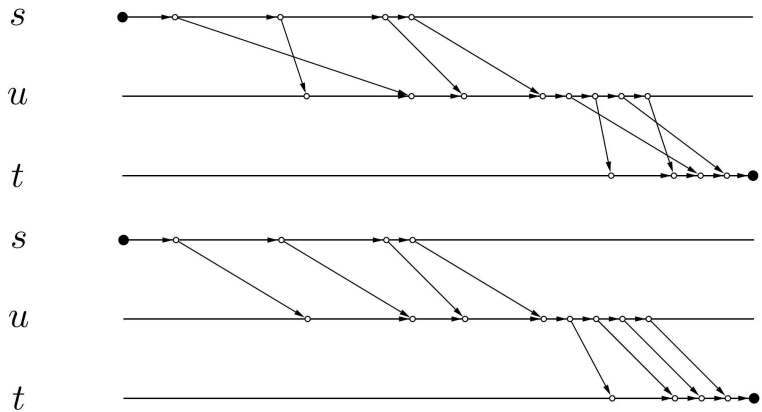

Fig. 5. Two examples of a network of length 2 with all-in-all-out schedules of capacity 4 (with and without re-ordering of received packets).

In the following, we derive a lower bound on the number of received packets $|\mathcal{I}_{v_\infty}|$ by any receiving node $v$ among which with a given probability, $k$ packets are innovative. (There exists an innovative collection $\mathcal{N}_v$ at node $v$ such that $|\mathcal{N}_v| = k$).

Over an edge $e = (u_\tau, v_{\tau'})$, a packet $\boldsymbol{y}_e$ is a function of $\{\mu_{e,j} : j \in \mathcal{M}_s\}$ and for all $j \in \mathcal{M}_s$, $\mu_{e,j}$ is itself a function of $\{\lambda_{e,i}\mu_{i,j} : i \in \mathcal{I}_{u_\tau}\}$. For all $i \in \mathcal{I}_{u_\tau}$, $\lambda_{e,i}$'s are i.u.d., and for all $j \in \mathcal{M}_s$, $\mu_{i,j}$'s are known at node $u$, and no longer random. Thus, for all $j \in \mathcal{M}_s$, $\mu_{e,j}$'s, and thus the entries of $Q_v$ are uniformly but not necessarily independently distributed. Let $r(Q_v)$ represent the *rank* of the matrix $Q_v$ over $\mathbb{F}_2$. Clearly, $r(Q_v)$ is a random variable. We are interested in finding lower bounds on the probability that $r(Q_v) = k$, i.e., the probability of receiving an innovative collection of size $k$ of packets by any receiving node $v$, when $n$ packets are received at the node. While finding such lower bounds is rather simple where $Q_v$'s entries are all i.u.d., the same cannot be said about the cases where $Q_v$'s entries are not necessarily i.u.d.. Our approach is thus to derive the bounds by just focusing on a subset of the packets in each node for which the entries of the global encoding vectors are i.u.d. To perform this, we remove a minimal collection of rows in $Q_v$, so that the remaining rows all have i.u.d. entries. We denote the set of remaining rows by $Q'_v$; a sub-matrix of $Q_v$. Clearly, $r(Q'_v) \leq r(Q_v)$.

Let $Q$ be an $n \times k$ matrix over $\mathbb{F}_2$. A maximal collection of rows in $Q$ with i.u.d. entries is called *dense*, and $Q$ is called a *dense matrix* if all its rows form a dense collection. We refer to the number of rows in a dense collection of rows in $Q$ as the *density* of $Q$, denoted by $\mathcal{D}(Q)$, and refer to each row in this collection as a *dense row*.

Let $Q_v$ be a decoding matrix of size $n \times k$ (at a receiving node $v$). The set of in-edges of node $v$ pertaining to the $Q_v$'s dense rows is denoted by $\mathcal{D}_v$; and $Q_v$ restricted to its $\mathcal{D}(Q_v)$ dense rows is denoted by $Q'_v$ of size $\mathcal{D}(Q_v) \times k$. The packets whose global encoding vectors are in the set of dense rows are called *dense packets*.

It should be clear that any packet sent by node $s$ is dense, i.e., $\mathcal{D}(Q_{v_1}) = n$. The density of the decoding matrices at the other nodes further down in the sequence of network nodes (including the sink) may however be less than $n$.

The collection of rows of $Q_v$ which are not in the collection of dense rows is denoted by a sub-matrix $\mathcal{P}_v$ of $Q_v$; i.e., the rows of $\mathcal{P}_v$ are $\{\boldsymbol{\mu}_e : e \in \mathcal{I}_{v_\infty} \setminus \mathcal{D}_v\}$. The minimal collection of rows of $Q_v \setminus \mathcal{P}_v$ whose removal would create a sub-matrix with linearly independent rows is denoted by a sub-matrix $\mathcal{P}'_v$ of $Q_v$; i.e., the rows of $\mathcal{P}'_v$ are $\{\boldsymbol{\mu}_e : e \in \mathcal{D}_v \setminus \mathcal{N}_v\}$, where $\mathcal{N}_v$ is a maximal innovative collection of packets in $\mathcal{D}_v$. When it causes no confusion, we adopt the same notation $\mathcal{P}_v$ (or $\mathcal{P}'_v$) for the set of in-edges pertaining to the rows in $\mathcal{P}_v$ (or $\mathcal{P}'_v$).

We denote the maximal innovative collection of packets in $\mathcal{I}_{v_\infty}$ by $\mathcal{I}_v$. The number $|\mathcal{I}_{v_\infty} \setminus \mathcal{I}_v|$ of non-innovative packets at every receiving node $v$ can be bounded above by the sum of the number of rows in $\mathcal{P}_v$ and $\mathcal{P}'_v$, i.e., $|\mathcal{I}_{v_\infty} \setminus \mathcal{D}_v| + |\mathcal{D}_v \setminus \mathcal{N}_v|$. The number of rows in $\mathcal{P}_v$ and $\mathcal{P}'_v$ are random variables and we shall give upper-bounds on $|\mathcal{I}_{v_\infty} \setminus \mathcal{D}_v|$ and $|\mathcal{D}_v \setminus \mathcal{N}_v|$ which hold with a given probability.

For two adjacent nodes $u$ and $v$ connected by the link $(u, v)$ in the network, we have $Q_v = T_u Q_u$, where $T_u$ is an $n \times n$ matrix over $\mathbb{F}_2$, whose rows are local encoding vectors $\{\boldsymbol{\lambda}_e : e \in \mathcal{I}_{u_\infty}\}$. For all $k \in \mathcal{I}_{u_\infty}$, $(T_u)_{e,k} = \lambda_{e,k}$. We call $T_u$ the *transfer matrix* at node $u$.

The following lemmas provide a lower bound on $\mathcal{D}(Q_v) = \mathcal{D}(T_u Q_u)$. The proofs are given in Appendix I.

*Lemma 1:* Let $\boldsymbol{u}$ be a column-vector of length $d$ whose entries are independently and uniformly drawn from $\mathbb{F}_2$, and $T$ be any $h \times d$ ($h \le d$) matrix over $\mathbb{F}_2$ so that $r(T) = h$. Then, the entries of $T\boldsymbol{u}$ are i.u.d. random variables over $\mathbb{F}_2$.

It follows from Lemma 1 that a set of linear combinations of i.u.d. random variables over $\mathbb{F}_2$ are i.u.d., so long as the coefficient vectors of the linear combinations are linearly independent.

*Lemma 2:* Let $M$ be a $d \times k$ ($k \le d$) dense matrix whose entries are drawn from $\mathbb{F}_2$, and let $T$ be a matrix over $\mathbb{F}_2$ with $d$ columns. Suppose that $r(T) \ge h$. Then $\mathcal{D}(TM) \ge h$.

Lemma 2 is applicable if the decoding matrix $Q_u$, at the transmitting node $u$, is dense, i.e., $\mathcal{D}(Q_u) = n$. The density of $Q_u$ might however be less than $n$, i.e., $Q_u$ might not be dense.

*Remark 1:* Lemma 2 implies that if at a transmitting node $u$, the transfer matrix $T_u$ is not full row-rank,

at the receiving node $v$, there are rows in the decoding matrix $Q_v$ which are not dense. This means that a packet sent by node $u$ would not be in the collection of dense packets at node $v$ if its local encoding vector, and thus its global encoding vector, is linearly dependent to those of dense packets at node $v$.

Suppose that $\mathcal{D}(Q_u) = d$, or equivalently, $|\mathcal{D}_u| = d$. We rewrite $T_u Q_u$ with respect to $Q'_u$ being $Q_u$ restricted to its dense rows. This results in $Q_v = T'_u Q'_u$, where $Q'_u$ is a dense matrix of size $d \times k$, and $T'_u$ is a matrix of size $n \times d$, so that for all $e \in \mathcal{O}_{u_\infty}$, and all $k \in \mathcal{D}_u$, $(T'_u)_{e,k} = \lambda_{e,k} + \sum_{i \in \mathcal{I}_{u_\infty} \backslash \mathcal{D}_u} \lambda_{e,i} \lambda_{i,k}$. We call $T'_u$ the *modified transfer matrix* at node $u$.

Every row of $T_u$ is the local encoding vector of a packet transmitted by node $u$, and every entry of a local encoding vector is either zero or chosen independently and uniformly at random from $\mathbb{F}_2$. Thus, the entries of $T_u$ are each either zero or an i.u.d. random variable. For each $i \in \mathcal{O}_{u_\infty}$, and each $j \in \mathcal{I}_{u_\infty}$, let the sets $\mathcal{T}^{(i)}_{u_{\text{row}}}$ and $\mathcal{T}^{(j)}_{u_{\text{col}}}$ denote the label sets of i.u.d. entries in the $i$th row and the $j$th column of $T_u$, respectively.

For worst-case schedules, there are at least $i$ packets received by an interior node $u$ by the time that the $i$th coded packet is to be transmitted. Thus, for all $i \in \mathcal{O}_{u_\infty}$, $|\mathcal{T}^{(i)}_{u_{\text{row}}}| \geq i$, and in particular the first $i$ entries of the $i$th row of $T_u$ are i.u.d. random variables. It can also be seen that for all $j \in \mathcal{I}_{u_\infty}$, $|\mathcal{T}^{(j)}_{u_{\text{col}}}| \geq n-j+1$, and in particular the last $n-j+1$ entries of the $j$th column of $T_u$ are i.u.d. random variables.

We now consider the modified transfer matrix $T'_u$. The $i$th row of $T'_u$ is representing the labels of dense packets received by node $u$ which are contributing to generate the $i$th coded packet to be transmitted. The $j$th column of $T'_u$ is also representing the labels of coded packets in which the $j$th dense packet at node $u$ is contributing. For all $e \in \mathcal{O}_{u_\infty}$, and all $k \in \mathcal{D}_u$, $(T'_u)_{e,k}$ is either zero or an i.u.d. random variable.

For each $i \in \mathcal{O}_{u_\infty}$, and each $j \in \mathcal{D}_u$, we use the notations $\mathcal{T}'^{(i)}_{u_{\text{row}}}$ and $\mathcal{T}'^{(j)}_{u_{\text{col}}}$ to denote the label sets of i.u.d. entries in the $i$th row and the $j$th column of $T'_u$, respectively.

One can see that for all $i \in \mathcal{O}_{u_\infty}$, $|\mathcal{T}'^{(i)}_{u_{\text{row}}}| \geq [i-n+d]^+$, and in particular the first $[i-n+d]^+$ entries of the $i$th row of $T'_u$ are i.u.d.[8] Also, for all $j \in \mathcal{I}_{u_\infty}$, $|\mathcal{T}'^{(j)}_{u_{\text{col}}}| \geq d-j+1$, and in particular the last $d-j+1$ entries of the $j$th column of $T'_u$ are i.u.d..

In what follows, we derive lower bounds on the rank of $T'_u$. It is worth noting that the rank property of such a matrix is highly dependent on the schedule. Thus the analysis of one-in-one-out and all-in-all-out worst-case schedules are given separately in the following.

---

[8] For every integer $x$, $[x]^+ = x$, if $x \geq 0$, and $[x]^+ = 0$, otherwise.

*One-In-One-Out Worst-Case Schedules:* In this setting, for all $i \in \mathcal{O}_{u_\infty}$, $|\mathcal{T}u_{\text{row}}^{(i)}| = i$, and for all $j \in \mathcal{I}_{u_\infty}$, $|\mathcal{T}_{u_{\text{col}}}^{(j)}| = n - j + 1$. Also, for all $i \in \mathcal{O}_{u_\infty}$, $|\mathcal{T'}_{u_{\text{row}}}^{(i)}| \geq (i - n + d)^+$, and for all $j \in \mathcal{I}_{u_\infty}$, $|\mathcal{T'}_{u_{\text{col}}}^{(j)}| \geq d - j + 1$.[9]

*Lemma 3:* Let $T$ be an $n \times d$ ($d \leq n$) matrix over $\mathbb{F}_2$, so that for any $1 \leq j \leq d$, at least $d - j + 1$ entries of its $j$th column are i.u.d. random variables. The rest of the entries are set to zero. For any integer $0 \leq \gamma \leq d - 1$, then

$$\Pr[r(T) < d - \gamma] \leq (d - \gamma)2^{-(\gamma + 1)}.$$

The proof of Lemma 3 is given in Appendix I.

Lemma 3 is applicable to $T'_u$, and since $Q_v = T'_u Q'_u$, and $Q'_u$ is dense, Lemma 2 gives $\mathcal{D}(Q_v) \geq h$, for every $h$ such that $r(T'_u) \geq h$. Combining Lemmas 2 and 3, we can hence give a lower bound on $\mathcal{D}(Q_v)$.

*Lemma 4:* For any $\epsilon > 0$, applying a dense coding scheme over a network with any one-in-one-out worst-case schedule of capacity $n$, for any network link $(u, v)$, the inequality

$$\mathcal{D}(Q_v) \geq \mathcal{D}(Q_u) - \log \mathcal{D}(Q_u) - \log(1/\epsilon)$$

fails with probability (w.p.) bounded above by (b.a.b.) $\epsilon$.

The proof of Lemma 4 is given in Appendix I.

For any network link $(u, v)$, $\mathcal{D}(Q_u) - \mathcal{D}(Q_v)$ is called *density loss* at node $v$, or over the link $(u, v)$. Lemma 4 gives an upper bound on the density loss at node $v$ with respect to $\mathcal{D}(Q_u)$, i.e., w.p. bounded below by (b.b.b.) $1 - \epsilon$, $\mathcal{D}(Q_u) - \mathcal{D}(Q_v) \leq \log \mathcal{D}(Q_u) + \log(1/\epsilon)$.

The density of the decoding matrix at node $t$, $\mathcal{D}(Q_t)$, can be bounded from below by subtracting the density losses over the network links from the density of the decoding matrix at the first receiving node. The proof of the following is given in Appendix I.

*Lemma 5:* For any $\epsilon > 0$, applying a dense code over a network of length $l$ with any one-in-one-out worst-case schedule of capacity $n$,

$$\mathcal{D}(Q_t) \geq n - l \log(nl/\epsilon)$$

w.p. b.b.b. $1 - \epsilon$.

Lemma 5 provides an upper bound on the number of rows in $\mathcal{P}_t$, as $|\mathcal{I}_{t_\infty} \setminus \mathcal{D}_t| = n - \mathcal{D}(Q_t)$.

Now, let $Q'_t$ be a sub-matrix of $Q_t$, so that $Q'_t$ includes all $\mathcal{D}(Q_t)$ dense rows in $Q_t$. By Lemma 5, the probability of $\mathcal{D}(Q_t) < k$ is upper bounded by $\epsilon$, if $k \leq n - l \log(nl/\epsilon)$. Thus, for every $k$ satisfying this inequality, w.p. no larger than $\epsilon$, $Q'_t$ fails to be a $\mathcal{D}(Q_t) \times k$ ($k \leq \mathcal{D}(Q_t)$) dense matrix. Finally, the following lemma gives an upper bound on the probability that a dense matrix fails to have rank $k$ (see Appendix I for the proof).

---

[9]The results for the matrices $T$ and $T'$ are consistent, in that the results for $T$ are the special case of the results for $T'$, where $d = n$.

*Lemma 6:* Let $M$ be a $d \times k$ ($k \leq d$) dense matrix over $\mathbb{F}_2$. Then, for every $\epsilon > 0$,

$$\Pr[r(M) < k] \leq \epsilon,$$

so long as $k \leq d - \log(1/\epsilon)$.

Lemma 5 together with Lemma 6 give an upper bound on the total number of packets at node $t$ not belonging to a maximal innovative collection of packets, i.e., $|\mathcal{I}_{t_\infty} \setminus \mathcal{I}_t|$, by respectively providing upper bounds on the number of rows in $\mathcal{P}_t$ and $\mathcal{P}'_t$, i.e., $|\mathcal{I}_{t_\infty} \setminus \mathcal{D}_t|$ and $|\mathcal{D}_t \setminus \mathcal{N}_t|$. This is used in the proof of the following theorem.

*Theorem 1:* For any $\epsilon > 0$, a dense code over a vector space $\mathbb{F}_2^L$, for any integer $L$, fails to be capacity-achieving for a network of length $l$ with $k$ message packets under any one-in-one-out schedule of capacity $n$, w.p. no larger than $\epsilon$, so long as

$$k \leq n - l \log(nl/\epsilon) - \log(1/\epsilon) - l - 1,$$

and $l \log(l/\epsilon) = o(n)$.

*Proof:* The decoding fails if the number of innovative packets at node $t$ is less than $k$. Let $\acute{\epsilon}$ denote $\epsilon/2$, for the simplicity of exposition. Replacing $\epsilon$ with $\acute{\epsilon}$ in Lemma 5, the number of dense packets at node $t$ is less than $n - l \log(nl/\acute{\epsilon})$ w.p. b.a.b. $\acute{\epsilon}$. Given that at least $n - l \log(nl/\acute{\epsilon})$ dense packets are received by node $t$, based on Lemma 6, there is less than $k$ innovative packets among the set of dense packets w.p. b.a.b. $\acute{\epsilon}$, so long as $k \leq n - l \log(nl/\epsilon) - \log(1/\epsilon) - l - 1$. Hence, the probability of decoding failure is b.a.b. $\epsilon$. Thus, for this scenario, a dense code is capacity-achieving, i.e., the code rate $k/n$ goes to 1, as $n$ goes to infinity, so long as $l \log(l/\epsilon) = o(n)$. ∎

*All-In-All-Out Worst-Case Schedules:* In all-in-all-out worst-case schedules, for all $i \in \mathcal{O}_{u_\infty}$, $|\mathcal{T}_{u_{\text{row}}}^{(i)}| = n$, and for all $j \in \mathcal{I}_{u_\infty}$, $|\mathcal{T}_{u_{\text{col}}}^{(j)}| = n$. Similarly, for all $i \in \mathcal{O}_{u_\infty}$, $|\mathcal{T'}_{u_{\text{row}}}^{(i)}| = d$, and for all $j \in \mathcal{I}_{u_\infty}$, $|\mathcal{T'}_{u_{\text{col}}}^{(j)}| = n$. Since these conditions on $|\mathcal{T'}_{u_{\text{row}}}^{(i)}|$ and $|\mathcal{T'}_{u_{\text{col}}}^{(j)}|$ satisfy those required in Lemma 3, one can use the result of Lemma 3 to upper bound $\Pr[r(T') < d - \gamma]$, for any integer $0 \leq \gamma \leq d-1$. We, however, derive a tighter bound for this setting in Lemma 7.[10] The proof of the following lemmas can be found in Appendix I.

*Lemma 7:* Let $T$ be a $n \times d$ ($d \leq n$) dense matrix over $\mathbb{F}_2$. For any integer $1 \leq \gamma \leq d - 1$,

$$\Pr[r(T) < d - \gamma] \leq 2^{-\gamma}.$$

We, now, give a lower bound on $\mathcal{D}(Q_v) = \mathcal{D}(T'_u Q'_u)$ by using Lemmas 2 and 7.

*Lemma 8:* For any $\epsilon > 0$, applying a dense coding scheme over a network with any all-in-all-out worst-case schedule of capacity $n$, for any network link $(u, v)$, the inequality

$$\mathcal{D}(Q_v) \geq \mathcal{D}(Q_u) - \log(1/\epsilon)$$

fails to hold w.p. b.a.b. $\epsilon$.

Taking a union bound over the number of links in the network, and subtracting the density losses over the network links from the density of the decoding matrix at the first receiving node, a lower bound can be given on the density of the decoding matrix at node $t$, i.e., $\mathcal{D}(Q_t)$. The proof of the following is similar to that of Lemma 5, except that we use the result of Lemma 8 instead of Lemma 4.

*Lemma 9:* For any $\epsilon > 0$, applying a dense code over a network of length $l$ with any all-in-all-out worst-case schedule of capacity $n$,

$$\mathcal{D}(Q_t) \geq n - l \log(l/\epsilon)$$

w.p. b.b.b. $1 - \epsilon$.

Applying Lemmas 6 and 9 yields the following main result (The proof is similar to that of Theorem 1, and is thus omitted).

*Theorem 2:* For any $\epsilon > 0$, a dense code over a vector space $\mathbb{F}_2^L$, for any integer $L$, fails to be capacity-achieving for a network of length $l$ with $k$ message packets under any all-in-all-out worst-case schedule of capacity $n$, w.p. no larger than $\epsilon$, so long as

$$k \leq n - l \log(l/\epsilon) - \log(1/\epsilon) - l - 1,$$

and $l \log(l/\epsilon) = o(n)$.

---

[10]In the case of $\gamma = 0$, $\Pr[r(T) < d - \gamma]$ can also be bounded by Lemma 6; in this case, the result of Lemma 7 is not as tight as that of Lemma 6. However, Lemma 6 cannot be generalized to the cases of $0 < \gamma \leq d - 1$, and hence Lemma 7 can be considered as a complement to Lemma 6.

*Coding Costs:* The worst case with regards to the encoding cost at any interior node $u$ (node $s$) occurs if for any time $\tau$ at which node $u$ transmits a packet, $|\mathcal{I}_{u_\tau}| = n$, and for all $i \in \mathcal{I}_{u_\tau}$, $\lambda_{e,i}$'s (for all $j \in \mathcal{M}_s$, $\mu_{e,j}$'s) are chosen to be nonzero. Thus the number of packet operations for encoding at any interior node (the source node) is $O(n^2)$ ($O(kn)$). Thus the encoding cost is $O(n)$.[11]

To solve the system of linear equations at node $t$ using Gauss-Jordan elimination, node $t$ requires $O(wn)$ row operations, where $w$ is the widest bandwidth of a row vector of $Q_t$. Note that a row operation is equivalent to at most $O(k)$ field operations along with at most one packet operation.[12] Thus node $t$ requires $O(wn)$ packet operations together with $O(wkn)$ field operations [9].

The worst case regarding the decoding cost at node $t$ occurs if $w$ equals $k$. Thus the number of operations for decoding is $O(k^2 n)$ field operations and $O(kn)$ packet operations. Thus the decoding cost is also $O(n)$.

*Theorem 3:* The encoding and decoding costs of a dense coding scheme over a line network with any worst-case schedule of capacity $n$ are $O(n)$.

We have shown that the class of dense codes is capacity-achieving over networks with arbitrary schedules (Theorems 1 and 2) but at the cost of large computational complexity (Theorem 3).

To reduce the coding costs, the method of generating coded packets has to be modified so that the global encoding vectors have smaller bandwidth. The smaller is the bandwidth of the global encoding vector of a packet, however, the smaller is its randomness and the larger is its probability to be linearly dependent on the global encoding vectors of the other packets sharing the same band. Thus the probability of a packet being innovative may become smaller in general. The problem is therefore to design coding schemes in which every global encoding vector has a small bandwidth but the bands are set up in a way to compensate for the reduction in the randomness of the global encoding vectors.

To have global encoding vectors with smaller bandwidth, the general approach is to apply a dense code to a *chunk*, a smaller sub-message of the original message. In fact, for the so-called chunked codes (CC) [8], the set of message packets is partitioned into chunks of equal size, and each chunk is transmitted by a dense code. In a general context, the design of a chunk-based code has to deal with the following issues: (i) how to divide the message packets into the chunks at the source node, (ii) how to schedule the chunks to be coded and transmitted by the network nodes, and (iii) how to recover each chunk at the sink node.

---

[11]The number of packet operations for encoding is at most $(l-1)n^2 + nk$, and hence the encoding cost is at most $(l-1)n^2/kl + n/l = O(n)$, since $n/k = O(1)$, as $n$ goes to infinity.

[12]The narrowest window (in an end-around fashion) within which the non-zero entries of a vector lie is called *band*. The length of a band is called *bandwidth*.

In the following, we review the CC scheme of [8], which provably performs well over the networks with arbitrary schedules.

## B. Chunked Codes

Suppose that node $s$ is given a set $\mathcal{M}$ of $k$ message packets $\{\boldsymbol{x}_i : 1 \leq i \leq k\}$, each of which is a vector in vector space $\mathbb{F}_2^L$. Let $\mathcal{M}_s$ be the set of labels of packets in $\mathcal{M}$.

A CC operates by dividing the set $\mathcal{M}$ into $q$ disjoint subsets $\{\mathcal{M}_\omega : \omega \in [q]\}$, called *chunks*, each of size $k/q$ (i.e., $|\mathcal{M}_\omega| = k/q$), so that $\mathcal{M}_\omega$ includes the last $k/q$ contiguous message packets whose labels are equal to or less than $\omega k/q$. The *size* of chunks is referred to as the *aperture size*.

*Encoding:* Every transmitting node $u$ at any time $\tau$, randomly chooses a chunk, say $\omega$, and constructs a coded packet, called an $\omega$-*packet*, by randomly linearly combining all the $\omega$-packets received earlier by node $u$ and transmits it over an out-edge of node $u_\tau$, i.e., each chunk is transmitted by using a dense code.

Let $\mathcal{I}_{u_\tau}^{(\omega)}$ ($\mathcal{O}_{u_\tau}^{(\omega)}$) be the set of (labels of) in-edges (out-edges) of node $u$ prior to time $\tau$ over which $\omega$-packets are received (sent), and let $\mathcal{M}_s^{(\omega)}$ be the set of labels of message packets in chunk $\omega$.

Therefore, for every out-edge $e$ of node $u_\tau$, a chunk is randomly chosen, say $\omega$, and an $\omega$-packet $\boldsymbol{y}_e$ is sent so that $\boldsymbol{y}_e = \sum_{i \in \mathcal{I}_{u_\infty}^{(\omega)}} \lambda_{e,i} \boldsymbol{y}_i$ if $u$ is interior, and $\boldsymbol{y}_e = \sum_{j \in \mathcal{M}_s^{(\omega)}} \mu_{e,j} \boldsymbol{x}_j$ if $u$ is the source, where for all $i \in \mathcal{I}_{u_\infty}^{(\omega)} \setminus \mathcal{I}_{u_\tau}^{(\omega)}$, $\lambda_{e,i}$ is 0; and for all $i \in \mathcal{I}_{u_\tau}^{(\omega)}$, and all $j \in \mathcal{M}_s^{(\omega)}$, $\lambda_{e,i}$ and $\mu_{e,j}$ are symbols independently and uniformly drawn from $\mathbb{F}_2$.

Let $\mathcal{N}_v$ be a subset of $\mathcal{I}_{v_\infty}$. Suppose a collection $\{\boldsymbol{y}_e : e \in \mathcal{N}_v\}$ of packets at a receiving node $v$. We refer to such a collection of packets by its set of in-edges $\mathcal{N}_v$. Let $\mathcal{N}_v^{(\omega)}$ be the set of all $\omega$-packets in $\mathcal{N}_v$. We say that a collection $\mathcal{N}_v^{(\omega)}$ (or $\mathcal{N}_v$) is $\omega$-*innovative* (or *innovative*) if the global encoding vectors of the packets therein are linearly independent. Since for all $j \in \mathcal{M}_s \setminus \mathcal{M}_s^{(\omega)}$, the $j$th entry of an $\omega$-packet's global encoding vector is 0, $\mathcal{N}_v$ is innovative iff $\mathcal{N}_v^{(\omega)}$ is $\omega$-innovative, for every $\omega$ such that $\mathcal{N}_v$ contains at least one $\omega$-packet.

*Decoding:* For every $\omega$, node $t$ has to solve a system of linear equations $\{\boldsymbol{y}_e = \sum_{j \in \mathcal{M}_s^{(\omega)}} \mu_{e,j} \hat{\boldsymbol{x}}_j : e \in \mathcal{I}_{t_\infty}^{(\omega)}\}$ for $k/q$ packets $\{\hat{\boldsymbol{x}}_j : j \in \mathcal{M}_s^{(\omega)}\}$. The system is uniquely solvable for every $\omega$, and for all $j \in \mathcal{M}_s^{(\omega)}$, $\hat{\boldsymbol{x}}_j$ equals $\boldsymbol{x}_j$ if there exists an $\omega$-innovative collection $\mathcal{N}_t^{(\omega)}$ at node $t$ such that $|\mathcal{N}_t^{(\omega)}| = k/q$. Thus, a CC succeeds if for every $\omega$, node $t$ receives a collection of $k/q$ $\omega$-packets which form an $\omega$-innovative collection.

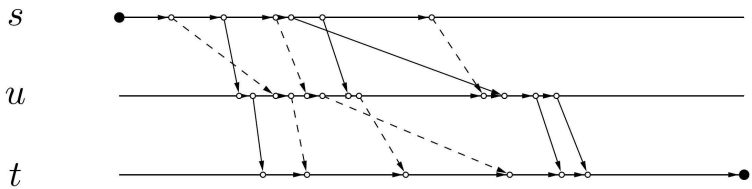

Fig. 6. An example of a network of length 2 with a schedule of (linear coding) capacity 6. Consider a CC with two chunks $\omega_1$, and $\omega_2$. The solid (dashed) lines are the edges over which $\omega_1$-packets ($\omega_2$-packets) are sent. The capacity of the flow by $\omega_1$-packets is 3, and that by $\omega_2$-packets is 2.

Let $Q_v$ be the decoding matrix at a receiving node $v$. For every $\omega$, let $Q_v^{(\omega)}$, called the $\omega$-*decoding matrix* at node $v$, be a $|\mathcal{I}_{v_\infty}^{(\omega)}| \times |\mathcal{M}_s^{(\omega)}|$ sub-matrix of $Q_v$ whose rows are the global encoding vectors of $\omega$-packets at node $v$, and whose columns are labeled with $\mathcal{M}_s^{(\omega)}$.

There exists an innovative collection of $k$ packets at a receiving node $v$, iff, for every $\omega$, there exists a collection of $k/q$ linearly independent rows in $Q_v^{(\omega)}$.

Suppose that for every $\omega$, an $\omega$-innovative collection $\mathcal{N}_t^{(\omega)}$ of $k/q$ $\omega$-packets is received by node $t$ (i.e., $r(Q_t^{(\omega)}) = k/q$). For every $\omega$, node $t$ applies Gauss-Jordan elimination to the extended matrix $[Q_t^{(\omega)}|Y_\omega]$ and after removing all-zero rows, it returns a new matrix $[I|Z_\omega]$, where $Y_\omega$ is the $|\mathcal{I}_{t_\infty}^{(\omega)}| \times 1$ vector of $\omega$-packets at node $t$, $I$ is a $k/q \times k/q$ identity matrix and $Z_\omega$ is a $k/q \times 1$ vector each of whose entries is a vector in $\mathbb{F}_2^L$. For all $j \in \mathcal{M}_s^{(\omega)}$, $\hat{x}_j$ equals the $j$th entry of vector $Z_\omega$.

*Analysis:* We need to derive a lower bound on the probability of receiving an $\omega$-innovative collection of $k/q$ $\omega$-packets by the sink node $t$, for every $\omega$.

Since for every $\omega$, chunk $\omega$ is of size $k/q$ and is transmitted by a dense code, we can use the result of Theorems 1 and 2 by replacing $k$ and $n$, respectively, with $k/q$ and the capacity of the schedule pertaining to the $\omega$-packets. We refer to such a schedule as $\omega$-*schedule*. The capacity of the $\omega$-schedule is a random variable. This capacity is equal to the number of paths between $s_0$ and $t_\infty$ whose (traffic) edges, which carry $\omega$-packets, are disjoint. Such paths are referred to as *flow paths* (see Figure 6).

To deal with the randomness of $\omega$-schedules, we derive an upper bound on the number of $\omega$-packets received by any interior node that cannot be "matched up" with an $\omega$-packet which is transmitted subsequently by the node and is not yet coupled with an $\omega$-packet received earlier by the node. Those $\omega$-packets received by the node whose (traffic) edges do not contribute in a maximal collection of flow paths are called *unusable*, and the rest of $\omega$-packets are called *usable*. The term "unusable" reflects the fact that these packets are not part of the flow paths that contribute to the capacity of the $\omega$-schedule.

*One-In-One-Out Worst-Case Schedules:* For this scenario, we take a union bound over interior nodes by subtracting the number of unusable $\omega$-packets received by every node from the number of usable $\omega$-packets sent by node $s$. This allows us to derive a lower bound on the number of usable $\omega$-packets at node $t$ yielding a lower bound on the capacity of the $\omega$-schedule.[13]

*Lemma 10:* For a CC with $q$ chunks, over a network of length $l$ with any one-in-one-out worst-case schedule of capacity $n$, for every $\omega$, the capacity of the $\omega$-schedule is less than

$$\varphi = \left(1 - O\left(\left(l^3(q/n)\ln(ln/\epsilon)\right)^{1/3}\right)\right).(n/q) \tag{1}$$

w.p. b.a.b. $\epsilon$, so long as

$$l^3 q \ln \frac{ln}{\epsilon} = o\left(n\right). \tag{2}$$

*Proof:* Please refer to Appendix II. ∎

We, now, derive a lower bound on the capacity of schedule, $n$, so that with a given probability, for every $\omega$, there is an innovative collection of $k/q$ $\omega$-packets at node $t$. For this, we use Theorem 1. To modify the result of Theorem 1 to be applicable to the transmission of a given chunk $\omega$ over a line network by a dense code, we replace $k$ and $n$ with $k/q$ and $\varphi$, respectively. Then the following is immediate.

*Lemma 11:* For any $\epsilon > 0$, in a CC with $q$ chunks, a given chunk $\omega$ (of size $k/q$) fails to be successfully sent through a line network of length $l$ with any one-in-one-out worst-case schedule of capacity $n$, by using a dense code, w.p. no larger than $\epsilon/q$, so long as

$$k/q \leq \varphi - l\log(nl/\epsilon) - \log(q/\epsilon) - l - 1,$$

and $l^3 q \ln(nl/\epsilon) = o(n)$, where $\varphi$ is the capacity of the $\omega$-schedule.

The decoding at node $t$ is successful if and only if every chunk can be decoded. Taking a union bound over the number of chunks ($q$), we upper bound the number of message packets ($k$) at node $s$, such that with a given probability, for every chunk $\omega$, there exist at least $n/q - O\left((l^3(n/q)^2 \ln(ln/\epsilon))^{1/3}\right)$ $\omega$-packets at node $t$, such that $k/q$ $\omega$-packets form an $\omega$-innovative collection. This yields the following for CC over networks with one-in-one-out worst-case schedules.

*Theorem 4:* For any $\epsilon > 0$, a CC with $q$ chunks, over any vector space $\mathbb{F}_2^L$, is capacity-achieving for a network of length $l$ with $k$ message packets under any one-in-one-out worst-case schedule of capacity $n$,

---

[13]One should note that Lemma 10 is different from [8, Theorem 4.1] in the sense that the lower bound derived here is tighter, though the proofs have generally a similar structure.

with failure probability no larger than $\epsilon$, so long as

$$k \leq q\varphi - ql \log(nl/\acute{\epsilon}) - q \log(q/\acute{\epsilon}) - ql - q,$$

and $l^3 q \ln(nl/\epsilon) = o(n)$, where

$$\varphi = n/q - O\left((l^3(n/q)^2 \ln(ln/\epsilon))^{1/3}\right).$$

*Proof:* Replacing $\epsilon$ with $\acute{\epsilon} = \epsilon/2$ in Lemmas 10 and 11, we obtain (i) for a chunk $\omega$, the capacity of the $\omega$-schedule fails to be at least $n/q - O\left((l^3(n/q)^2 \ln(ln/\epsilon))^{1/3}\right)$, w.p. b.a.b. $\acute{\epsilon}$, so long as $ql^3 \ln(nl/\epsilon) = o(n)$; and (ii) given that every chunk $\omega$ has been allocated a flow with capacity $\varphi$ of at least $n/q - O\left((l^3(n/q)^2 \ln(ln/\epsilon))^{1/3}\right)$, a dense code fails to transmit the chunk $\omega$ successfully over the network w.p. b.a.b. $\acute{\epsilon}$, so long as $k \leq \varphi q - ql \log(nl/\acute{\epsilon}) - q \log(q/\acute{\epsilon}) - ql - q$. Thus, the probability of failure, when either (i), or (ii) occurs, is b.a.b. $\epsilon$. ∎

*All-In-All-Out Worst-Case Schedules:* By the nature of the all-in-all-out schedule, it is easy to see that the capacity of an $\omega$-schedule is equal to the minimum number of $\omega$-packets transmitted over all the links in the network. Using this, we obtain the following lower bound on the capacity of an $\omega$-schedule.

*Lemma 12:* For a CC with $q$ chunks, over a network of length $l$ with any all-in-all-out worst-case schedule of capacity $n$, for every $\omega$, $\omega$-packets fail to form a flow of capacity $\varphi$ larger than

$$\left(1 - O\left(((q/n) \ln(lq/\epsilon))^{1/2}\right)\right).(n/q) \tag{3}$$

w.p. b.a.b. $\epsilon$, so long as

$$q \ln \frac{lq}{\epsilon} = o(n). \tag{4}$$

*Proof:* Please refer to Appendix III. ∎

The proofs of the following results are similar to that of Lemma 11 and Theorem 4, except that in this case, we use Lemma 12 and Theorem 2 instead of Lemma 10 and Theorem 1, respectively.

*Lemma 13:* For any $\epsilon > 0$, in a CC with $q$ chunks, a given chunk $\omega$ (of size $k/q$) fails to be successfully transmitted through a line network of length $l$ with any all-in-all-out worst-case schedule of capacity $n$, given that the $\omega$-packets form a flow of capacity $\varphi$ larger than $n/q - O\left(((n/q) \ln(lq/\epsilon))^{1/2}\right)$, by using a dense code, w.p. no larger than $\epsilon/q$, so long as

$$k/q \leq \varphi - l \log(lq/\epsilon) - \log(q/\epsilon) - l - 1,$$

and $q \ln(lq/\epsilon) = o(n)$.

*Theorem 5:* For any $\epsilon > 0$, a CC with $q$ chunks, over any vector space $\mathbb{F}_2^L$, is capacity-achieving for a network of length $l$ with $k$ message packets under any all-in-all-out worst-case schedule of capacity $n$, with failure probability no larger than $\epsilon$, so long as

$$k \le q\varphi - ql\log(lq/\dot\epsilon) - q\log(q/\dot\epsilon) - ql - q,$$

and $lq \ln(lq/\epsilon) = o(n)$, where

$$\varphi = n/q - O\left(((n/q)\ln(lq/\epsilon))^{1/2}\right).$$

*Coding Costs:* The worst-case with regards to the encoding cost occurs if for every $\omega$, any transmitting interior node $u$ (or the source node $s$) sends all its successful $\omega$-packets after receiving all $\omega$-packets sent by the node in the upper layer, and if all the local encoding coefficients are chosen to be non-zero. In this case, the number of packet operations for encoding at any interior node $u$ (or node $s$) is $O(n^2/q)$ (or $O(kn/q)$). (There are $O(n/q)$ $\omega$-packets sent and received by any interior node $u$ and there are $O(n/q)$ $\omega$-packets sent by node $s$.) Thus the encoding cost is $O(n/q)$.[14]

To solve the system of linear equations $\{\boldsymbol{y}_e = \sum_{j \in \mathcal{M}_s^{(\omega)}} \mu_{e,j} \hat{\boldsymbol{x}}_j : e \in \mathcal{I}_{t_\infty}^{(\omega)}\}$ for $k/q$ packets $\{\hat{\boldsymbol{x}}_j : j \in \mathcal{M}_s^{(\omega)}\}$ using Gauss-Jordan elimination, node $t$ requires $O(kn/q)$ row operations for every $\omega$ (the bandwidth of a row in an $\omega$-decoding matrix is at most $k/q$). Thus, node $t$ requires $O(k^2n/q^3)$ field operations along with $O(kn/q^2)$ packet operations (the $Q_t^{(\omega)}$'s bandwidth and size are $k/q$ and $O(n/q) \times k/q$, respectively). The total number of operations for decoding at node $t$ is $O(k^2n/q^2)$ field operations and $O(kn/q)$ packet operations, and thus the decoding cost is $O(n/q)$.

*Theorem 6:* The encoding and decoding costs of a CC with $q$ chunks are $O(n/q)$.

## C. Comparison of Dense Codes and Chunked Codes: Making the Case for Overlapping Chunks

By comparing the results of Theorems 1 and 2 with Theorems 4 and 5, respectively, one can observe that, for each type of worst-case schedules, the overhead per message, $(n-k)/k$, for CC is larger by a factor of $q$ compared to dense codes. This is the price that one has to pay for the lower coding complexity of CC. The question now is whether the tradeoff between the speed of convergence and coding complexity of CC can be improved.

---

[14]The number of packet operations for encoding is $O((l-1)n^2/q) + O(kn/q)$, and hence the encoding cost is $O((l-1)n^2/qkl) + O(n/ql) = O(n/q)$, since $n/k = O(1)$, as $n$ goes to infinity.

The analysis of CC over networks with arbitrary schedules shows that the performance of CC is affected by (i) the capacity of the $\omega$-schedule for every $\omega$; (ii) the number of innovative packets in the collection of dense packets, and (iii) the condition of decoding completion, i.e., for all $\omega$, the $\omega$-decoding matrix $Q_t^{(\omega)}$ at node $t$ needs to have $k/q$ linearly independent rows (i.e., $r(Q_t^{(\omega)}) = k/q$).

In the scenario of interest in this work, where the network nodes are blind to the schedule, and there is no feedback information available to the nodes, random coding appears to be the best strategy as far as issues (i) and (ii) are concerned. There is however room for improvement in the performance of CC by modifying the chunking scheme to speed up the decoding process. The main idea is to devise a chunking scheme such that for every $\omega$, $r(Q_t^{(\omega)})$ does not necessarily have to be $k/q$ for $r(Q_t) = k$. We demonstrate that this can be achieved by allowing chunks to overlap.

To explain the idea, we start by the simple case of a single "erasure channel" with an arbitrary schedule. It is important to note that in this case, the lack of interior nodes significantly simplifies the analysis as one does not need to consider density losses over the network links, i.e., all the received packets at node $t$ are dense.[15] The following results are simple to prove.

*Theorem 7:* For any $\epsilon > 0$, a dense code over a vector space $\mathbb{F}_2^L$, is capacity-achieving over an erasure channel with $k$ message packets under any schedule of capacity $n$, with failure probability no larger than $\epsilon$, so long as $k \leq n - \log(1/\epsilon)$ and $\log(1/\epsilon) = o(k)$. The encoding and decoding costs are $O(n)$.

*Theorem 8:* For any $\epsilon > 0$, a CC with $q$ chunks, over any vector space $\mathbb{F}_2^L$, is capacity-achieving over an erasure channel with $k$ message packets under any schedule of capacity $n$, with failure probability no larger than $\epsilon$, so long as $k \leq q\varphi - q\log(q/\epsilon) - q$, and $q\log(q/\epsilon) = o(n)$, where $\varphi = n/q - O\left(((n/q)\ln(q/\epsilon))^{1/2}\right)$. The encoding and decoding costs are $O(n/q)$.

The comparison of the results of the two theorems shows that for a single erasure channel also, CC have a slower convergence to capacity than dense codes. This is the cost for having a lower complexity by a factor of $q$.

Recently, Studholme and Blake [9] introduced a class of erasure codes called *windowed erasure codes*, which is similar to CC except that for windowed erasure codes, the chunks are allowed to overlap. The codes are used in [9] to deliver $k$ message packets over an erasure channel with any schedule of capacity $n$. To perform this task, windowed erasure codes operate on $k$ chunks of aperture size $\alpha$, where any two contiguous chunks overlap in all but one message packet in an end-around fashion. Similar to CC, in

---

windowed erasure codes, the chunks are scheduled at random, i.e., the source node at any time instant randomly chooses a chunk, constructs and transmits a coded packet by randomly linearly combining the message packets in the chosen chunk. The decoding of windowed erasure codes is, however, similar to that of dense codes, not that of CC, i.e., the sink node uses Gauss-Jordan elimination to solve the system of linear equations for all the chunks together.

The following theorem shows that windowed erasure codes can achieve the capacity of erasure channels with arbitrary schedules at the same speed as dense codes.

*Theorem 9:* For any $\epsilon > 0$, a windowed erasure code with any aperture size $\alpha \geq 2\sqrt{k}$, over a vector space $\mathbb{F}_2^L$, is capacity-achieving over an erasure channel with $k$ message packets under any schedule of capacity $n$, with failure probability no larger than $\epsilon$, so long as $k \leq n - \log(1/\epsilon)$, and $\log(1/\epsilon) = o(k)$. The encoding and decoding costs are $O(\alpha)$.

Prior to proving Theorem 9, we introduce two categories of structured random matrices.

*Banded Random Binary Matrices:* Let $n, k, \alpha$ and $\gamma$ be integers ($k \leq n$, $\gamma < \alpha$), so that $\alpha - \gamma$ is a divisor of $k$. Let $\chi$ be $k/(\alpha - \gamma)$, and $I$ be the set of integers in $[k]$. We divide $I$ into $\chi$ subsets $I_i$'s, for all $i \in [\chi]$, where $I_i$ (the $i$th *aperture* of size $\alpha$) is the set of $\alpha$ contiguous integers in $I$ in an end-around fashion, starting from $(i-1)(\alpha - \gamma) + 1$.

We construct an $n \times k$ matrix as follows: (i) for each row, an index, say $i$, is randomly chosen from the set of integers $[\chi]$, and (ii) the row's entries indexed by the $i$th aperture are independently and uniformly chosen from $\mathbb{F}_2$, and the rest of the entries are set to zero. We call such a matrix a $(\gamma, \alpha)$ *irregular symmetric banded matrix* of size $n \times k$. Now, consider a similar construction, with the difference that $\alpha - \gamma$ is now a divisor of $k - \gamma$ (not $k$), and $\chi$ is $(k - \gamma)/(\alpha - \gamma)$ (not $k/(\alpha - \gamma)$). The resulting matrix in this case is called a $(\gamma, \alpha)$ *irregular asymmetric banded matrix* of size $n \times k$. Also, consider a matrix constructed as each of the above two procedures, except that in part (i), each index in $[\chi]$ is assigned to $n/\chi$ rows ($\chi$ has to be a divisor of $n$). We call such a matrix a $(\gamma, \alpha)$ "regular" symmetric or asymmetric banded matrix of size $n \times k$.

*Proof of Theorem 9:* Each packet received by node $t$ pertains to a randomly chosen chunk. Each chunk contains a set of $\alpha$ contiguous indices in an end-around fashion from the set of integers $[k]$. Each packet's global encoding vector has random entries in the positions indexed by the aperture pertaining to the chosen chunk, and the rest of the entries are zero. Thus, the decoding matrix $Q_t$ at node $t$ is an $(\alpha - 1, \alpha)$ symmetric banded random binary matrix of size $n \times k$.

The decoding at node $t$ will be successful if $Q_t$ has rank $k$. The following result which is a short version of [9, Conjecture 4.2] is useful to upper bound the probability that $Q_t$ fails to have rank $k$, i.e., $\Pr[r(Q_t) < k]$.

*Conjecture 1:* Let $M$ be an $(\alpha - 1, \alpha)$ (regular/irregular) symmetric banded random binary matrix of size $n \times k$ ($k \leq n$). For any $\epsilon > 0$, and for sufficiently large $k$, $\Pr[r(M) < k] \leq \epsilon$, so long as

$$k \leq n - \log(1/\epsilon),$$

and $\alpha \geq 2\sqrt{k}$.

By Conjecture 1, the probability of failure of a windowed erasure code is b.a.b. $\epsilon$, i.e., $\Pr[r(Q_t) < k] \leq \epsilon$, so long as $k \leq n - \log(1/\epsilon)$. By definition, a coding scheme is capacity-achieving if the ratio $k_{n_{\max}}/n$ goes to 1, as $n$ goes to infinity. Here, $k_{n_{\max}} = n - \log(1/\epsilon)$, and $k_{n_{\max}}/n$ goes to 1, as $n$ goes to infinity, so long as $\log(1/\epsilon) = o(n)$. This completes the proof of the first part of the theorem. We prove the second part regarding the coding costs in two steps. The proof that the encoding cost is $O(\alpha)$ is similar to that of Theorem 6. To prove the decoding cost, it suffices to recall that applying Gauss-Jordan elimination to the matrix $Q_t$ of bandwidth of $\alpha$, node $t$ requires at most $O(\alpha n)$ row (or packet) operations. Thus the decoding cost is $O(\alpha)$, since $n/k = O(1)$, as $n$ goes to infinity. ∎

The comparison of the results of Theorem 9 with those of Theorem 7 indicates that for the transmission over a single erasure channel, windowed erasure codes with sufficiently large apertures achieve the capacity at the same speed as dense codes do but with lower complexity. This motivates the application of chunked codes with overlapping chunks, referred to as *overlapped chunked codes* (OCC), to the problem of information transmission over erasure networks with arbitrary schedules.

### D. Overlapped Chunked Codes

Suppose that node $s$ is given a set $\mathcal{M}$ of $k$ message packets $\{\boldsymbol{x}_i : 1 \leq i \leq k\}$, each of which is a vector in $\mathbb{F}_2^L$, for any integer $L$.

An OCC operates by dividing the set $\mathcal{M}$ into $q$ overlapping chunks $\{\mathcal{M}_\omega : \omega \in [q]\}$, each of size $\alpha$ (i.e., $|\mathcal{M}_\omega| = \alpha$), so that any two contiguous chunks overlap by $\gamma = \alpha - k/q$ message packets in an end-around fashion. The set of labels of the message packets in a given chunk $\omega$, $\mathcal{M}_s^\omega$ is called the *aperture* of chunk $\omega$.

To ensure that all the message packets appear in the same number of chunks, $(\alpha - \gamma)$ must be a divisor of $\alpha$. For the simplicity of exposition, we consider $\gamma = \alpha/\tau_e$, where $\tau_e = \tau/(\tau - 1)$, for any divisor $\tau$ of $\alpha$ ($1 \leq \tau \leq \alpha$).[16] For instance, there is no overlap between chunks when $\tau$ is equal to 1; and the overlap becomes larger as $\tau$ increases; namely, when $\tau$ is equal to $\alpha$, any two contiguous chunks overlap in all but one message packet. We call $\tau$ the *overlap parameter*.

*Encoding:* The encoding is performed similar to CC; i.e., every transmitting node $u$ at any time instant $\tau'$ chooses a chunk at random, say $\omega$; and constructs an $\omega$-packet, by randomly linearly combining the $\omega$-packets already received by node $u$ and transmits it over an out-edge of node $u_{\tau'}$.

It is important to note that for OCC, unlike CC, the $\omega$-packets for different values of $\omega$ with overlapping apertures, are not necessarily linearly independent.

*Decoding:* The decoding is performed similar to dense codes; i.e., node $t$ has to solve a system of linear equations $\{\boldsymbol{y}_e = \sum_{j \in \mathcal{M}_s} \mu_{e,j} \hat{\boldsymbol{x}}_j : e \in \mathcal{I}_{t\infty}\}$ for $k$ packets $\{\hat{\boldsymbol{x}}_j : j \in \mathcal{M}_s\}$. The system is uniquely solvable, and for all $j \in \mathcal{M}_s$, $\hat{\boldsymbol{x}}_j$ is equal to $\boldsymbol{x}_j$, if there exists an innovative collection of $k$ packets at node $t$.[17]

Suppose that a collection of $k$ innovative packets are received by node $t$, i.e., $r(Q_t) = k$. To solve the system of linear equations node $t$ applies Gaussian elimination to the extended matrix $[Q_t|Y]$ and after removing all-zero rows it returns a new matrix $[I|Z]$, where $Y$ is the $n \times 1$ vector of the packets at node $t$, $I$ is a $k \times k$ identity matrix and $Z$ is a $k \times 1$ vector each of whose entries is a vector in $\mathbb{F}_2^L$. For all $j \in \mathcal{M}_s$, $\hat{\boldsymbol{x}}_j$ is equal to the $j$th entry of vector $Z$.

*Analysis:* Similar to CC, we analyze OCC over the two extremal types of worst-case schedules separately.

*One-In-One-Out Worst-Case Schedules:* The chunks are scheduled in OCC similar to CC. Thus, for every $\omega$, w.p. b.a.b. $\epsilon$, the capacity of the flow by $\omega$-packets ($\omega$-schedule) fails to be larger than the lower bound given in Lemma 10. However, for successful decoding, unlike CC, in OCC, a given chunk $\omega$ does not have to be necessarily recoverable in isolation (i.e., $r(Q_t^{(\omega)})$ does not need to be $\alpha$). This is because the decoding is performed on the set of all the packets received by node $t$. The goal is to derive an upper bound on $k$, such that $r(Q_t)$ equals $k$, w.h.p., so long as $n$ packets are received by node $t$.

We first lower bound the total number of dense $\omega$-packets, given that $n$ packets are received by node

---

[16] The parameter $\tau$, defined as the overlap parameter here, should not be mistaken with the same notation for the transmission time instances used earlier.

[17] It is worth noting that in prior related works, the chunks are to be decoded in isolation. However, by performing the decoding algorithm on the set of all the chunks simultaneously, the decoding of the OCC may be successful even when none of the chunks are recoverable in isolation. Thus, a smaller number of packets at the sink node is sufficient to ensure successful decoding with a given probability of success.

$t$, so that, for all $\omega$, the lower bound fails to hold w.p. b.a.b. $\epsilon$.[18] We then derive an upper bound on $k$, so that, w.h.p., for all $\omega$, there exists an innovative collection of $k$ packets among the union set of maximal collections of dense $\omega$-packets.

*Lemma 14:* For any $\epsilon > 0$, applying a dense code to a given chunk $\omega$ over a network of length $l$ with any one-in-one-out worst-case schedule of capacity $n$, given that the capacity of the flow by $\omega$-packets is $\varphi$, the number of dense $\omega$-packets at node $t$ is less than $\varphi - l \log(\varphi l/\epsilon)$ w.p. b.a.b. $\epsilon$.

*Proof:* This is a direct result of Lemma 5, by replacing $n$ with $\varphi$. ∎

*Lemma 15:* Let $l, n, q, \varphi$ and $\epsilon$ be defined as above. Then the node $t$ fails to receive at least $\varphi - l \log(q\varphi l/\acute{\epsilon})$ dense $\omega$-packets, for every $\omega$, w.p. b.a.b. $\acute{\epsilon}$, and $\varphi$ fails to be at least

$$\left(1 - O\left(\left(l^3(q/n)\ln(\ln/\epsilon)\right)^{1/3}\right)\right).(n/q),$$

w.p. b.a.b. $\acute{\epsilon}$, so long as

$$l^3 q \ln \frac{\ln}{\epsilon} = o\left(n\right). \tag{5}$$

*Proof:* Replacing $\epsilon$ with $\acute{\epsilon}/q$ in Lemma 14, for taking a union bound over $q$, the first part of the lemma follows. Replacing $\epsilon$ with $\acute{\epsilon}$ in Lemma 10 proves the second part of the lemma. ∎

Thus, so long as $l^3 q \ln(\ln/\epsilon) = o(n)$, w.p. b.b.b. $1 - \acute{\epsilon}$, there exist more than $n/q - O\left((l^3(n/q)^2 \ln(\ln/\epsilon))^{1/3}\right)$ $\omega$-packets at node $t$, for every $\omega$, and hence there will be more than $n/q - O\left((l^3(n/q)^2 \ln(\ln/\epsilon))^{1/3}\right) - l \log(nl/\acute{\epsilon})$ dense $\omega$-packets at node $t$ w.p. b.b.b. $1 - \epsilon$.

Let $Q_t'$ be $Q_t$ restricted to its rows pertaining to the dense $\omega$-packets, for all $\omega$. Thus, $Q_t'$ is of size $h \times k$, for some $h$ smaller than $n - O\left(q(l^3(n/q)^2 \ln(\ln/\epsilon))^{1/3}\right) - ql \log(nl/\acute{\epsilon})$, w.p. b.a.b. $\epsilon$. Now, the problem is to derive an upper bound on the probability that $Q_t'$ fails to have rank $k$. By the structure of OCC, it should be clear that $Q_t'$ is a symmetric banded random binary matrix with aperture size $\alpha$, where the overlap size between any two rows pertaining to any two contiguous chunks is $\gamma$. Lemma 15 shows that the number of dense $\omega$-packets, for every $\omega$, fails to be larger than $h$, w.p. b.a.b. $\acute{\epsilon}$, where

$$h = n/q - O\left((l^3(n/q)^2 \ln(\ln/\epsilon))^{1/3}\right) - l \log(nl/\acute{\epsilon}).$$

Let $Q_t''$ be a sub-matrix of $Q_t'$ of size $m \times k$ ($k \leq m$), so that $Q_t''$ includes $h$ ($= m/q$) rows pertaining to each chunk. By the above argument, w.p. b.b.b. $1 - \acute{\epsilon}$, such a sub-matrix $Q_t''$ of $Q_t'$ exists, and $Q_t''$ is an $m \times k$ $(\gamma, \alpha)$ regular symmetric banded matrix.

---

[18]Note that a collection of $\omega$-packets is dense if the entries of their global encoding vectors' entries indexed by the aperture of chunk $\omega$ are i.u.d. r.v.'s, and a given $\omega$-packet is dense if it belongs to a maximal dense collection of $\omega$-packets.

We shall derive an upper bound on the probability that $Q_t''$ fails to have rank $k$ in order to upper bound the probability that $Q_t'$ fails to have rank $k$. The results of Conjecture 1 cannot be applied to our setting in general, since we do not restrict the overlap $\gamma$ to be $\alpha - 1$. Surprisingly, similar result also holds for more general settings as it can be seen through our simulation results in Section V (no formal proof is known yet). We formalize this observation in a conjecture as follows.[19]

*Conjecture 2:* Let $n, k, \alpha$ and $\gamma$ be integers ($k \leq n$, $\gamma < \alpha$). Let $M$ be a $(\gamma, \alpha)$ (irregular/regular) symmetric or asymmetric banded random binary matrix of size $n \times k$. For any $\epsilon > 0$, and for sufficiently large $k$, $\Pr[r(M) < k] \leq \epsilon$, so long as

$$k \leq n - \log(1/\epsilon),$$

and $\gamma \geq 2\sqrt{k}$, or $\gamma \geq \tau_e \tau \sqrt{k}$, respectively, where $\gamma = \alpha/\tau_e$, and $\tau_e = \tau/(\tau - 1)$, for any constant divisor $\tau$ of $\alpha$.

Lemma 15 together with Conjecture 2 (symmetric case) yield the following.[20]

*Theorem 10:* For any $\epsilon > 0$, an OCC with $q$ chunks of size $\alpha$, and overlap $\gamma \geq 2\sqrt{k}$, over any vector space $\mathbb{F}_2^L$, is capacity-achieving for a network of length $l$ with $k$ message packets under any one-in-one-out worst-case schedule of capacity $n$, with failure probability no larger than $\epsilon$, so long as

$$k \leq qh - \log(1/\acute{\epsilon}),$$

and $l^3 q \ln(nl/\epsilon) = o(n)$, where $h = n/q - O\left((l^3(n/q)^2 \ln(ln/\epsilon))^{1/3}\right) - l \log(nl/\acute{\epsilon}) - l$.

*Proof:* Replacing $\epsilon$ with $\acute{\epsilon}$ in Lemma 15 and Conjecture 2, (i) the capacity of the $\omega$-schedule, for any $\omega$, fails to be at least $(1 - O((l^3(q/n) \ln(ln/\epsilon))^{1/3})).(n/q)$, w.p. b.a.b. $\acute{\epsilon}/2$, so long as $l^3 q \ln(nl/\epsilon) = o(n)$; (ii) given that every $\omega$ is allocated a flow of capacity $(1 - O((l^3(q/n) \ln(ln/\epsilon))^{1/3})).(n/q)$, there does not exist $n/q - O((l^3(n/q)^2 \ln(ln/\epsilon))^{1/3}) - l \log(nl/\acute{\epsilon})$ dense $\omega$-packets at node $t$, w.p. b.a.b. $\acute{\epsilon}/2$; and (iii) given that node $t$ receives at least the above number of dense $\omega$-packets, for every $\omega$, the matrix $Q_t'$ fails to have rank $k$, w.p. b.a.b. $\acute{\epsilon}$, so long as $k \leq n - O(q(l^3(n/q)^2 \ln(ln/\epsilon))^{1/3}) - ql \log(nl/\acute{\epsilon}) - ql - \log(1/\acute{\epsilon})$; Thus, the probability of failure, when either (i), or (ii), or (iii) occurs, is b.a.b. $\epsilon$. ∎

Let $\tau^*$ be the smallest integer divisor of a given aperture size $\alpha$, such that by choosing $\tau = \tau^*$, we get

$\gamma \geq 2\sqrt{k}$. (We are assuming that $\alpha$ is chosen in a way that such a $\tau^*$ exists. The case for which there is no such $\tau^*$ will be discussed later.) For all $\tau > \tau^*$, $\gamma \geq 2\sqrt{k}$, and Theorem 10 holds. The larger is the value of $\tau$, however, the larger is the number of chunks and the slower would be the speed of convergence of OCC to the capacity for a given $\alpha$ (note that fixing $\alpha$ implies fixing the coding costs).[21] Thus, an OCC with the overlap parameter $\tau^*$ has the largest speed of convergence to the capacity, for a given aperture size, in one-in-one-out worst-case schedules.

Now suppose that there is no such $\tau^*$ for a given $\alpha$ (i.e., for any integer divisor $\tau$ of $\alpha$, $\gamma < 2\sqrt{k}$), but there exists at least an integer divisor $\tau$ of $\alpha$, so that $q$ $(= \tau k/\alpha)$ satisfies condition (5). Since we are assuming that $\gamma < 2\sqrt{k}$, Conjecture 2 is no longer useful. The probability that the rank of an $m \times k$ $(k \leq m)$ $(\gamma, \alpha)$ regular symmetric banded matrix with $\gamma < 2\sqrt{k}$ is less than $k$ is an open problem for an arbitrary choice of $\tau$; however, this probability can be given a trivial upper-bound by the probability that all the sub-matrices corresponding to different apertures of the underlying regular symmetric banded matrix have full rank. The following theorem summarizes the above discussion.[22]

*Theorem 11:* For any $\epsilon > 0$, an OCC with $q$ chunks of size $\alpha$, and overlap $\gamma < 2\sqrt{k}$, over any vector space $\mathbb{F}_2^L$, is capacity-achieving for a network of length $l$ with $k$ message packets under any one-in-one-out worst-case schedule of capacity $n$, with failure probability no larger than $\epsilon$, so long as

$$k \leq q\varphi - ql \log(nl/\acute{\epsilon}) - q \log(q/\acute{\epsilon}) - ql - q, \tag{6}$$

and $l^3 q \ln(nl/\epsilon) = o(n)$, where

$$\varphi = n/q - O\left((l^3(n/q)^2 \ln(ln/\epsilon))^{1/3}\right).$$

Furthermore, the larger is the overlap size, the looser would be the upper bound on $k$.

Since $k_{n_{\max}} = n - O\left(q(l^3(n/q)^2 \ln(ln/\epsilon))^{1/3}\right) - ql \log(nl/\acute{\epsilon}) - q \log(q/\acute{\epsilon}) - ql - q$ is a decreasing function of $q$ (when the other parameters are fixed), the larger is the number of chunks for a given aperture size, the smaller is the speed of convergence to the capacity (the ratio $k_{n_{\max}}/n$ becomes smaller with increasing $q$, for fixed $l, n$, and $\epsilon$). Thus, similarly as before, an OCC with overlap $\gamma < 2\sqrt{k}$ provides the largest speed of convergence to the capacity when its overlap is the smallest possible value. This implies that among OCC with smaller overlaps, CC has the fastest convergence to capacity for given coding costs.

---

[21]The ratio $k_{n_{\max}}/n = (n - O(q(l^3(n/q)^2 \ln(ln/\epsilon))^{1/3}) - ql \log(nl/\acute{\epsilon}) - ql - \log(1/\acute{\epsilon}))/n$ is a decreasing function of $q$. Hence, for larger values of $q$, this ratio becomes smaller when the other parameters $l, n$ and $\epsilon$ are fixed, which implies a lower speed of convergence.

[22]Part of the reason that the result of Theorem 11 is not tight is that it is based on the assumption that each chunk has to be decodable in isolation which is a sufficient but not a necessary condition in the case of overlapping chunks. Indeed, our analysis in this case is sub-optimal, and one can expect an OCC with sufficiently large apertures, and smaller overlap to perform even better than what Theorem 11 presents.

*All-In-All-Out Worst-Case Schedules:* Corresponding to Lemmas 14 and 15, we have the following results for all-in-all-out schedules.

*Lemma 16:* For any $\epsilon > 0$, applying a dense code to a given chunk $\omega$ over a network of length $l$ with any all-in-all-out worst-case schedule of capacity $n$, given that the capacity of the $\omega$-schedule is $\varphi$, the number of dense $\omega$-packets at node $t$ is less than $\varphi - l \log(l/\epsilon)$ w.p. b.a.b. $\epsilon$.

*Proof:* This is a direct result of Lemma 9, by replacing $n$ with $\varphi$. ∎

*Lemma 17:* Let $l, n, q, \varphi$ and $\epsilon$ be defined as above. Then the node $t$ fails to receive at least $\varphi - l \log(ql/\acute{\epsilon})$ dense $\omega$-packets, for every $\omega$, w.p. b.a.b. $\acute{\epsilon}$, and $\varphi$ fails to be larger than

$$\left(1 - O\left(((q/n)\ln(lq/\epsilon))^{1/2}\right)\right).(n/q),$$

w.p. b.a.b. $\acute{\epsilon}$, so long as

$$q \ln \frac{lq}{\epsilon} = o\left(n\right).$$

*Proof:* Replacing $\epsilon$ with $\acute{\epsilon}/q$ in Lemma 16 (for taking a union bound over $q$), the first part of the lemma follows. Replacing $\epsilon$ with $\acute{\epsilon}$ in Lemma 12 proves the second part of the lemma. ∎

Lemma 17 along with Conjecture 2 yield the following results.

*Theorem 12:* For any $\epsilon > 0$, an OCC with $q$ chunks of size $\alpha$, and overlap $\gamma \geq 2\sqrt{k}$, over any vector space $\mathbb{F}_2^L$, is capacity-achieving for a network of length $l$ with $k$ message packets under any all-in-all-out worst-case schedule of capacity $n$, with failure probability no larger than $\epsilon$, so long as

$$k \leq qh - \log(1/\acute{\epsilon}),$$

and $q \ln(lq/\epsilon) = o(n)$, where

$$h = n/q - O\left(((n/q)\ln(lq/\epsilon))^{1/2}\right) - l \log(lq/\acute{\epsilon}) - l.$$

Moreover, an OCC with the above description but with $\gamma < 2\sqrt{k}$ is capacity-achieving for a network scenario as above, w.f.p. no larger than $\epsilon$, so long as

$$k \leq qh - q \log(q/\acute{\epsilon}) - q.$$

*Coding Costs:* The worst-case with regards to the encoding cost at any transmitting interior node $u$ (or the source node $s$) occurs if node $u$ sends all its successful $\omega$-packets after receiving all the $\omega$-packets sent by the node in the upper layer, and for all $i \in \mathcal{I}_{u_t}^{(\omega)}$, $\lambda_{e,i}$'s (or for all $j \in \mathcal{M}_s^{(\omega)}$, $\mu_{e,j}$'s) are chosen to be nonzero. The number of packet operations for encoding at any such node $u$ (or node $s$) is $O(n^2/q)$ (or $O(n\alpha)$). (There are $O(n/q)$ $\omega$-packets sent and received by any such node $u$ and there are $O(n/q)$ $\omega$-packets sent by node $s$.) Thus the encoding cost is $O(\alpha)$.[23]

To solve the system of linear equations $\{\boldsymbol{y}_e = \sum_{j \in \mathcal{M}_s} \mu_{e,j} \hat{\boldsymbol{x}}_j : e \in \mathcal{I}_{t\infty}\}$ using Gauss-Jordan elimination, node $t$ requires $O(kn\alpha)$ field operations and $O(n\alpha)$ packet operations ($Q_t$ has a bandwidth and a size of $\alpha$ and $n \times k$, respectively). Thus the decoding cost is $O(\alpha)$, since $n/k = O(1)$, as $n$ goes to infinity.

*Theorem 13:* The encoding and decoding costs of an OCC with an aperture size $\alpha$ are each $O(\alpha)$.

### E. Comparison

We now compare CC and OCC with sufficiently large apertures over one-in-one-out worst-case schedules in the asymptotic regime. Similar results can be shown while comparing these codes over all-in-all-out worst-case schedules.

Consider a one-in-one-out worst-case schedule of capacity $n$ over a network of length $l$. Suppose that the $k$ message packets are divided into $\tau q$ chunks of size $\alpha$ ($= k/q$).[24]

We compare Theorem 5 with Theorems 10 and 11. Note that Theorem 5 for CC is a special case of Theorem 11, where $\tau$ is set to 1. We study the tradeoff between the probability of failure (referred to as the "message error rate" (MER)) and the speed of convergence. In particular, CC and OCC with similar aperture size (similar coding costs) are compared with respect to their speed of convergence to the capacity when the MER is given.

*Theorem 14:* For a given MER and for sufficiently large apertures, OCC with larger overlap has smaller speed of convergence than OCC with smaller overlap.

*Proof:* It follows from Theorems 10 and 11 that

$$k_{n_{\max}}/n = 1 - O\left(l\tau q \log(nl/\epsilon)/n\right),$$

---

[23] The number of packet operations for encoding is $O((l-1)n^2/q) + O(n\alpha)$, and hence the encoding cost is $O((l-1)n^2/qkl) + O(n\alpha/kl)$, i.e., $O((l-1)n/ql) + O(\alpha/l) = O(\alpha)$, since $n/k = O(1)$, and $k/q = O(\alpha)$, as $n$ goes to infinity.

[24] We assume that the number of chunks in OCC with overlap parameter $\tau$ is $\tau q$, not $q$, as was the case in the previous sections. The reason is to be consistent in the definition of the aperture size $\alpha$ as $k/q$, for both CC and OCC, as we compare CC and OCC with different overlap parameters for a similar aperture size. Thus, $q$ needs to be replaced with $\tau q$ in the results presented earlier for an OCC with overlap parameter $\tau$.

TABLE I
Comparison of various network codes over line networks with worst-case schedules.

| Network Codes | Schedule Type | $(n-k)/k$ | MER | PER[25] | Chunk Size $(\alpha)$[26] | Constraints |
|---|---|---|---|---|---|---|
| Dense Codes | One-In-One-Out | $O\left(\frac{1}{k}\log\left(\frac{k}{\epsilon}\right)\right)$ | $\epsilon$ | – | $k$ | – |
| | All-In-All-Out | $O\left(\frac{1}{k}\log\left(\frac{1}{\epsilon}\right)\right)$ | $\epsilon$ | – | $k$ | – |
| CC: Large $\alpha$ | One-In-One-Out | $O\left(\frac{1}{\alpha}\log\left(\frac{k}{\epsilon}\right)\right)$ | $\epsilon$ | – | $\Omega\left(\ln\left(\frac{k}{\epsilon}\right)\right)$ | – |
| | All-In-All-Out | $O\left(\frac{1}{\alpha}\log\left(\frac{k}{\alpha\epsilon}\right)\right)$ | $\epsilon$ | – | $\Omega\left(\ln\left(\frac{k}{\alpha\epsilon}\right)\right)$ | – |
| CC: Relatively Small $\alpha$ | One-In-One-Out | $\lambda$ | $\epsilon k/\alpha$ | – | $\Omega\left(\frac{1}{\lambda^3}\ln\left(\frac{1}{\lambda\epsilon}\right)\right)$ | $\epsilon = o\left(\frac{1}{k}\ln\left(\frac{1}{\epsilon}\right)\right)$ |
| | All-In-All-Out | $\lambda$ | $\epsilon k/\alpha$ | – | $\Omega\left(\frac{1}{\lambda^2}\ln\left(\frac{1}{\epsilon}\right)\right)$ | $\epsilon = o\left(\frac{1}{k}\ln\left(\frac{1}{\epsilon}\right)\right)$ |
| CC: Very Small $\alpha$ | One-In-One-Out | $\lambda$ | – | $\epsilon$ | $\Omega\left(\frac{1}{\lambda^3}\ln\left(\frac{1}{\lambda}\right)\right)$ | $\epsilon = O(1)$ |
| | All-In-All-Out | $\lambda$ | – | $\epsilon$ | $\Omega\left(\frac{1}{\lambda^2}\ln\left(\frac{1}{\epsilon}\right)\right)$ | $\epsilon = O(1)$ |
| OCC: Large $\alpha$ | One-In-One-Out | $O\left(\frac{\tau}{\alpha}\log\left(\frac{k}{\epsilon}\right)\right)$ | $\epsilon$ | – | $\Omega\left(\tau\ln\left(\frac{k}{\epsilon}\right)\right)$ | – |
| | All-In-All-Out | $O\left(\frac{\tau}{\alpha}\log\left(\frac{k}{\alpha\epsilon}\right)\right)$ | $\epsilon$ | – | $\Omega\left(\tau\ln\left(\frac{k}{\alpha\epsilon}\right)\right)$ | – |
| OCC: Relatively Small $\alpha$ | One-In-One-Out | $\lambda$ | $p\tau k/\alpha$ | – | $\Omega\left(\frac{\tau}{\lambda^3}\ln\left(\frac{\tau}{\lambda\epsilon}\right)\right)$ | $\epsilon = o\left(\frac{1}{k}\ln\left(\frac{1}{\epsilon}\right)\right)$ |
| | All-In-All-Out | $\lambda$ | $p\tau k/\alpha$ | – | $\Omega\left(\frac{\tau}{\lambda^2}\ln\left(\frac{\tau}{\lambda\epsilon}\right)\right)$ | $\epsilon = o\left(\frac{1}{k}\ln\left(\frac{1}{\epsilon}\right)\right)$ |
| OCC: Very Small $\alpha$ | One-In-One-Out | $\lambda$ | – | $p$ | $\Omega\left(\frac{\tau}{\lambda^3}\ln\left(\frac{\tau}{\lambda\epsilon}\right)\right)$ | $\epsilon = O(1)$ |
| | All-In-All-Out | $\lambda$ | – | $p$ | $\Omega\left(\frac{\tau}{\lambda^2}\ln\left(\frac{\tau}{\lambda\epsilon}\right)\right)$ | $\epsilon = O(1)$ |

for OCC with aperture size $k/q$, and overlap parameter $\tau$, so long as $q = o(n/(l^3\ln(nl/\epsilon)))$. Thus, for given $n, l, q$ and $\epsilon$, the larger is $\tau$, the smaller is the ratio $k_{n_{\max}}/n$. ∎

Noting that CC is a special case of OCC with zero overlap size, i.e., $\tau$ is equal to 1, the following result is a corollary of Theorem 14.

*Corollary 1:* For a given MER and for sufficiently large apertures, CC has higher speed of convergence to the capacity compared to OCC with any non-zero overlap.

Table I summarizes the performance of the network codes discussed in this paper over two types of worst-case schedules. In particular, the performance measures are the overhead per message, $(n-k)/k$, and the message or packet error rate (MER/PER), for different chunk sizes ($\alpha$). For example, for each type of schedules, the comparison of the corresponding rows of Table I for CC: large $\alpha$ and OCC: large $\alpha$ shows that for a given overhead (similar speed of convergence) and a given MER, the chunk size $\alpha$ for OCC must be larger than that for CC by a factor of $\tau$. This implies that for the same speed of convergence and MER, the coding cost of OCC is $\tau$ times that of CC. Moreover, for both CC and OCC with large $\alpha$, the lower bound on $\alpha$ is a super-logarithmic function of $k$.

Therefore, our asymptotic results so far indicate that for sufficiently large apertures, which guarantee the convergence to capacity, CC (OCC with zero overlap size) is superior to OCC with any non-zero overlap size. One may then wonder about whether there is any provable advantage in using OCC in the asymptotic regime. We will answer this question in the next section, and the following provides a motivation.

---

[25]The parameter $p$ lies in the interval $(\epsilon^{\chi+\tau-1}, \epsilon^2)$, where $\chi$ is an arbitrary constant integer sufficiently larger than $(\tau-1)/\lambda$.

[26]The coding costs of the codes are linear in the size of chunks, i.e., $O(\alpha)$.

From Table I, it can be seen that for CC and OCC with large $\alpha$ to be capacity-achieving (i.e., for $(n-k)/k$ to go to zero, as $k$ tends to infinity), the chunk size $\alpha$ needs to be bounded from below by a super-logarithmic function of $k$. However, this lower bound might be larger than the affordable coding costs in many practical scenarios. Therefore, in Section IV, we study CC and OCC with chunks of smaller sizes and show that in such cases, OCC can outperform CC in the asymptotic regime (e.g., for relatively smaller aperture sizes, comparing an OCC and a CC with similar speed of convergence and coding costs, the former has a smaller probability of failure).

## IV. Capacity-Approaching Codes: Towards Linear-Time Network Codes

Focusing on the design of CC and OCC with smaller coding costs when compared to those with sufficiently large aperture sizes discussed in Section III, in this section, we analyze CC and OCC with smaller aperture sizes.

### A. CC with Small Apertures

*One-In-One-Out Worst-Case Schedules:* Suppose that a CC with $q$ chunks of size $\alpha$ is applied to the $k$ message packets at node $s$ of a network of length $l$ with an arbitrary one-in-one-out worst-case schedule of capacity $n = (1+\lambda)k$, for an arbitrarily small but constant $\lambda > 0$, as $k$ goes to infinity. Further, suppose that $\alpha$ is selected small enough such that it violates the condition (2). In this case, from Lemma 10 follows that for a CC with such a small aperture size $\alpha$, there exists at least one chunk, say $\omega$, so that the $\omega$-packets do not form a flow of capacity of at least $n/q - O\left((l^3(n/q)^2 \ln(\ln/\epsilon))^{1/3}\right)$ w.p. b.a.b. $\epsilon$. Thus Lemma 10 and Theorems 4 and 5 do not apply to the underlying setting.

Our approach in this case is to fix a particular chunk, say $\omega$, and give a lower bound on the capacity of the $\omega$-schedule. We are then able to lower bound the probability of receiving a collection of $\alpha$ innovative $\omega$-packets by node $t$.

The expected capacity of the $\omega$-schedule is $\mu = n/q = (1+\lambda)\alpha$. The deviation of the actual capacity of the $\omega$-schedule from its expectation is upper bounded as follows.

*Lemma 18:* For any $\epsilon > 0$, for a CC with aperture size $\alpha$, over a network of length $l$ with any one-in-one-out worst-case schedule of capacity $(1+\lambda)k$, for a given $\omega$, the capacity of the $\omega$-schedule is less than

$$\varphi = \left(1 - O\left(\left((l^3/\mu)\ln(l\mu/\epsilon)\right)^{1/3}\right)\right) \cdot \mu \tag{7}$$

w.p. b.a.b. $\dot{\epsilon}/2$, where $\mu = (1+\lambda)\alpha$.

*Proof:* The proof is similar to that of Lemma 10, except that we do not need to take a union bound over all the chunks. This comes from the fact that in this setting, we need to find the capacity of the flow allocated to one chunk alone. ∎

*Lemma 19:* For any $\epsilon > 0$, applying a dense code (to a chunk $\omega$) over a network of length $l$ with any one-in-one-out worst-case schedule, the number of dense $\omega$-packets is not larger than $\varphi - l \log(l\varphi/\dot{\epsilon}) - l$ w.p. b.a.b. $\dot{\epsilon}/2$, where $\varphi$ is the capacity of the $\omega$-schedule.

*Proof:* Note that the actual $\omega$-schedule is not necessarily a one-in-one-out worst-case schedule. However, since we are interested in analyzing the worst-case scenario, the $\omega$-schedule can also be assumed to be a one-in-one-out worst-case schedule. Then, Lemma 19 follows from Lemma 5, by replacing $n$ and $\epsilon$ with $\varphi$ and $\dot{\epsilon}/2$, respectively. ∎

Lemma 18 together with Lemma 19 show that there are less than $\varphi - l \log(l\varphi/\dot{\epsilon}) - l$ dense $\omega$-packets w.p. b.a.b. $\dot{\epsilon}$; and by applying Lemma 6, the probability of decoding failure of a given chunk $\omega$ can be upper bounded as follows.

*Lemma 20:* For any $\epsilon > 0$, and $\lambda > 0$, applying a CC with aperture size $\alpha$, over a network of length $l$ with any one-in-one-out worst-case schedule of capacity $(1 + \lambda)k$, the probability that a given chunk $\omega$ fails to be decoded is b.a.b. $\epsilon$, so long as

$$\alpha \leq \varphi - l \log(l\varphi/\dot{\epsilon}) - \log(1/\epsilon) - l - 1, ^{[27]} \tag{8}$$

where the capacity of the $\omega$-schedule is not less than

$$\varphi = \left(1 - O\left(\left((l^3/\mu) \ln(l\mu/\epsilon)\right)^{1/3}\right)\right) \cdot \mu,$$

and $\mu = (1 + \lambda)\alpha$.

*Proof:* Replacing $k$, $h$ and $\epsilon$ in Lemma 6 with $\alpha$, $\varphi - l \log(l\varphi/\dot{\epsilon}) - l$ and $\dot{\epsilon}$, respectively, Lemma 20 follows by a similar argument as in the proof of Theorem 1, except that in this case, we focus on the decoding failure probability of a given chunk alone, not all the chunks. ∎

Substituting (7) into (8), we obtain

$$\alpha = \Omega\left(\frac{l^3}{\lambda^3} \ln\left(\frac{l}{\lambda\epsilon}\right)\right). \tag{9}$$

---

[27]One should note that, despite its appearance, inequality (8) imposes a lower bound on $\alpha$ (the right hand side of inequality (8) is itself a function of $\alpha$).

The details of the derivation of (9) are deferred to Appendix IV. The above result shows that, over any one-in-one-out worst-case schedule, for a CC with an aperture size satisfying (9), the probability of decoding failure for a given chunk is b.a.b. $\epsilon$. Therefore, the expected fraction of undecodable chunks is b.a.b. $\epsilon q$. For sufficiently small choice of $\epsilon$, so long as $\epsilon q$ goes to zero, as $k$ goes to infinity, not all the chunks would be decodable w.p. b.a.b. $\epsilon q$, and the following is immediate.

*Theorem 15:* For any $\epsilon > 0$, when $\epsilon$ goes to $0$ sufficiently fast, as $k$ goes to infinity,[28] applying a CC with aperture size $\alpha = \Omega((l^3/\lambda^3)\ln(l/\lambda\epsilon))$, to $k$ message packets over a network of length $l$ with any one-in-one-out worst-case schedule of capacity $(1 + \lambda)k$, all the chunks are decodable w.p. b.b.b. $1 - \epsilon q$.

In some cases, however, such small choices of $\epsilon$ might not be practical in that the corresponding aperture size $\alpha$ (and the coding costs) may be too large (the lower bound on $\alpha$ is a logarithmic function of $1/\epsilon$). Let us assume larger values of $\epsilon$ up to a constant. We shall show that for any such $\epsilon$, the actual fraction of undecodable chunks is tightly concentrated around its expectation.[29]

*Theorem 16:* For any $\epsilon > 0$, and $\lambda > 0$, in a CC with an aperture size $\alpha = \Omega\left((l^3/\lambda^3)\ln(l/\lambda\epsilon)\right)$, over a network of length $l$ with any one-in-one-out worst-case schedule of capacity $(1 + \lambda)k$, for any $\gamma_a > 0$, the fraction of undecodable chunks deviates farther than $\gamma_a$ from $\epsilon$, w.p. b.a.b. $e^{-ck}$, for some positive constant $c = O(\gamma_a^2 \epsilon^2/\alpha^2)$.

*Proof:* Please refer to Appendix V. ∎

Therefore, with high probability, for any $\gamma_a > 0$, a CC with an aperture size as above fails to recover at most $(1 + \gamma_a)\epsilon k$ message packets. This is because the number of chunks is $q$, and that any undecodable chunk accounts for at most $k/q$ unrecovered message packets.

In such cases, therefore, the expected fraction of undecodable chunks and unrecovered message packets are bounded away from zero. Thus, a CC, alone, does not recover all the message packets. Now, a question is how a CC with small apertures can recover all the message packets? One solution is to devise a proper precoding scheme in order to guarantee the completion of decoding of all the message packets. The combination of a CC and a precode is a *chunked code with precoding* (CCP).

The precode works as follows. Suppose that the $k$ message packets are the input of a precode of rate $R$. The coded packets at the output of the precode are called the *intermediate packets*. The number of intermediate packets is $k/R$. The intermediate packets are sent through the network by using a CC. The number of the intermediate packets that cannot be recovered at the output of the CC decoder is at most $pk/R$, where $p = (1+\gamma_a)\epsilon$, for any arbitrarily small constant $\gamma_a > 0$ (as shown in Theorem 16). Therefore, there are at least $(1-p)k/R$ intermediate packets recovered by the CC decoder. These packets constitute the input symbols of the precode decoder. The precode thus needs to recover the $k$ message packets from the set of recovered intermediate packets. This implies that the precode has to be an erasure-correcting code of dimension $k$, block length $k/R$ (rate $R$), that is capable of recovering $p$ fraction of erasures. [30]

---

[28]We say that $\epsilon$ goes to 0 "sufficiently fast," if $\epsilon q$ tends to zero as $k$ goes to infinity, i.e., $\epsilon/\ln(1/\epsilon) = o(1/k)$.

[29]We consider the worst-case assuming that the probability of decoding failure of any chunk is the largest possible, i.e., $\epsilon$ (as shown in Lemma 20), and hence the expected fraction of undecodable chunks is $\epsilon$.

TABLE II
Comparison of various linear-time erasure-correcting codes

| Erasure Codes | Coding Costs | Probability of Failure | Error-Exponent |
|---|---|---|---|
| Tornado/LDPC Codes in [20]–[22] | $O\left(\log(1/\gamma_p)\right)$ | poly$(1/k)$ | 0 |
| LDPC Codes in [23] | $O\left((1/\gamma_p)\log(1/\gamma_p)\right)$ | poly$(1/k)$ | 0 |
| Codes in [24] | $O\left(1/\gamma_p^2 \log(1/\gamma_p)\right)$ | $e^{-\Omega(k)}$ | poly$(\gamma_p)$ |
| | $O\left(1/\gamma_p^4 \log(1/\gamma_p)\right)$ | 0 | – |

We are, further, interested in a CCP scheme with linear-time encoding/decoding algorithms (i.e., with constant coding costs with respect to $k$). Therefore, both the CC and the precode must be linear-time codes. It should be clear that CC is linear-time so long as the aperture size is a constant, and therefore, the problem is to look for a precode with constant coding costs.[31]

Table II lists a number of linear-time erasure-correcting codes in the literature which provide the best known tradeoffs between the computational complexity and the probability of failure. In Table II, each code has dimension $k$, rate $R$, and is able to recover a fraction of erasures up to $p$. The coding costs in Table II are expressed in terms of the parameter $\gamma_p \triangleq (1-p)(1+R) - 1$.

As it can be seen in Table II, from top to bottom, for given $k$ and $\gamma_p$, the speed of convergence of the failure probabilities to zero and the coding costs of the codes increase. By choosing $R$ and $p$ to be constant, the coding costs will also be constant. For a CCP, the precode needs to recover a fraction of erasures up to $p = (1 + \gamma_a)\epsilon$, for an arbitrarily small constant $\gamma_a > 0$, where $\epsilon$ is a constant as well. The codes listed in Table II are each able to recover a fraction of erasures up to $p$ requiring constant coding costs, and thus are each a good candidate to be combined with CC.

*All-In-All-Out Worst-Case Schedules:* The results of CC with small apertures over all-in-all-out worst-case schedules are quite similar to those over one-in one-out schedules, except that the lower bound on the aperture size differs. The difference arises from the difference between the conditions (2) and (4) on the aperture size in Lemmas 10 and 12, respectively. Thus, for brevity, we only state and prove the main lemmas; and the resulting theorems would be similar to Theorems 15 and 16 (and not repeated), except that the lower bound on the aperture size needs to be replaced with a new lower bound derived in the following.

---

[30]The precode does not have to be capacity-achieving and hence the rate of the precode $R$ does not have to tend to $1 - p$.

[31]The specifications of all the precodes discussed in the rest of this paper are the same as those discussed here, and hence not repeated for brevity.

*Lemma 21:* For any $\epsilon > 0$, for a CC with aperture size $\alpha$, over a network of length $l$ with any all-in-all-out worst-case schedule of capacity $(1 + \lambda)k$, for a given $\omega$, the capacity of the $\omega$-schedule is less than

$$\varphi = \left(1 - O\left((\ln(l/\epsilon)/\mu)^{1/2}\right)\right) \cdot \mu \tag{10}$$

w.p. b.a.b. $\dot{\epsilon}/2$, where $\mu = (1 + \lambda)\alpha$.

*Proof:* The proof is similar to that of Lemma 12, except that we do not take a union bound over all the chunks. This is because, in this setting, we are interested in the capacity of the flow allocated to one chunk only. ∎

*Lemma 22:* For any $\epsilon > 0$, applying a dense code (to a chunk $\omega$) over a network of length $l$ with any all-in-all-out worst-case schedule, the number of dense $\omega$-packets is less than $\varphi - l \log(l/\dot{\epsilon}) - l$ w.p. b.a.b. $\dot{\epsilon}/2$, where $\varphi$ is the capacity of the $\omega$-schedule.

*Proof:* In this case, the $\omega$-schedule is itself always an all-in-all-out worst-case schedule. Then, the result follows from Lemma 9 by replacing $n$ and $\epsilon$ with $\varphi$ and $\dot{\epsilon}/2$, respectively. ∎

Lemma 21 together with Lemma 22 show that the number of dense $\omega$-packets is less than $\varphi - l \log(l/\dot{\epsilon}) - 1$ w.p. b.a.b. $\dot{\epsilon}$; and the following gives an upper bound on the probability of decoding failure of a given chunk $\omega$.

*Lemma 23:* For any $\epsilon > 0$, and $\lambda > 0$, applying a CC with aperture size $\alpha$, over a network of length $l$ with any all-in-all-out worst-case schedule of capacity $(1 + \lambda)k$, the probability of a given chunk $\omega$ to be undecodable is b.a.b. $\epsilon$, so long as

$$\alpha \leq \varphi - l \log(l/\dot{\epsilon}) - \log(1/\epsilon) - l - 1, \tag{11}$$

where

$$\varphi = \left(1 - O\left((\ln(l/\epsilon)/\mu)^{1/2}\right)\right) \cdot \mu,$$

and $\mu = (1 + \lambda)\alpha$.

*Proof:* Replacing $k$, $h$ and $\epsilon$ in Lemma 6 with $\alpha$, $\varphi - l \log(l/\acute{\epsilon}) - l$ and $\acute{\epsilon}$, respectively, the result follows from an argument similar to that of Theorem 2, while focusing on the decoding failure probability of a given chunk alone, not all the chunks. ∎

Substituting (10) into (11), we obtain

$$\alpha = \Omega \left( \frac{l}{\lambda^2} \ln \left( \frac{l}{\epsilon} \right) \right). \tag{12}$$

The details are similar to that of (9), and hence omitted. Thus, over any all-in-all-out worst-case schedule, for a CC with an aperture size as given in (12), the probability of decoding failure of a given chunk is b.a.b. $\epsilon$.

## B. OCC with Small Apertures

*One-In-One-Out Worst-Case Schedules:* Suppose an OCC with $q$ $(= k\tau/\alpha)$ chunks, each of size $\alpha$, and overlap $\gamma = \alpha/\tau_e$ (when $\tau_e = \tau/(\tau - 1)$, and $\tau$ is a constant integer divisor of $\alpha$), applied to the $k$ message packets at node $s$ of a network of length $l$ with an arbitrary one-in-one-out worst-case schedule of capacity $n = (1 + \lambda)k$.

The same approach used for the analysis of CC with small apertures is not applicable to OCC with small apertures. In particular, unlike CC, in OCC, a martingale argument alone does not provide a tight upper bound on the the fraction of unrecoverable message packets. The reason is that in OCC the chunks are to be decoded together. (In CC, upper bounding the fraction of undecodable chunks yields a trivial upper bound on the fraction of unrecovered message packets as the chunks do not share any message packets.) A new approach is thus required for the analysis of OCC.

In the following, we provide a sketch of our analysis. Consider the decoding matrix $Q_t$ at node $t$, and let $Q'_t$ be $Q_t$ restricted to its dense rows. In the underlying setting, the result of Lemma 15 is not applicable to lower bound the probability that there exists a set of rows in $Q'_t$, so that these rows form a $(\gamma, \alpha)$ regular symmetric banded matrix with $k$ columns and a sufficiently large number of rows. Consequently, the result of Conjecture 2 (symmetric case) is no longer useful.

Let $\chi > 1$ be an integer sufficiently smaller than the number of chunks. Consider a particular set of $\chi$ contiguous chunks (we will specify the precise choice of $\chi$ later). Focus on the set of dense packets pertaining to the given set of chunks. We first lower bound the probability that the set of rows pertaining to these chunks in $Q'_t$ includes a $(\gamma, \alpha)$ regular asymmetric banded matrix with $\chi(\alpha - \gamma) + \gamma$ columns (the number of distinct message packets in $\chi$ contiguous chunks) and a sufficiently large number of rows. By using Conjecture 2 (asymmetric case), we next upper bound the probability that such a set fails to be decodable. Studying all such sets of $\chi$ contiguous chunks (lower bounding the probability that any such set is decodable), the fraction of recoverable message packets can be lower bounded. (All the message packets in a chunk belonging to a decodable set of chunks are recoverable).

It is worth noting that our analysis is sub-optimal in the sense that there might be some recoverable message packets that we declare as unrecoverable. This is because the decoding, in our setting, is performed on the set of all the chunks together, not on the subsets of chunks in isolation.

We formalize the above process as follows. We call each set of $\chi$ contiguous chunks, in an end-around fashion, a *hyperchunk* (the first hyperchunk includes the first $\chi$ chunks $\omega \in \{1, ..., \chi\}$, and the second hyperchunk includes the chunks $\omega \in \{2, ..., \chi + 1\}$, and so on).[32] The hyperchunks are overlapping in chunks and regardless of the choice of $\chi$, there are $q$ hyperchunks. We also call each (disjoint) set of $\alpha/\tau = k/q$ contiguous message packets a *block* (the first block starts from the first message packet, the second block starts from the message packet next to the last message packet in the first block, and so forth). By definition, each hyperchunk consists of $\ell := \chi + \tau - 1$ contiguous blocks.[33] We say that a given hyperchunk is not decodable (called a *bad hyperchunk*) if it fails to be decoded *in isolation*. We also say that a given block is not recoverable (called a *bad block*) if it does not belong to any decodable hyperchunk.

We shall upper bound the probability that a given hyperchunk fails to be decodable (by lower bounding the probability of receiving an innovative collection of $\chi k/q$ packets belonging to this hyperchunk). Lemma 19 serves this purpose when the capacity of the flow by the packets pertaining to any chunk in this hyperchunk is given. We lower bound this capacity in the following.

The expected number of $\omega$-packets for a given chunk $\omega$ is $\mu = (1 + \lambda)\alpha/\tau$. The capacity of the $\omega$-schedule is a random variable and its deviation from the expectation is upper bounded as follows.

*Lemma 24:* For any $\epsilon > 0$, $\lambda > 0$, and any constant integer $\chi > 0$, an OCC with an aperture size $\alpha$, and overlap parameter $\tau$, over a network of length $l$ with any one-in-one-out worst-case schedule of capacity $(1 + \lambda)k$ fails to provide an $\omega$-schedule of capacity larger than

$$\varphi = \left(1 - O\left(\left((l^3/\mu)\ln(l\mu\chi/\epsilon)\right)^{1/3}\right)\right) \cdot \mu, \tag{13}$$

---

[32] We will determine the optimal value of $\chi$ that results in the tightest possible bounds as part of our analysis.

[33] Since $\chi > 1$, and in the case of OCC, we have $\tau > 1$, then, each hyperchunk contains at least two contiguous blocks (i.e., $\ell > 1$).

for a given $\omega$, w.p. b.a.b. $\dot{\epsilon}/2\chi$, where $\mu = (1+\lambda)\alpha/\tau$.

*Lemma 25:* For any $\epsilon > 0$, and any constant integer $\chi > 0$, applying a dense code to a given chunk $\omega$ over a network of length $l$ with any one-in-one-out worst-case schedule, the number of dense $\omega$-packets is smaller than $\varphi - l\log(l\varphi\chi/\dot{\epsilon}) - l$, w.p. b.a.b. $\dot{\epsilon}/2\chi$, where $\varphi$ is the capacity of the $\omega$-schedule.

Lemmas 24 and 25 readily follow from Lemmas 18 and 19, by replacing $\epsilon$ with $\epsilon/\chi$.

Lemma 24 together with Lemma 25 show that, for a given $\omega$, w.p. b.a.b. $\dot{\epsilon}/\chi$, there are less than $\varphi - l\log(l\varphi\chi/\dot{\epsilon}) - l$ dense $\omega$-packets at node $t$. Taking a union bound over a set of $\chi$ chunks (a hyperchunk), it can be seen that, w.p. b.a.b. $\dot{\epsilon}$, there is not a subset of dense packets pertaining to a hyperchunk that form a $(\gamma, \alpha)$ regular asymmetric banded matrix of size $(\chi(\alpha - \gamma) + \gamma) \times (\chi\varphi - \chi l\log(l\varphi\chi/\dot{\epsilon}) - \chi l)$. Then, by applying Conjecture 2, the probability that a given hyperchunk is not decodable can be upper bounded as follows.

*Lemma 26:* For any $\epsilon > 0$, and $\lambda > 0$, applying an OCC with aperture size $\alpha$, and overlap parameter $\tau$, over a network of length $l$ with any one-in-one-out worst-case schedule of capacity $(1+\lambda)k$, a given hyperchunk of size $\chi$ is bad (fails to be decoded) w.p. b.a.b. $\epsilon$, so long as

$$r\alpha \leq \chi\varphi - \chi l\log(l\varphi\chi/\dot{\epsilon}) - \log(1/\dot{\epsilon}) - \chi l, \tag{14}$$

and $\gamma \geq \tau_e \tau \sqrt{r\alpha}$, where

$$\varphi = \left(1 - O\left(\left((l^3/\mu)\ln(l\mu\chi/\epsilon)\right)^{1/3}\right)\right) \cdot \mu,$$

$\mu = (1+\lambda)\alpha/\tau$, and $r = (\chi - 1)/\tau + 1$.

*Proof:* The result follows from Conjecture 2, by replacing $n$ and $k$ with the number

$$\left(1 - O\left(\left((l^3/\mu)\ln(l\mu\chi/\epsilon)\right)^{1/3}\right)\right)\chi\mu - \chi l\log(l\mu\chi/\dot{\epsilon}) - \chi l$$

of rows pertaining to the given hyperchunk that constitute a regular asymmetric banded matrix and the number $\chi(\alpha - \gamma) + \gamma = r\alpha$ of message packets in the given hyperchunk, respectively. ∎

Substituting (13) into (14), we obtain

$$\alpha = \Omega\left(\frac{l^3}{\lambda^3}\tau\ln\left(\frac{l}{\lambda\epsilon}\tau\right)\right), \tag{15}$$

by choosing $\chi$ to be an arbitrary constant integer sufficiently larger than $(\tau - 1)/\lambda$. The details are deferred to Appendix VI.

Thus, applying an OCC with an aperture size as given in (15), over any one-in-one-out worst-case schedule, the probability that a given hyperchunk is bad is b.a.b. $\epsilon$; and thus the expected fraction of bad hyperchunks is upper bounded by $\epsilon$. Similarly as before, by using a martingale argument over the hyperchunks, it can be shown that with high probability the actual fraction of bad hyperchunks does not deviate far from its expected value.

*Theorem 17:* For any $\epsilon > 0$, and $\lambda > 0$, in an OCC with an aperture size $\alpha = \Omega\left((l^3/\lambda^3)\tau \ln\left((l/\lambda\epsilon)\tau\right)\right)$, and overlap parameter $\tau$, over a network of length $l$ with any one-in-one-out worst-case schedule of capacity $(1 + \lambda)k$, for any $\gamma_a > 0$, the fraction of hyperchunks that are not decodable deviates farther than $\gamma_a$ from $\epsilon$, w.p. b.a.b. $e^{-ck}$, for some positive constant $c = O((\gamma_a^2\epsilon^2/\alpha^2)(\lambda\tau))$.

*Proof:* Please refer to Appendix VII. ∎

Now, the problem is to upper bound the fraction of message packets (or blocks) that are not recoverable. The fraction of bad blocks however depends on the relative location of bad hyperchunks.[34] It should be clear that among different positioning of bad hyperchunks, the one in which all the bad hyperchunks are adjacent results in the largest fraction of bad blocks. Since the expected fraction of bad hyperchunks is $\epsilon$, the expected fraction of bad blocks is upper bounded by $\epsilon$. Each block, itself, contains $k/q$ message packets, and hence, the expected number of message packets that are unrecoverable is upper bounded by $\epsilon k$.

This, however, is not a tight bound since the probability that a large number of bad hyperchunks are adjacent is very small. To derive a tighter bound, the distribution of bad hyperchunks would be required. This however is too complex to obtain. We, instead, analyze two extremal types of dependency structures of hyperchunks as defined below.

Let $I$ be the set of integers in $[q]$. For all $i \in I$, let $\mathcal{G}_i$ ($\mathcal{B}_i$) be the set of indices of the message packets in the $i$th hyperchunk (block). We use the same notation $\mathcal{G}_i$ ($\mathcal{B}_i$) to refer to the $i$th hyperchunk (block) unless there is a danger of confusion. We, further, let $G_i$ ($B_i$) be the event that $\mathcal{G}_i$ ($\mathcal{B}_i$) is not decodable (recoverable).

Let $G_I$ be the set of events $\{G_i\}_{i\in I}$. Let $I_i$ be an arbitrary non-empty subset of $I \setminus \{i\}$. For any $I_i$, we write $i \prec I_i$, $i \succ I_i$, or $i \asymp I_i$, respectively, if $\Pr[G_i|\wedge_{j\in I_i} G_j] < \Pr[G_i]$, $\Pr[G_i|\wedge_{j\in I_i} G_j] > \Pr[G_i]$, or $\Pr[G_i|\wedge_{j\in I_i} G_j] = \Pr[G_i]$. For a given probability measure on the set $G_I$, for any $i$, and $I_i$, either $i \prec I_i$, $i \succ I_i$, or $i \asymp I_i$. We call the set of relationships $\{(i \prec I_i) \vee (i \succ I_i) \vee (i \asymp I_i)\}_{i \in I, I_i \subset I \setminus \{i\}}$, the characteristic set of the given probability measure on $G_I$. The set of all probability measures on $G_I$, whose characteristic set is the same, is said to have the same type of dependency, and hence any characteristic set defines a *type of dependency*.

---

[34] We explain this dependency through an example. Consider a case with only two bad hyperchunks. Suppose that these two bad hyperchunks share a given block which does not belong to any other hyperchunk. This block would therefore be a bad block. Now, consider the case where these two bad hyperchunks either do not share any blocks, or any block that they share is in some other hyperchunks as well. In such cases, there is no bad block, because each block belongs to at least one hyperchunk which is not bad.

For $i \in I$, let $N_{\mathcal{G}}(i)$ be an ordered set (in an increasing cyclic order) of indices of hyperchunks that overlap with the $i$th hyperchunk, and $I_i$ be an arbitrary subset of $I \setminus \{i\}$. The first dependency type is the one that the occurrence of any subset of $G_j$'s, for all $j \in N_{\mathcal{G}}(i)$ $(j \neq i)$, increases the probability that $G_i$ occurs, i.e., for all $I_i$ such that $I_i \setminus N_{\mathcal{G}}(i) = \emptyset$, either $i \succ I_i$, or $i \asymp I_i$. The second is the one that, for all $I_i$ such that $I_i \setminus N_{\mathcal{G}}(i) = \emptyset$, either $i \prec I_i$, or $i \asymp I_i$. In the case of both dependency types, for all $I_i \cap N_{\mathcal{G}}(i) = \emptyset$, we have either $i \prec I_i$, or $i \asymp I_i$, and this is, indeed, the case for any possible type of dependency between the hyperchunks.[35] For any other $I_i$, no consideration is made. We refer to the first (second) type as the *dependency with positively (negatively) dependent neighborhoods*.

We upper bound, for each type of dependency, (i) the probability that not all the blocks are recoverable, and (ii) the probability that a block is unrecoverable. Such bounds are "outer" upper bounds for the class of dependency structures of the underlying type in that they hold for any dependency structure in the class.

We say that an outer bound is "tight" over the class of dependency structures of a given type, if, in the limit of interest (as $k$ goes to infinity, $\epsilon$ goes to zero sufficiently fast, or it is bounded away from zero), the outer bound is tight for any worst-case structure in the class.

We derive tight outer upper bounds for each type, by studying the worst case, i.e., given that any arbitrary subset of hyperchunks is not decodable, the conditional probability of undecodability of any given hyperchunk is the largest possible. For each of the probabilities (i) and (ii), these (tight) outer upper bounds act as the two limits of an interval that for any possible type of dependency, a tight outer upper bound lies within. We prove the following theorems. (Proofs are provided in Appendix VIII.)[36]

*Theorem 18:* For any $\epsilon > 0$, when $\epsilon$ goes to $0$ sufficiently fast,[37] as $k$ tends to infinity, and for any $\lambda > 0$, applying an OCC with aperture size $\alpha = \Omega((l^3/\lambda^3)\tau \ln((l/\lambda\epsilon)\tau))$, and overlap parameter $\tau$, over a network of length $l$ with any one-in-one-out worst-case schedule of capacity $(1 + \lambda)k$, for any type of dependency between hyperchunks, there exists a tight outer upper bound on the probability that not all the blocks are recoverable. This bound is between $\epsilon^{\chi+\tau-1}q$, and $\epsilon^2 q$, for arbitrary constant integer $\chi$ sufficiently larger than $(\tau-1)/\lambda$.

---

[35]The undecodability of a subset of hyperchunks that does not share any chunk with a given hyperchunk $\mathcal{G}$, increases the probability that a chunk in $\mathcal{G}$ has been allocated a sufficiently large number of dense packets. This, therefore, increases the probability of decodability of $\mathcal{G}$.

[36]Theorems 18 and 19 correspond to Theorems 15 and 16, respectively. The proofs have similar structure, yet, the bounds on the aperture size and the probability of failure of CC are replaced with those for OCC.

[37]For definition, see Footnote 28.

*Theorem 19:* For any $\epsilon > 0$, and $\lambda > 0$, for an OCC with aperture size $\alpha = \Omega\left((l^3/\lambda^3)\tau\ln((l/\lambda\epsilon)\tau)\right)$, and overlap parameter $\tau$, over a network of length $l$ with any one-in-one-out worst-case schedule of capacity $(1+\lambda)k$, for any type of dependency between hyperchunks, there exists a tight outer upper bound on the probability that a block is unrecoverable. This bound is between $\epsilon^{\chi+\tau-1}$, and $\epsilon^2$, for arbitrary constant integer $\chi$ sufficiently larger than $(\tau-1)/\lambda$.

*All-In-All-Out Worst-Case Schedules:* The results of OCC with small apertures over all-in-all-out worst-case schedules are similar to those over one-in-one-out worst-case schedules, except that the lower bound on the aperture size needs to be revised due to the difference between the conditions (2) and (4) on the aperture size in Lemmas 15 and 17, respectively. For the sake of brevity, in the following, we only present the main lemmas along with their proofs. The resulting theorems are similar to Theorems 18, and 19, with the only difference being the lower bound on the aperture size, which needs to be modified accordingly.

*Lemma 27:* For any $\epsilon > 0$, applying an OCC with aperture size $\alpha$, and overlap parameter $\tau$, to $k$ message packets over a network of length $l$ with any all-in-all-out worst-case schedule of capacity $(1+\lambda)k$, the node $t$ fails to receive at least $\varphi - l\log(l/\acute{\epsilon}) - l$ dense $\omega$-packets, for a given $\omega$, w.p. b.a.b. $\acute{\epsilon}/2$, and $\varphi$ fails to be larger than

$$\left(1 - O\left(((1/\mu)\ln(l/\epsilon))^{1/2}\right)\right) \cdot \mu, \tag{16}$$

w.p. b.a.b. $\acute{\epsilon}/2$, where $\mu = (1+\lambda)\alpha/\tau$.

*Proof:* The proof is similar to that of Lemma 17, yet, no union bound is taken over the chunks since we need the capacity of the flow allocated to one chunk only. ∎

*Lemma 28:* For any $\epsilon > 0$, and $\lambda > 0$, applying an OCC with aperture size $\alpha$, and overlap parameter $\tau$, to $k$ message packets over a network of length $l$ with any all-in-all-out worst-case schedule of capacity $(1+\lambda)k$, the probability of a given hyperchunk of size $\chi$ to be bad is b.a.b. $\epsilon$, so long as

$$r\alpha \leq \chi\varphi - \chi l\log(l\chi/\acute{\epsilon}) - \log(1/\acute{\epsilon}) - \chi l, \tag{17}$$

and $\gamma \geq \tau_e\tau\sqrt{r\alpha}$, where

$$\varphi = \left(1 - O\left(((1/\mu)\ln(l/\epsilon))^{1/2}\right)\right) \cdot \mu,$$

$\mu = (1 + \lambda)\alpha/\tau$, and $r = (\chi - 1)/\tau + 1$.

*Proof:* The proof follows from Conjecture 2, similar to that of Lemma 20, by replacing $n$ and $k$ with the lower bound given on the number of rows pertaining to the chunks in the given hyperchunk which constitute a regular asymmetric banded matrix and the number of message packets in the given hyperchunk, respectively. ∎

Substituting (16) into (17), we obtain

$$\alpha = \Omega\left(\frac{l}{\lambda^2}\tau \ln\left(\frac{l}{\lambda\epsilon}\tau\right)\right),\tag{18}$$

by choosing $\chi \gg (\tau - 1)/\lambda$. Over any all-in-all-out worst-case schedule, therefore, for an OCC with an aperture size as given in (18), the probability of decoding failure of a given hyperchunk is b.a.b. $\epsilon$.

## C. Comparison

Now, we compare our analytical results for CC and OCC with small chunks over one-in-one-out worst-case schedules. Similar comparisons are also valid for all-in-all-out worst-case schedules.

We consider a one-in-one-out worst-case schedule of capacity $n = (1 + \lambda)k$, for a given $\lambda > 0$, over a network of length $l$. Similar to the notations used in Section III-E, let the $k$ message packets be divided into $\tau q$ chunks of size $\alpha$ $(= k/q)$. We partition CC and OCC (with constant overlap parameter $\tau$) with small chunks into two categories based on their aperture size depending on $k, l, \tau, \lambda$ and $\epsilon$.[38] We say that a code with $\alpha = \Omega((l/\lambda)^3 \tau \ln((l/\lambda\epsilon)\tau))$ has "relatively" or "very" small aperture, if $\epsilon\tau q$ goes to zero or does not, as $k$ goes to infinity, respectively.

We compare (i) Theorem 15 with Theorem 18 for CC and OCC with relatively small apertures, and (ii) Theorem 16 with Theorem 19 for CC and OCC with very small apertures. We focus on the tradeoff between the probability of failure (MER) or the expected fraction of unrecoverable message packets (referred to as the "packet error rate" (PER)) and the speed of convergence.

*Theorem 20:* For a sufficiently small given MER and for relatively small apertures, OCC with larger overlap has higher speed of convergence than OCC with smaller overlap.

In the following, for the ease of notation, in the case of OCC, let $\theta$ be 2, or $\chi + \tau - 1$, for any constant integer $\chi$ sufficiently larger than $(\tau - 1)/\lambda$, through the analysis of the dependency types with positively or negatively dependent neighborhoods, respectively. Further, in the case of CC, let $\theta$ be 1.

---

[38]The overlap parameter $\tau$ of CC is 1, i.e., there is no overlap between the chunks.

*Proof of Theorem* 20: For positive $\epsilon, \epsilon', \lambda$ and $\lambda'$, and for some positive constants $c$ and $c'$, consider two OCC's with aperture sizes $\alpha = c(l/\lambda)^3 \tau \ln((l/\lambda\epsilon)\tau)$ and $\alpha' = c'(l/\lambda')^3 \tau' \ln((l/\lambda'\epsilon')\tau')$, and constant overlap parameters $\tau$ and $\tau'$ ($\tau' < \tau$), over two networks of length $l$ with two schedules of capacities $(1+\lambda)k$, and $(1+\lambda')k$, respectively. By the result of Theorem 18, it can be seen that for a given message error rate (i.e., $\epsilon^\theta \tau q = \epsilon^{\theta'} \tau' q$), and for a given aperture size (i.e., $\alpha = \alpha'$), the speed of convergence of OCC with overlap parameter $\tau$ is larger than that of OCC with overlap parameter $\tau'$ (i.e, $\lambda < \lambda'$), so long as

$$\epsilon < \left[ l^{\mu' - \mu} \frac{(\tau'^2/c')^{\mu'/3}}{(\tau^2/c)^{\mu/3}} \left( \frac{\tau'}{\tau} \right)^{\mu'} \right]^{\frac{1}{\kappa(\mu - \mu') + \mu'(\theta/\theta' - 1)}},$$

and $c'/c > \tau/\tau'$, for some constant $\kappa > 0$, where $\mu = c\tau/\lambda^3$, and $\mu' = c'\tau'/\lambda'^3$. ∎

The following result is a corollary of Theorem 20.

*Corollary 2:* For a sufficiently small given MER and for relatively small apertures, OCC has higher speed of convergence compared to CC.

The method of proving the following results are similar to that of Theorem 20 and Corollary 2, and hence omitted. The only difference is that, instead of Theorems 15 and 18, Theorems 16 and 19 are needed to study the tradeoff between the PER and the speed of convergence.

*Theorem 21:* For a sufficiently small given PER and for very small apertures, OCC with larger overlap has higher speed of convergence than OCC with smaller overlap.

*Corollary 3:* For a sufficiently small given PER and for very small apertures, OCC has higher speed of convergence compared to CC.

These results for CC and OCC with relatively/very small chunks over two types of worst-case schedules are also summarized in Table I. The comparison of the rows corresponding to CC and OCC with small chunks, as shown in Theorems 20 and 21, reveals that while $\tau$ increases, the speed of decreasing MER/PER is higher than the speed of increasing the lower bound on $\alpha$. Thus, for a given $\alpha$, and for sufficiently large $\lambda$, the MER/PER of OCC with larger overlap size (larger $\tau$) are smaller than that of OCC with smaller overlap size (e.g., CC). Moreover, for a given constant $\lambda$, and for relatively small $\alpha$ (i.e., when $\epsilon$ goes to zero sufficiently fast, as $k$ tends to infinity), the lower bound on $\alpha$ is logarithmic in $k$. For very small $\alpha$ (i.e., when $\epsilon$ is a constant with respect to $k$), however, the lower bound on $\alpha$ is constant with respect to $k$. This implies that for very small $\alpha$, the coding costs are constant with respect to $k$, and hence the code is linear-time. The above comparison therefore shows that linear-time OCC can outperform linear-time CC over networks with worst-case schedules.

## V. Simulation Results

In this section, we present our simulation results in two parts. The first part is concerned about the probability that banded random binary matrices are full rank; and the second part is concerned about the performance of dense codes, chunked codes and overlapped chunked codes over worst-case schedules.

### A. Banded Random Binary Matrices

We study both irregular symmetric and asymmetric banded random binary (BRB) matrices. The results for regular symmetric and regular asymmetric cases are similar to those for irregular symmetric and irregular asymmetric cases, respectively, and hence not presented.

*1) Setup:* In symmetric case, we consider matrices with $k = 128, 256, 512$ columns, and $n = k + m$ rows, where $m = 0, 1, ..., 10$. For $k = 128, 256$, we consider the aperture sizes $\alpha = k/8, k/4, k/2$, and for $k = 512$, we consider the aperture sizes $\alpha = k/16, k/8, k/4$. In asymmetric case, we consider matrices with $k = 512, 1024, 2048$ columns, and $n = k + m$ rows, where $m = 0, 1, 2, 4, 8, 16$; and for each $k$, we consider the aperture sizes $\alpha = k/8, k/4, k/2$. For a given $k$, the lower bound on the overlap size in asymmetric cases is larger than that in symmetric cases, and hence we need to consider larger matrices for asymmetric cases. In both symmetric and asymmetric cases, we consider the overlap sizes $\gamma = \alpha/\tau_e$, where $\tau_e = \tau/(\tau - 1)$, for the overlap parameters $\tau = 2, 4, 8$.

We simulate 10000 irregular symmetric or asymmetric BRB matrices for each $k, n, \alpha$, and $\gamma$, and plot the relative frequency of the number of times that the simulated matrices have full column-rank (the probability that a $(\gamma, \alpha)$ irregular (symmetric or asymmetric) banded random binary matrix of size $n \times k$ has rank $k$). For each $k$ and $n$, for comparison, we also plot the theoretical (asymptotic) result (derived in [25]) for the probability that a fully random matrix of size $n \times k$ has rank $k$ (in a fully random matrix, each entry is chosen uniformly at random from $\mathbb{F}_2$).

*2) Results:* Figures 7-9 show that for fixed $k$ and $n$, so long as the overlap size $\gamma \geq 2\sqrt{k}$, a $(\gamma, \alpha)$ irregular symmetric BRB matrix behaves similar to a fully random matrix of the same size in terms of the probability of being full column-rank.

Figures 10-12 show similar results for irregular asymmetric BRB matrices. The results show that the probability that such matrices have full column-rank is similar to that of fully random matrices when the aperture size $\gamma \geq \tau_e \tau \sqrt{k}$.

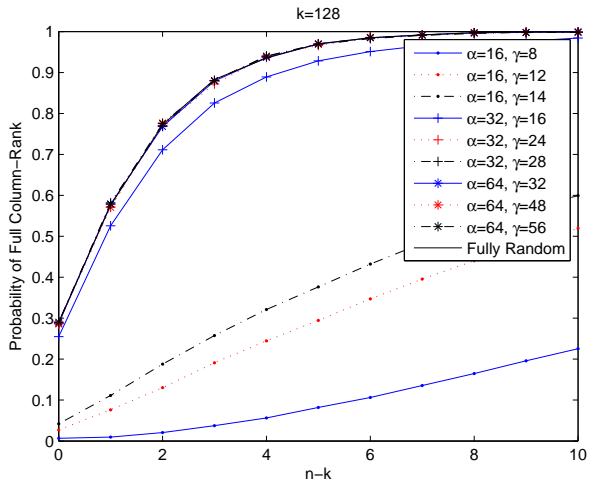

Fig. 7. The rank property of symmetric BRB matrices: $k = 128$.

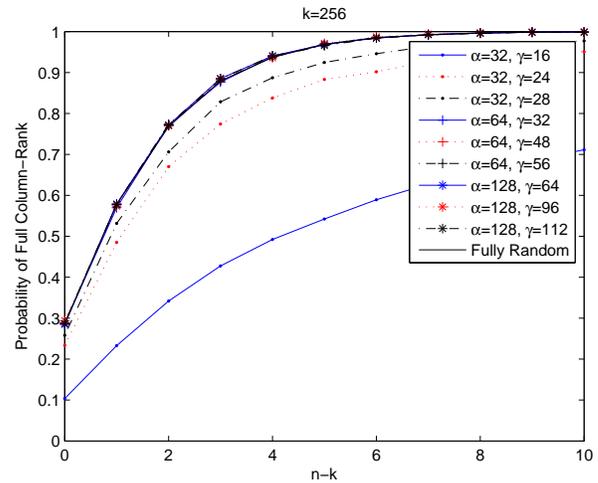

Fig. 8. The rank property of symmetric BRB matrices: $k = 256$.

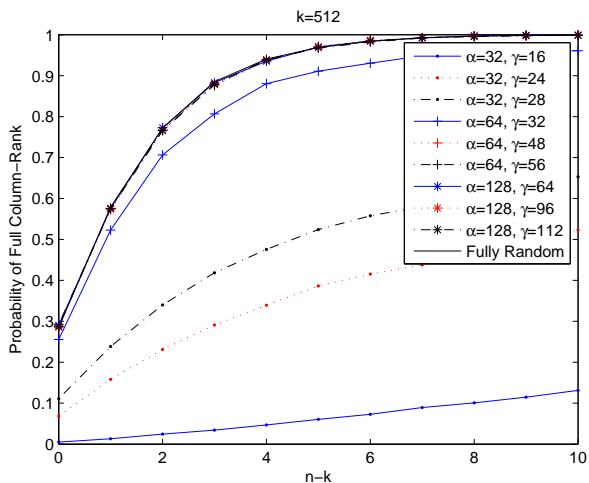

Fig. 9. The rank property of symmetric BRB matrices: $k = 512$.

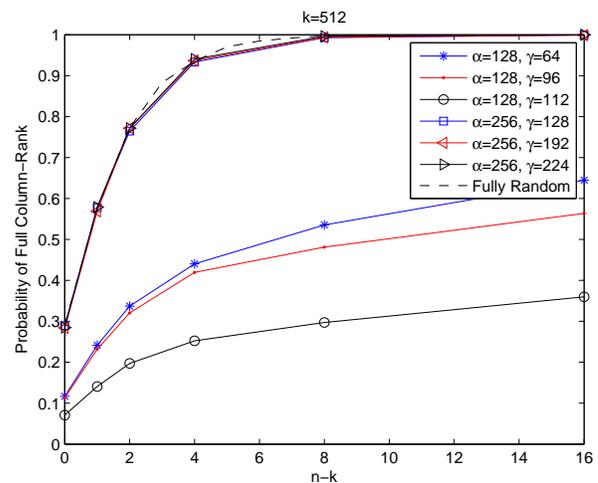

Fig. 10. The rank property of asymmetric BRB matrices: $k = 512$.

## B. Dense Codes, CC and OCC: Comparison

We compare the performance of finite-length dense codes, chunked codes and overlapped chunked codes over one-in-one-out and all-in-all-out worst-case schedules. The comparisons are in terms of the tradeoff between (i) the overhead per message $\lambda := (n - k)/k$, simply called "overhead," and the probability of decoding failure (message error rate, MER), and (ii) the overhead and the expected fraction of unrecoverable message packets (packet error rate, PER). The variables in these comparisons are the message, aperture and overlap sizes as well as the network length and the schedule type.

*1) Setup:* We consider line networks of lengths $l = 2, 4$. The networks are simulated with random codes over one-in-one-out or all-in-all-out worst-case schedules of capacity $n = (1 + \lambda)k$, for some $0 \leq \lambda \leq 3$, and with the message sizes $k = 64, 256$. For each $k$, we consider the aperture sizes $\alpha = k/8, k/4, k/2, k$, and the overlap sizes $\gamma = \alpha/\tau_e$, where $\tau_e = \tau/(\tau - 1)$, for the overlap parameters $\tau = 1$ (i.e., CC), $2$, and $4$. We are interested in the MER and PER, each as a function of the overhead, for a given aperture size (given coding costs). We would also like to investigate how each of these functions changes with the aperture and overlap sizes as well as the length of the network and the type of the worst-case schedule. For each $l, n, k, \alpha$ and $\gamma$, each coding scheme is applied to the underlying networks until 1000 decoding failures occur.

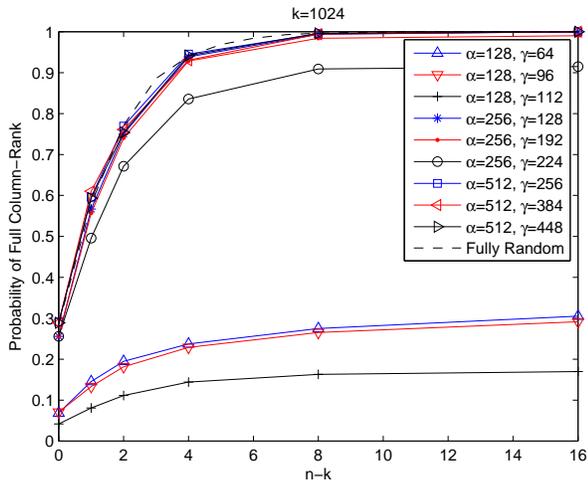 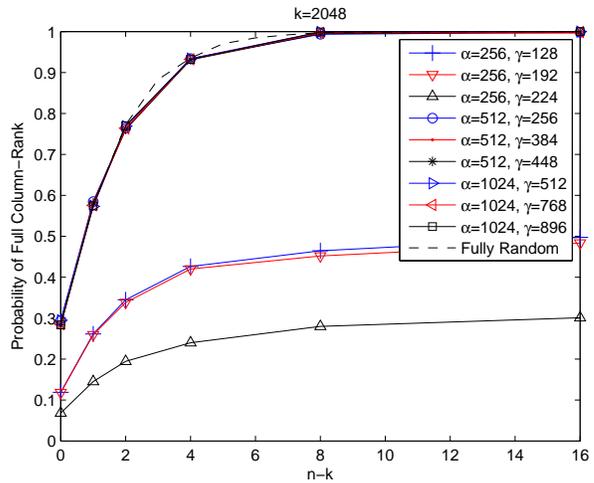

Fig. 11. The rank property of asymmetric BRB matrices: $k = 1024$. Fig. 12. The rank property of asymmetric BRB matrices: $k = 2048$.

*2) One-In-One-Out Worst-Case Schedules:* Figures 13 and 14 depict the MER vs. the overhead $\lambda$ for the cases where $l = 2$, and $k$ is $64$, and $256$, respectively (for different aperture and overlap sizes). Figures 15 and 16, respectively, depict the same scenarios as in Figures 13 and 14, only for longer networks of length $l = 4$. Similar results for PER, instead of MER, are presented in Figures 17-20. Since the trends for MER and PER are similar, in the following, we just discuss the MER results.

For given $k$ and $l$, investigating the results in each of the Figures 13-16 shows that for the same aperture size (same coding costs), OCC with larger overlap is more efficient (requires less overhead) than CC (OCC with smaller overlap), for sufficiently small MER. More detailed analysis of the simulation results is provided in the following.

*The Effect of the Message Size:* A comparison of Figures 13 and 14 shows that for a given aperture size $\alpha$, and a given network length, the MER below which OCC (OCC with larger overlap) outperforms CC (OCC with smaller overlap) is an increasing function of the message size. So, OCC are more advantageous over CC for a given coding cost (a given aperture size), when the message size is larger. For example, in Figure 13, it can be seen that for $k = 64$, and $\alpha = 32$, for MERs below about $0.35$, OCC with $\tau = 2$ requires a smaller overhead compared to CC ($\tau = 1$). However, as can be seen in Figure 14, for larger message size $k = 256$, and similar aperture size $\alpha = 32$, OCC with $\tau = 2$ is superior to CC, for MERs below about 1 (for almost all MERs).

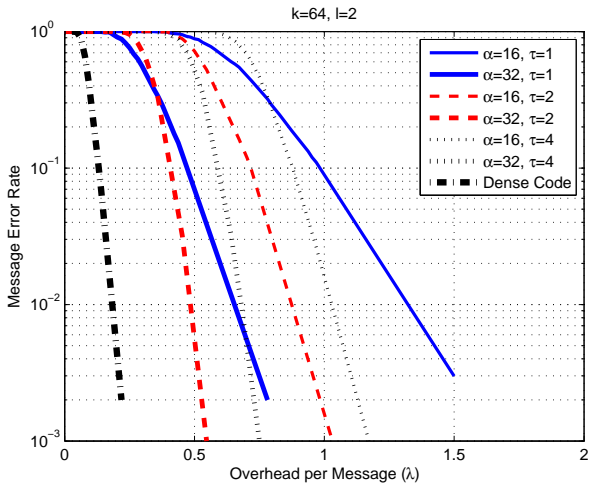

Fig. 13. CC and OCC over one-in-one-out schedules: smaller message size and shorter network.

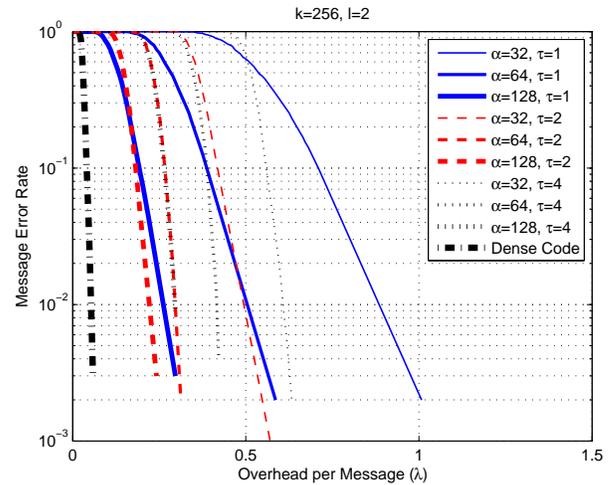

Fig. 14. CC and OCC over one-in-one-out schedules: larger message size and shorter network.

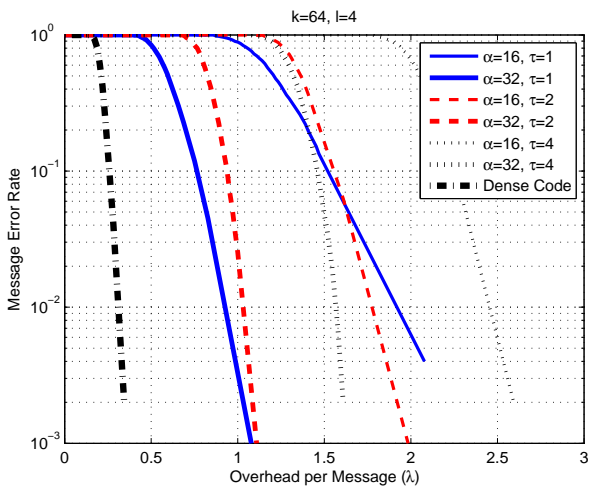

Fig. 15. CC and OCC over one-in-one-out schedules: smaller message size and longer network.

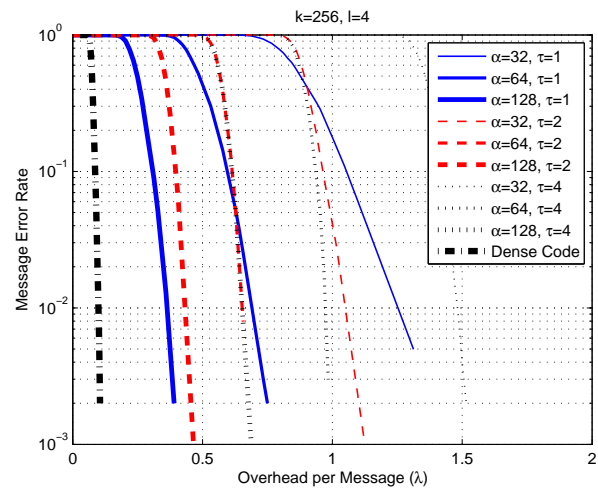

Fig. 16. CC and OCC over one-in-one-out schedules: larger message size and longer network.

By comparing Figures 15 and 16, similar behavior can be observed for longer networks.

*The Effect of the Network Length:* Comparison between Figures 13 and 15, or between Figures 14 and 16 demonstrates that OCC (OCC with larger overlap) are more advantageous over CC (OCC with smaller overlap) for shorter networks. For example, Figure 13 (shorter network) shows that for $\alpha = 16$, OCC with $\tau = 2$ is superior to CC for MERs below 0.3; however, as can be seen in Figure 15 (longer network), this occurs for MERs below 0.035.

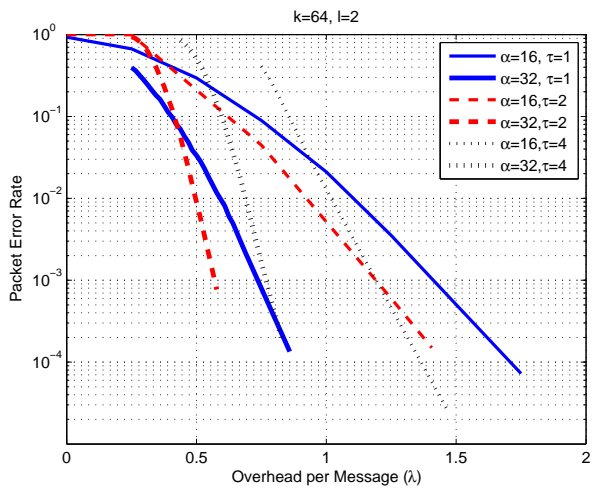

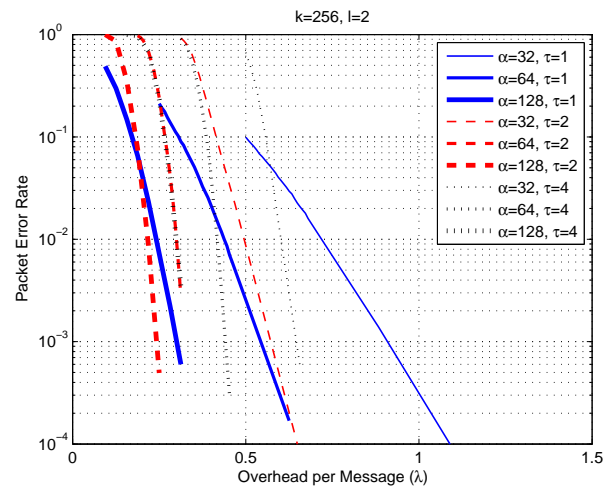

Fig. 17.  CC and OCC over one-in-one-out schedules: smaller message size and shorter network.

Fig. 18.  CC and OCC over one-in-one-out schedules: larger message size and shorter network.

*The Effect of the Overlap Size:* In Figures 13-16, one can see that for a given aperture size $\alpha$, the MER below which OCC (OCC with larger overlap) outperforms CC (OCC with smaller overlap) is a decreasing function of the overlap size. Therefore, for a given coding cost, and sufficiently small MERs, OCC with larger overlaps are more advantageous. For example, as can be seen in Figure 14, for $\alpha = 32$, OCC with $\tau = 4$ (larger overlap size) outperforms CC for MERs below about $0.6$; and OCC with $\tau = 2$ (smaller overlap size) is superior to CC for MERs below about $1$. Moreover, the slope of the curves for OCC with $\tau = 4$ and OCC with $\tau = 2$ shows that the former crosses (outperforms) the latter for a MER somewhere below $10^{-3}$.

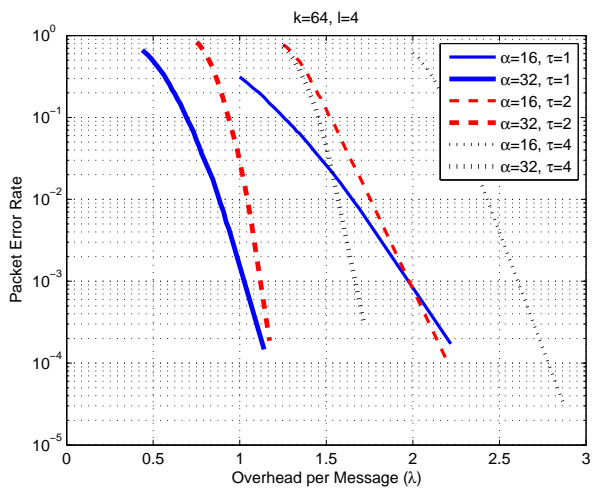

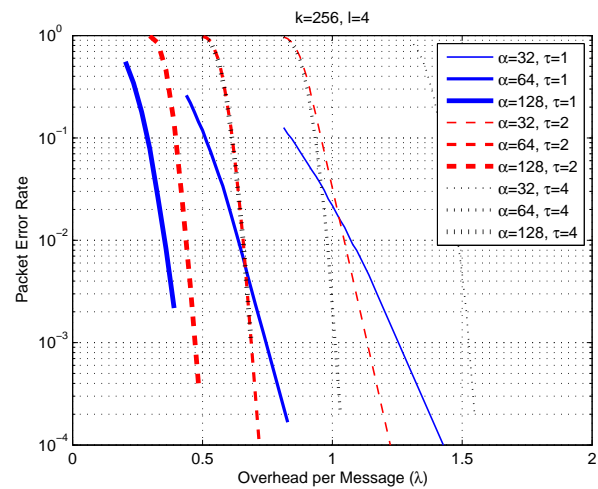

Fig. 19.  CC and OCC over one-in-one-out schedules: smaller message size and longer network.

Fig. 20.  CC and OCC over one-in-one-out schedules: larger message size and longer network.

*The Effect of the Aperture Size:* It can be seen in Figures 13-16 that for a given overlap parameter, and given message size and network length, the smaller the aperture size (the smaller the coding costs), the larger the overhead.

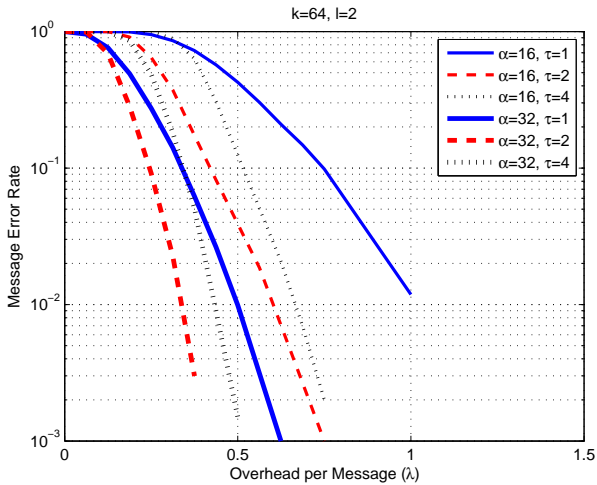

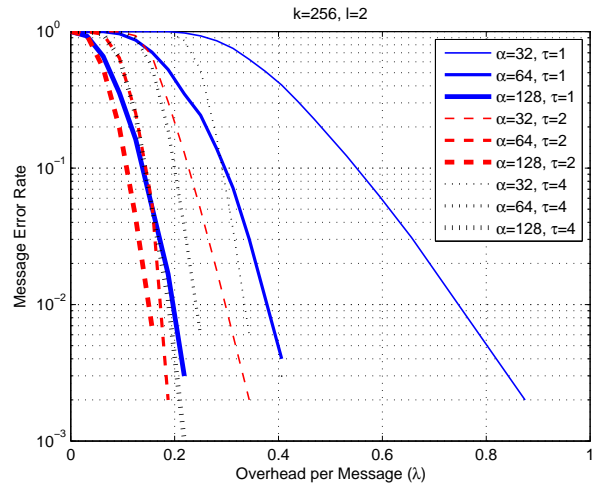

Fig. 21. CC and OCC over all-in-all-out schedules: smaller message size and shorter network.

Fig. 22. CC and OCC over all-in-all-out schedules: larger message size and shorter network.

*3) All-In-All-Out Worst-Case Schedules:* Figures 21-24 and 25-28 represent similar scenarios to those in Figures 13-16 and 17-20, for MER and PER, respectively. While the general trends for all-in-all-out schedules are similar to those of one-in-one-out schedules, discussed in the previous part, those trends are more pronounced for the all-in-all-out schedules. As an example, comparisons between Figures 13 and 21 reveal that OCC with $\alpha = 16$ and $\tau = 4$ outperforms CC of the same aperture size for MERs below $0.3$ for one-in-one-out schedules. This crossover MER for the all-in-all-out schedules is improved to $0.85$. The corresponding values of MER if $\alpha$ is increased to $32$ are $0.005$ and $0.9$, for the two categories of schedules, respectively. Therefore, for a randomly generated worst-case schedule, one expects to see improvements in the performance/complexity tradeoff of OCC versus CC that are in between those reported for one-in-one-out and all-in-all-out schedules.

*Example 1:* Consider downloading a 1MB file from a file server $4$ hops away ($l = 4$). Assuming the worst-case scenario, the underlying network is under a worst-case schedule. Suppose that packets of length 4KB are used for the transmission. This implies that the number of message packets $k = 256$. Consider a target PER of $10^{-4}$, and two possible transmission scenarios: (a) using a CC with $\alpha = 64$, and (b) using an OCC with $\alpha = 64$ and $\tau = 2$ ($\gamma = 32$). From Figure 20 (for one-in-one-out schedules), one can see that the overhead $\lambda$ for the two scenarios (a) and (b) is about $0.85$ and $0.7$, respectively. From Figure 28 (for all-in-all-out schedules), it can be seen that the overhead $\lambda$ for the two scenarios is about $0.5$ and $0.35$, respectively. This implies that downloading the file by scenario (b) is from about $17\%$ to about $30\%$

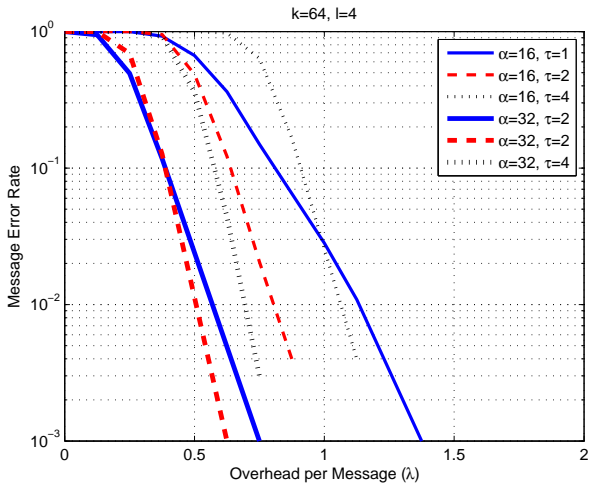

Fig. 23. CC and OCC over all-in-all-out schedules: smaller message size and longer network.

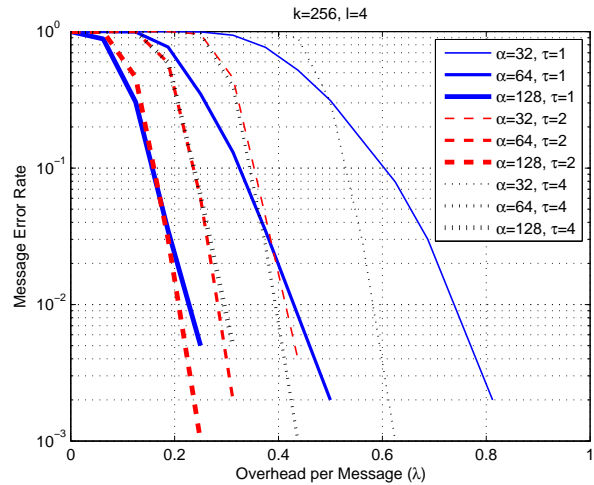

Fig. 24. CC and OCC over all-in-all-out schedules: larger message size and longer network.

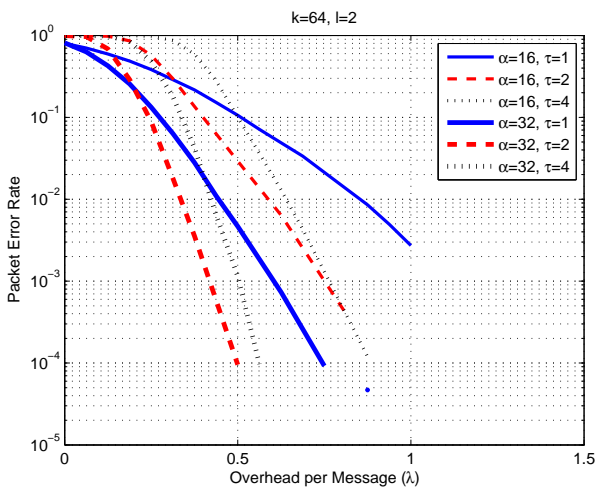

Fig. 25. CC and OCC over all-in-all-out schedules: smaller message size and shorter network.

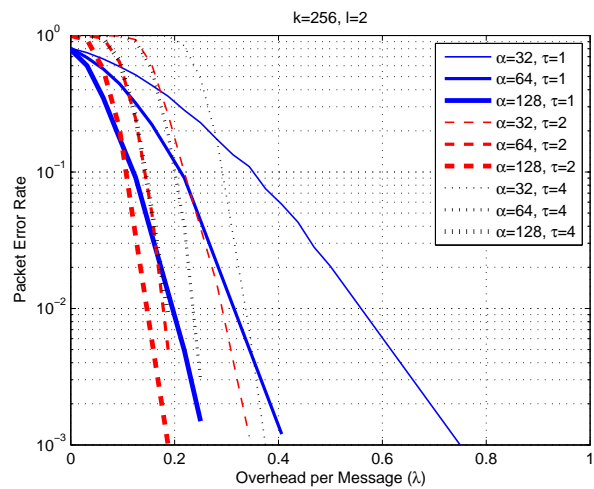

Fig. 26. CC and OCC over all-in-all-out schedules: larger message size and shorter network.

faster than scenario (a), depending on the type of the worst-case schedule under consideration.

## VI. CONCLUSION

In this paper, we analyzed and designed communicationally and computationally efficient network codes over line networks with worst-case schedules.

We first presented a detailed analysis of random linear network codes (dense codes). This analysis and its building blocks then served as part of our machinery to analyze chunk-based coding schemes. In particular, tighter bounds on the performance of "chunked codes" (CC) by Maymounkov *et al.* were derived. It was shown that for sufficiently large chunks (super-logarithmic aperture size in the message length), chunked codes asymptotically achieve the capacity, yet with a slower speed of convergence (smaller communicational efficiency) compared to dense codes. This is the price for the superior computational efficiency of CC.

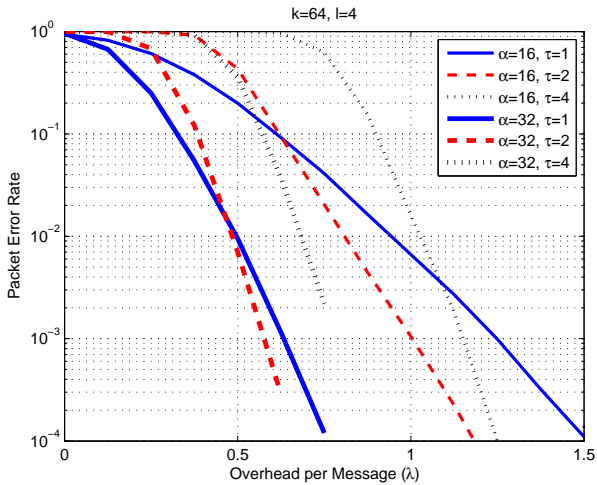

Fig. 27. CC and OCC over all-in-all-out schedules: smaller message size and longer network.

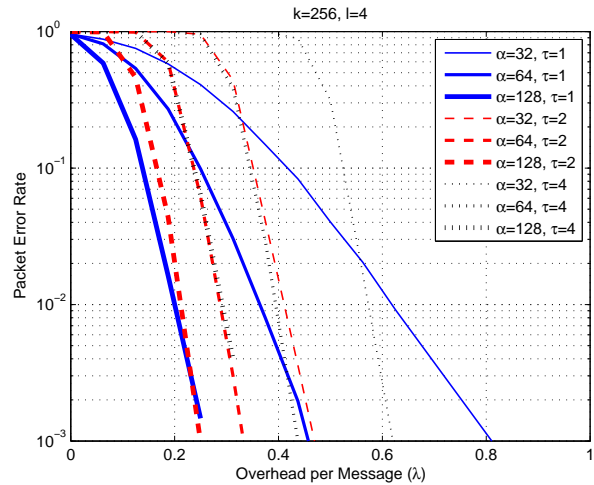

Fig. 28. CC and OCC over all-in-all-out schedules: larger message size and longer network.

Searching for codes that are more computationally efficient than dense codes and more communicationally efficient than chunked codes, we proposed and analyzed chunked codes with overlapping chunks, called "overlapped chunked codes" (OCC).

We showed that (i) for sufficiently large apertures (super-logarithmic in the message length), an OCC with a constant overlap parameter is asymptotically capacity-achieving, and (ii) for a given aperture size (similar coding costs), an OCC with sufficiently large apertures performs inferior to a CC (OCC with zero overlap) with regards to the tradeoff between the speed of convergence and the message error rate. We also proved that the larger is the overlap size in OCC with sufficiently large apertures, the worse is the tradeoff between the speed of convergence and the message error rate.

Targeting practical scenarios, where computational resources are scarce, we analyzed CC and OCC with smaller aperture sizes (logarithmic or sub-logarithmic in the message size down to some constant).

We showed that there exists a lower bound on the aperture size logarithmic in the message length, such that a code with relatively small apertures satisfying the given bound, yet not satisfying the bound for sufficiently large apertures, asymptotically approaches the capacity with an arbitrarily small constant non-zero gap. We also proved that for codes with relatively small aperture sizes, the larger is the overlap size, the better would be the tradeoff between the speed of convergence and the message error rate.

We, next, showed that there exists a lower bound on the aperture size constant in the message length,

such that a code with very small apertures, i.e., with an aperture size satisfying the given bound, but not satisfying the bound for relatively small apertures, asymptotically approaches the capacity with an arbitrarily small constant non-zero gap. In this case, the message error rate is bounded away from zero, yet still, the packet error rate can be made arbitrarily small. We also proved that for codes with very small aperture sizes, the larger is the overlap size, the better would be the tradeoff between the speed of convergence and the packet error rate. In fact, one of the main contributions of this work was to design linear-time network codes (OCC with very small apertures) that are both computationally and communicationally efficient, and that outperform linear-time CC.

In line with the asymptotic analytical results, we also presented finite-length simulation results which verified that OCC with larger overlaps are more efficient, both communicationally and computationally, when sufficiently small message or packet error rates are desired.

## APPENDIX I

### PROOF OF LEMMAS 1–9

*Proof of Lemma* 1: Let $\boldsymbol{v} = T\boldsymbol{u}$, and $\boldsymbol{v} = [v_1; ...; v_h]$. The proof of uniformity of $v_i$'s is straight-forward. To prove the independence of $v_i$'s, it suffices to show that for any integer $1 < j \leq h$, $v_j$ is independent of the set of $v_m$'s, for all $1 \leq m < j$. First, the independence of $v_2$ and $v_1$: The vector of coefficients in the linear combination of $v_1$ is linearly independent of that of $v_2$, due to the assumption of $T$ having full row-rank. Thus there exists at least one entry of $\boldsymbol{u}$ which only appears in either $v_1$, or $v_2$. Since the entries of $\boldsymbol{u}$ are all independent, this implies the independence of $v_2$ and $v_1$. Second, the independence of $v_3$ and $v_2, v_1$: The following lemma shows that $v_3$ is independent of $v_2, v_1$, if and only if $v_3$ is independent of any non-zero linear combination of $v_2$ and $v_1$, given that $v_2$ and $v_1$ are independent.

*Lemma 29:* [26] A discrete random variable $z$ is independent of a set of independent Bernoulli random variables if and only if $z$ is independent of any non-zero linear combination of the Bernoulli random variables with coefficients from $\mathbb{F}_2$.

Since $v_1$ and $v_2$ are independent, then by Lemma 29, it suffices to show that $v_3$ is independent of any non-zero linear combination of $v_2$ and $v_1$. The vector of coefficients in the linear combination of $v_3$ is linearly independent of those of $v_2$ and $v_1$, and hence is linearly independent of any non-zero linear combination of $v_2$ and $v_1$. Then, for any given non-zero linear combination $v_{1,2}$ of $v_2$ and $v_1$, there exists at least one entry of $u$, say $u_\ell$, for some $1 \leq \ell \leq h$, so that $u_\ell$ only appears in either $v_3$ or $v_{1,2}$. Thus, $v_3$ and $v_{1,2}$ are independent. Similarly, it can be shown that for all $1 < j \leq h$, $v_j$ is independent of the set of $v_m$'s, for all $1 \leq m < j$, as was to be shown. ∎

*Proof of Lemma* 2: By the assumption of $r(T) \geq h$, there exist at least $h$ linearly independent rows in $T$. Let $T'$ be $T$ restricted to these $h$ linearly independent rows in $T$, i.e., $T'$ is an $h \times d$ sub-matrix of $T$. Then, $T'M$ has $h$ rows which are a subset of rows in $TM$. Clearly $\mathcal{D}(TM) \geq \mathcal{D}(T'M)$. For $i \leq h$, and $j \leq k$, $(T'M)_{i,j} = \sum_{m=1}^{d} T'_{i,m} M_{m,j}$ and $\{M_{m,j} : m \leq d, j \leq k\}$ are all i.u.d. Then, for $i, i' \leq h$, $j, j' \leq k$ (and for $i \neq i'$, or $j \neq j'$), $(T'M)_{i,j}$ and $(T'M)_{i',j'}$ are i.u.d.[39] Thus the $h$ rows in $T'M$ are all dense, i.e., $\mathcal{D}(T'M) = h$, and hence $\mathcal{D}(TM) \geq h$. ∎

*Proof of Lemma* 3: For any integer $0 \leq \gamma \leq d-1$, let $T'$ be $T$ restricted to its first $d - \gamma$ columns, i.e., $T'$ is an $n \times (d-\gamma)$ sub-matrix of $T$. Suppose that there exists a non-zero column-vector $\boldsymbol{u}$ of length $(d-\gamma)$ whose entries are from $\mathbb{F}_2$ and that the column-vector $T'\boldsymbol{u}$ of length $n$ is an all-zero vector. This is the necessary and the sufficient condition for $r(T') < d - \gamma$. Suppose that the first non-zero entry of $\boldsymbol{u}$ is the $j$th entry. There are $2^{d-\gamma-j}$ such vectors. Since there exist at least $d - j + 1$ i.u.d. random entries in $j$th column of $T'$, there exist at least $d - j + 1$ i.u.d. random entries in the vector $T'\boldsymbol{u}$. The probability of all these entries being zero is $2^{-(d-j+1)}$, and thus the probability of $T'\boldsymbol{u}$ being zero given that $\boldsymbol{u}$ is a vector with the first non-zero entry being the $j$th is at most $2^{-(d-j+1)}$. Taking a union bound over all such vectors $\boldsymbol{u}$ whose first non-zero entry is the $j$th, the probability of $T'\boldsymbol{u}$ being an all-zero vector given that the first non-zero entry of $\boldsymbol{u}$ is its $j$th is at most $2^{d-\gamma-j} \times 2^{-(d-j+1)} = 2^{-(\gamma+1)}$. Taking a union bound over all $j$, (i.e., $1 \leq j \leq d - \gamma$), the probability of $T'\boldsymbol{u}$ being an all-zero vector is at most $(d-\gamma)2^{-(\gamma+1)}$. ∎

*Proof of Lemma* 4: Since $T'_u$ is a $n \times \mathcal{D}(Q_u)$ matrix over $\mathbb{F}_2$ such that for all $j \in \mathcal{I}_{u_\infty}$, $|\mathcal{T}'^{(j)}_{u_{\text{col}}}| \geq \mathcal{D}(Q_u) - j + 1$, Lemma 3 implies that $r(T'_u) \geq \mathcal{D}(Q_u) - \gamma$, for any integer $0 \leq \gamma \leq \mathcal{D}(Q_u) - 1$, fails to hold w.p. b.a.b. $(\mathcal{D}(Q_u) - \gamma)2^{-\gamma-1}$. Lemma 2 together with Lemma 3 give $\mathcal{D}(Q_v) = \mathcal{D}(T'_u Q'_u) \geq \mathcal{D}(Q_u) - \gamma$ fails to hold w.p. b.a.b. $(\mathcal{D}(Q_u) - \gamma)2^{-\gamma-1}$. Setting $(\mathcal{D}(Q_u) - \gamma)2^{-\gamma-1} \leq \epsilon$, we get

$$\gamma \geq \log(\mathcal{D}(Q_u) - \gamma) + \log(1/\epsilon) - 1. \tag{19}$$

Let $\gamma$ be the smallest integer equal to or larger than $\log \mathcal{D}(Q_u) + \log(1/\epsilon)$.[40] Thus, $\mathcal{D}(Q_v) \geq \mathcal{D}(Q_u) - \log \mathcal{D}(Q_u) - \log(1/\epsilon)$ fails to hold w.p. b.a.b. $\epsilon$. ∎

---

[39]For $j \neq j'$, $(T'M)_{i,j}$ and $(T'M)_{i',j'}$ are linear combinations of disjoint sets of i.u.d. random variables over $\mathbb{F}_2$ and hence they are i.u.d.; and for $j = j'$ and $i \neq i'$, $(T'M)_{i,j}$ and $(T'M)_{i',j'}$ are linear combinations of the same set of random variables over $\mathbb{F}_2$ with linearly independent vectors of coefficients, and hence they are i.u.d. (by Lemma 1).

[40]Such a choice of $\gamma$ satisfies (19) because for any $\gamma \geq 0$, $\log \mathcal{D}(Q_u) + \log(1/\epsilon) \geq \log(\mathcal{D}(Q_u) - \gamma) + \log(1/\epsilon) > \log(\mathcal{D}(Q_u) - \gamma) + \log(1/\epsilon) - 1$.

*Proof of Lemma* 5: Since $\mathcal{D}(Q_v) \leq n$ for every node $v$, replacing $\epsilon$ with $\epsilon/(l-1)$ in Lemma 4, we get

$$
\begin{aligned}
\mathcal{D}(Q_u) - \mathcal{D}(Q_v) &\leq \log \mathcal{D}(Q_u) + \log((l-1)/\epsilon) \\
&\leq \log(n(l-1)/\epsilon) \\
&\leq \log(nl/\epsilon),
\end{aligned}
$$

which fails to hold w.p. b.a.b. $\epsilon/(l-1)$. Thus,

$$
\sum_{i=1}^{l-1} \left( \mathcal{D}(Q_{v_i}) - \mathcal{D}(Q_{v_{i+1}}) \right) \leq (l-1)\log(nl/\epsilon)
$$

fails to hold w.p. b.a.b. $\epsilon$. Further,

$$
\mathcal{D}(Q_t) = \mathcal{D}(Q_{v_1}) - \sum_{i=1}^{l-1} \left( \mathcal{D}(Q_{v_i}) - \mathcal{D}(Q_{v_{i+1}}) \right).
$$

Since $\mathcal{D}(Q_{v_1}) = n$, we get

$$
\begin{aligned}
\mathcal{D}(Q_t) &= n - \sum_{i=1}^{l-1} \left( \mathcal{D}(Q_{v_i}) - \mathcal{D}(Q_{v_{i+1}}) \right) \\
&\geq n - (l-1)\log(nl/\epsilon) \\
&\geq n - l\log(nl/\epsilon)
\end{aligned}
$$

which holds w.p. b.b.b. $1 - \epsilon$. ∎

*Proof of Lemma* 6: Clearly, $r(M) < k$ if there exists a non-zero column vector $\boldsymbol{u}$ of length $k$, so that $M\boldsymbol{u} = \boldsymbol{0}$. For a given non-zero $\boldsymbol{u}$, the probability that $M\boldsymbol{u} = \boldsymbol{0}$ is equal to $2^{-d}$. Taking a union bound over all such vectors $\boldsymbol{u}$, $\Pr[r(M) < k]$ is upper bounded by $(2^k - 1)2^{-d} \leq 2^{k-d}$. Setting $2^{k-d} \leq \epsilon$, i.e., $k \leq d - \log(1/\epsilon)$, we get $\Pr[r(M) < k] \leq \epsilon$. ∎

*Proof of Lemma* 7: The following lemma is useful in order to give an upper bound on $\Pr[r(T) < d - \gamma]$, for any integer $0 \leq \gamma \leq d - 1$.

*Lemma 30:* [25, Theorem 1] Let $T$ be a $n \times d$ matrix over $\mathbb{F}_2$ whose entries are all i.u.d. random variables. Then, for any integer $1 \leq \gamma \leq d - 1$,

$$
\Pr[r(T) = d - \gamma] = c_\gamma \cdot 2^{-\gamma(n-d+\gamma)},
$$

for $c_\gamma = \prod_{i=n-d+\gamma+1}^\infty \left(1 - 2^{-i}\right) / \prod_{i=1}^\gamma \left(1 - 2^{-i}\right)$.

For any integer $1 \leq \gamma \leq d-1$, it can be shown by induction that $c_\gamma \leq 2^\gamma$. Thus,

$$\Pr[r(T) = d - \gamma] \leq 2^\gamma \cdot 2^{-\gamma(n-d+\gamma)} \leq 2^\gamma \cdot 2^{-\gamma^2} \leq 2^{-\gamma}.$$

The probability of $r(T) < d - \gamma$ is the sum of the probabilities of $r(T)$ being $d - (\gamma + 1)$, $d - (\gamma + 2)$, ..., or 1. Thus,

$$\begin{aligned}
\Pr[r(T) < d - \gamma] &= \sum_{i=\gamma+1}^{d-1} \Pr[r(T) = d - i] \\
&\leq \sum_{i=\gamma+1}^{d-1} 2^{-i} \\
&= 2^{-\gamma} - 2^{-(d-1)} \\
&\leq 2^{-\gamma}.
\end{aligned}$$

∎

*Proof of Lemma* 8: Since $T'_u$ is a $n \times \mathcal{D}(Q_u)$ matrix over $\mathbb{F}_2$, such that for all $j \in \mathcal{I}_{u_\infty}$, $|\mathcal{T}'^{(j)}_{u_{\text{col}}}| = n$, Lemma 7 implies that $r(T'_u) \geq \mathcal{D}(Q_u) - \gamma$ fails to hold w.p. b.a.b. $2^{-\gamma}$, for any integer $0 \leq \gamma \leq \mathcal{D}(Q_u) - 1$. Lemma 2 together with Lemma 7 then result in $\mathcal{D}(Q_v) = \mathcal{D}(T'_u Q'_u) \geq \mathcal{D}(Q_u) - \gamma$, which fails to hold w.p. b.a.b. $2^{-\gamma}$. Setting $2^{-\gamma} \leq \epsilon$, we get $\gamma \geq \log(1/\epsilon)$. Let $\gamma$ be the smallest integer equal to or larger than $\log(1/\epsilon)$. Thus, $\mathcal{D}(Q_v) \geq \mathcal{D}(Q_u) - \log(1/\epsilon)$ fails w.p. b.a.b. $\epsilon$. ∎

*Proof of Lemma* 9: The proof follows the same steps as the proof of Lemma 5, except that the bound of Lemma 4 is replaced by that of Lemma 8. ∎

# Appendix II

## Proof of Lemma 10

We show that with high probability for any $\omega$, the most of $\omega$-packets in any given subset of packets consecutively received by a node (when the size of the subset is chosen with some care) are matched up with the most of $\omega$-packets in a particular subset of packets consecutively transmitted afterwards by the node. We formalize this idea in the following.

Consider a non-sink (or non-source) node $u$ (or $v$). Consider the transmissions over the link $(u, v)$. We partition the $n$ packets sent (received) by the node $u$ (node $v$) into $b$ *departure* (*arrival*) buckets from the perspective of the node $u$ (node $v$) such that the first bucket includes the first $n/b$ packets sent (or received) by the node, the second bucket includes the second $n/b$ packets and so forth. The packets sent by the node $u$ might not be received in order by node $v$, and hence, for all $1 \leq j \leq b$, the packets in the $j$th departure bucket of node $u$ might be different from those in the $j$th arrival bucket of node $v$.

The expected number of $\omega$-packets in any departure or arrival bucket is $\mu = n/bq$, for any divisor $bq$ of $n$, and the expected number of $\omega$-packets in all the departure or arrival buckets is $n/q$. The chunks are however randomly scheduled by the nodes and hence the actual number of $\omega$-packets in any bucket is a random variable.

Let $Y$ denote the number of $\omega$-packets in a given bucket. Thus, $\mathrm{E}[Y] = \mu$. The set of labels of packets in the bucket is denoted by $I$. Thus, $|I| = n/b$. Every packet in the bucket happens to be an $\omega$-packet independent of any other packet. Random variable $Y$ is thus the sum of independent indicator random variables $\{X_i : i \in I\}$ where $X_i$ is 1, if the $i$th packet is an $\omega$-packet and, $X_i$ is 0, otherwise.

*Lemma 31:* [27, Corollary A.1.14] Let $Y$ be the sum of mutually independent indicator (arbitrarily distributed) random variables, and $\mu = \mathrm{E}[Y]$. For any $\delta > 0$, $\Pr[Y \geq (1+\delta)\mu] \leq e^{-\delta^2 \mu/2}$, and $\Pr[Y \leq (1-\delta)\mu] \leq e^{-\delta^2 \mu/2}$.

Let $\mu'$, and $\mu''$ be $\mu - c\sqrt{\mu}$, and $\mu + c\sqrt{\mu}$, respectively, for some positive $c$. Taking $\delta = c/\sqrt{\mu}$, and $c = \sqrt{2\ln(2/\epsilon)} = O(\sqrt{\ln(1/\epsilon)})$ in Lemma 31, we get $\Pr[Y > \mu''] < \epsilon/2$, and $\Pr[Y < \mu'] < \epsilon/2$, i.e., $Y$ is less than $\mu'$, or greater than $\mu''$, w.p. b.a.b. $\epsilon$. Thus the number of $\omega$-packets for a given $\omega$ in a given bucket fails to lie within $\mu'$ and $\mu''$ w.p. b.a.b. $\epsilon$.

We are interested in deriving an upper bound (or a lower bound) on the number of unusable (or usable) $\omega$-packets for a given $\omega$.

Fix a particular interior node $u$ ($u = v_i$, for $1 \leq i < l$). There exists at least one packet sent by node $u$ at some time $\tau' \geq \tau$, after receiving a packet at time $\tau$. Thus, all the packets in the $j$th arrival bucket of node $u$, for all $1 \leq j < b$, land at node $u$ before any of the packets in the $(j+1)$th departure bucket of node $u$ leaves. Thus, a given $\omega$-packet in the $j$th arrival bucket of node $u$ can be matched up with any $\omega$-packet in the $(j+1)$th departure bucket of node $u$.

Consider the $\omega$-packets over two tandem links $(w, u)$ and $(u, v)$ (note that node $w$ or node $v$ might be node $s$ or node $t$, respectively). We randomly choose $\mu'$ $\omega$-packets in each departure (arrival) bucket of nodes $w$ and $u$ (nodes $u$ and $v$) and call them *half-good* from the perspective of node $w$, or node $u$ (node $u$, or node $v$), respectively. Since each packet over these two links belongs to one departure bucket (of node $w$, or node $u$) and one arrival bucket (of node $u$, or node $v$), we call a packet *good* if it is half-good from the perspective of both nodes $w$ and $u$ (or nodes $u$ and $v$).

For all $1 < j \leq b$, there exists at least one particular good $\omega$-packet in the $(j-1)$th arrival bucket of node $u$ to be matched up with any given good $\omega$-packet in the $j$th departure bucket of node $u$. Thus, the set of good $\omega$-packets in all but the last arrival bucket of node $u$ are a subset of usable $\omega$-packets received by node $u$ and hence the number of good $\omega$-packets in these buckets gives a lower bound on the number of usable $\omega$-packets at node $u$.

Taking $c = o(\sqrt{\mu})$, the number of those $\omega$-packets in any given bucket of any node which are not good fails to be less than $\mu'' \cdot O(c/\sqrt{\mu}) = O(c\sqrt{\mu})$ w.p. b.a.b. $\epsilon$.[41] Taking a union bound over all but the last arrival bucket of node $u$, w.p. b.a.b. $(b-1)\epsilon$, the number of $\omega$-packets which are not good fails to be less than $(b-1) \cdot O(c\sqrt{\mu}) \leq b \cdot O(c\sqrt{\mu}) = O(bc\sqrt{n/bq}) = O(c\sqrt{nb/q})$.

The number of $\omega$-packets which are not usable due to landing at the $b$th arrival bucket of node $u$ is, w.p. b.a.b. $\epsilon$, larger than $\mu'' = \mu + c\sqrt{\mu} = (1 + o(1))\mu = O(\mu) = O(n/bq)$. Such packets are unusable because there is no departure bucket at node $u$, such that the packets therein can be matched up with the packets in the last arrival bucket of node $u$.

Thus by adding the number of packets in both groups of unusable packets together, i.e., $O(c\sqrt{nb/q})$, and $O(n/bq)$, the probability that there are more than $O(c\sqrt{nb/q} + n/bq)$ unusable $\omega$-packets at node $u$ is at most $b\epsilon$. Summing over all the interior nodes of the network ($v_i$'s for all $1 \leq i < l$), w.p. b.a.b. $(l-1)b\epsilon < lb\epsilon$, the total number of unusable $\omega$-packets at node $t$ fails to be less than

$$O((l-1)(c\sqrt{nb/q} + n/bq)) \leq O(l(c\sqrt{nb/q} + n/bq)). \tag{20}$$

We now specify $c$ so that for every $\omega$ the probability that the number of unusable $\omega$-packets at node $t$ is larger than (20) is at most $\epsilon$. Replacing $\epsilon$ by $\epsilon/lbq$ in Lemma 31, one can readily see that $c$ needs to be at least $O(\sqrt{\ln(lbq/\epsilon)})$. Let $c = O(\sqrt{\ln(ln/\epsilon)})$ ($bq$ is a divisor of $n$ and hence smaller than $n$). Minimizing the number of unusable $\omega$-packets at node $t$ with respect to $b$ (by setting the derivative of

---

[41]Hereafter, we operate under the assumption that the actual number of $\omega$-packets in any bucket lies within $\mu'$ and $\mu''$. Let $\mu^*$ be the actual number of $\omega$-packets (for a given $\omega$) in a given bucket. We randomly choose $\mu'$ packets from the set of $\mu^*$ packets (i.e., $\mu^* - \mu'$ packets will not be chosen). Therefore the probability that a given $\omega$-packet in the given bucket is not half-good from the perspective of its transmitting node is $(\mu^* - \mu')/\mu^*$. Since $\mu^* \leq \mu''$, the latter probability is upper bounded by $(\mu'' - \mu')/\mu''$. Similarly, from the perspective of its receiving node, the given $\omega$-packet fails to be half-good w.p. b.a.b. $(\mu'' - \mu')/\mu''$. Thus the probability that a given $\omega$-packet is not good (it is not half-good from the perspective of its transmitting or receiving node) is at most $2(\mu'' - \mu')/\mu'' \leq 2(\mu'' - \mu')/\mu' = 4c/(\sqrt{\mu} - c)$ (the inequality follows from $\mu' < \mu''$, and the equality follows from replacing $\mu'$ and $\mu''$, respectively, with $\mu - c\sqrt{\mu}$, and $\mu + c\sqrt{\mu}$). Taking $c = o(\sqrt{\mu})$, we have $4c/(\sqrt{\mu} - c) \leq 4c/[(1 - o(1))\sqrt{\mu}] \leq (1 + o(1))(4c/\sqrt{\mu}) = O(c/\sqrt{\mu})$.

(20) with respect to $b$ equal to zero and solving for $b$), we obtain

$$b = \left\lceil \left( \frac{n}{q \ln(ln/\epsilon)} \right)^{1/3} \right\rceil,$$

where the ceiling $\lceil \cdot \rceil$ has no significant effect provided that $q \ln(ln/\epsilon) = o(n)$. Such choices of $c$ and $b$ ensure that $c = o(\sqrt{\mu}) = o(\sqrt{n/bq})$, because $o(\sqrt{n/bq}) = o((n/q)^{1/3} \ln^{1/6}(ln/\epsilon))$, and $c = O(\sqrt{\ln(ln/\epsilon)}) = o((n/q)^{1/3} \ln^{1/6}(ln/\epsilon))$, so long as $q \ln(ln/\epsilon) = o(n)$. Substituting $c$ and $b$ into (20), the total number of unusable $\omega$-packets at node $t$ is at most

$$O\left( l \, (n/q)^{2/3} \ln^{1/3}(ln/\epsilon) \right),$$

w.p. b.b.b. $1 - \epsilon$. The expected number of $\omega$-packets received by node $t$ is $n/q$. The actual number however is lower bounded by $b\mu' = n/q - O((n/q)^{2/3} \ln^{1/3}(ln/\epsilon))$. The capacity $\varphi$ of the $\omega$-schedule, for every $\omega$, (i.e., the number of usable $\omega$-packets at node $t$) is therefore lower bounded by subtracting the number of unusable $\omega$-packets at node $t$ from the total number of $\omega$-packets at node $t$, i.e.,

$$\left( 1 - O\left( \left( l^3(q/n) \ln(ln/\epsilon) \right)^{1/3} \right) \right) . (n/q),$$

w.p. b.b.b. $1 - \epsilon$, so long as

$$l^3 q \ln \frac{ln}{\epsilon} = o\left( n \right).{}^{42}$$

The last condition is only needed to ensure that the number of unusable $\omega$-packets at node $t$ is asymptotically smaller than the total number of $\omega$-packets at node $t$.

## Appendix III

### Proof of Lemma 12

By applying Lemma 31, it can be easily shown that, for a given $\omega$, the number of $\omega$-packets over a given link fails to be larger than

$$\left( 1 - O\left( \left( (q/n) \ln(1/\epsilon) \right)^{1/2} \right) \right) . (n/q), \tag{21}$$

---

[42] Assuming $l^3 q \ln(ln/\epsilon) = o(n)$, one can easily conclude that $O((l^3(q/n) \ln(ln/\epsilon))^{1/3})$ is asymptotically $o(1)$ with respect to $n$.

w.p. b.a.b. $\epsilon$.[43] To lower bound the number of $\omega$-packets for all the chunks, and over all the links, we take a union bound over the number of links and the number of chunks by setting $\epsilon = \epsilon/lq$ in (21) and (22), yielding (3) and (4), respectively.

## Appendix IV

### Details of Equation (9)

Inequality (8) can be rewritten as

$$\mu \geq \alpha + O(l\mu^{2/3} \ln^{1/3}(l\mu/\epsilon))$$

$$+ O(l \log(l\varphi/\epsilon))$$

$$+ O(\log(1/\epsilon)).$$

By replacing $\mu$ with $(1 + \lambda)\alpha$, the latter inequality reduces to

$$\alpha = \Omega((l/\lambda)\alpha^{2/3} \ln^{1/3}(\alpha l/\epsilon))$$

$$+ \Omega((l/\lambda) \log(\alpha l/\epsilon))$$

$$+ \Omega((1/\lambda) \log(1/\epsilon)).$$

The first term in the above relationship dominates the other two, and hence we obtain

$$\alpha = \Omega((l^3/\lambda^3) \ln(l\alpha/\epsilon)), \tag{23}$$

which results in

$$\alpha = \Omega((l^3/\lambda^3) \ln(l/\lambda\epsilon)).$$

[43] For a given chunk $\omega$, let $X_i$, $1 \leq i \leq n$, be an indicator random variable, so that $X_i$ is 1, if the $i$th packet is an $\omega$-packet, and $X_i$ is 0, otherwise. Since each packet pertains to a randomly (uniformly) chosen chunk, $X_i$'s are independent Bernoulli random variables with $\Pr[X_i = 1] = 1/q$. Let $Y$ be the number of $\omega$-packets over a given link, i.e., $Y = \sum_{1 \leq i \leq n} X_i$. Then, $E[Y] = n/q$. By Lemma 31, for any $\delta > 0$, $\Pr[Y \leq (1 - \delta)n/q] \leq e^{-\delta^2 n/2q}$. Setting $e^{-\delta^2 n/2q} = \epsilon$, we get $\delta = \sqrt{2(q/n)\ln(1/\epsilon)} = O(\sqrt{(q/n)\ln(1/\epsilon)})$, and hence we can write $\Pr[Y \leq (1 - O(\sqrt{(q/n)\ln(1/\epsilon)})) \cdot (n/q)] \leq \epsilon$. For $\delta$ to be in the range $(0, 1)$ (for Chernoff bound to be valid), we need the condition

$$q \ln(1/\epsilon) = o(n). \tag{22}$$

APPENDIX V

PROOF OF THEOREM 16

Before giving the proof of Theorem 16, we provide a brief overview of matringales, which are useful in proving some concentration results (see, e.g., [27, Chapter 7]).

Let $L : \mathcal{A}^{\mathcal{B}} \to \mathbb{R}$ be a functional in a probability space $\Omega = \mathcal{A}^{\mathcal{B}}$, where $\Omega$ denotes the set of functions $f : \mathcal{B} \to \mathcal{A}$. We define a measure by setting the values of $\Pr[f(b) = a]$, where the values $f(b)$ are assumed to be mutually independent. We also fix a gradation $\{\mathcal{B}_i\}_{i=0}^m$, i.e.,

$$\emptyset = \mathcal{B}_0 \subset \mathcal{B}_1 \subset \cdots \mathcal{B}_m = \mathcal{B}.$$

We define a martingale $X_0, X_1, ..., X_m$ by setting

$$X_i(h) = \mathbf{E}[L(f) | f(b) = h(b) \text{ for all } b \in \mathcal{B}_i],$$

where $h$ is a function in $\Omega$. $X_0$ is a constant, the expected value of $L$ of the random $f$. $X_m$ is $L$ itself. We say the functional $L$ satisfies $\Delta$-*Lipschitz* (or *Lipschitz*) condition relative to the gradation so long as for some $0 \leq i < m$, if $h$ and $h'$ differ only on $\mathcal{B}_{i+1} \setminus \mathcal{B}_i$, then $|L(h') - L(h)| \leq \Delta$ (or $|L(h') - L(h)| \leq 1$).

Let $L$ satisfy $\Delta$-Lipschitz condition, and $\mu = \mathbf{E}[L(h)]$. Then, for any $\gamma_a > 0$, Azuma's inequality states that

$$\Pr\left[L(h) \geq (1 + \gamma_a)\mu\right] \leq e^{-\gamma_a^2 \mu^2 / 2m\Delta}.$$

*Proof of Theorem* 16: We define the sequence of $n$ ($= (1 + \lambda)k$) independent random variables $h_1, ..., h_n$ (denoted by vector $h$) such that $h_i$ represents the index of the chunk to which the $i$th packet received at node $t$ pertains. We define $L(h)$ as the number of not fully decodable chunks given a specific vector $h$. The sequence of $X_0, X_1, ..., X_n$ defined as $X_i = \mathbf{E}[L(h)|h_j$ for $1 \leq j \leq i]$ yields a standard

martingale.[44]

Further, the functional $L$, as defined above, has Lipschitz property in that the number of chunks that cannot be decoded can differ by at most one if two sets of the indices of the chunks to which the $n$ packets (at node $t$) pertain differ in only one index. The expected number of chunks that cannot be recovered is $\epsilon q$, and applying Azuma's inequality, by setting $m = (1 + \lambda)k$, and $\mu = \epsilon q$, for any $\gamma_a > 0$, we can write

$$\Pr[L(h) \geq (1 + \gamma_a)\epsilon q] \leq e^{-(\gamma_a^2 \epsilon^2 / 2\alpha^2)k}.$$

■

## Appendix VI

## Details of Equation (15)

Inequality (14) can be rewritten as

$$\mu \geq (r/\chi)\alpha + O(l\mu^{2/3} \ln^{1/3}(l\chi\mu/\epsilon))$$

$$+ O(l \log(l\chi\mu/\epsilon))$$

$$+ O((1/\chi) \log(1/\epsilon)).$$

By replacing $\mu$ with $(1 + \lambda)\alpha/\tau$, and letting $\rho = (\chi\lambda + 1)/\tau - 1$, we can write the inequality as

$$\alpha = \Omega(l(\chi/\rho)(\alpha/\tau)^{2/3} \ln^{1/3}(l\chi\alpha/\tau\epsilon))$$

$$+ \Omega(l(\chi/\rho) \log(l\chi\alpha/\tau\epsilon))$$

$$+ \Omega((1/\rho) \log(1/\epsilon)).$$

[44]Here, $\mathcal{B}_i$ (for $0 \leq i < n$) is defined as $\mathbf{h}_i = \{h_j\}_{j=1}^{i}$. We shall show $\mathbf{E}[X_{i+1}|\mathbf{X}_i] = X_i$, where $\mathbf{X}_i = \{X_j\}_{j=0}^{i}$. Using the independence of $h_j$'s, we can write

$$\mathbf{E}[X_{i+1}|\mathbf{X}_i] \overset{(a)}{=} \mathbf{E}\left[\mathbf{E}[L(h)|\mathcal{B}_{i+1}]|\mathbf{X}_i\right]$$

$$\overset{(b)}{=} \mathbf{E}\left[\mathbf{E}[L(h)|\mathcal{B}_{i+1}]|\mathcal{B}_i\right]$$

$$\overset{(c)}{=} \sum_{\mathcal{B}_{i+1}\backslash\mathcal{B}_i} \Pr[\mathcal{B}_{i+1}|\mathcal{B}_i]\mathbf{E}[L(h)|\mathcal{B}_{i+1}]$$

$$\overset{(d)}{=} \sum_{\mathcal{B}_{i+1}\backslash\mathcal{B}_i} \Pr[\mathcal{B}_{i+1}|\mathcal{B}_i] \sum_{\mathcal{B}_n\backslash\mathcal{B}_{i+1}} \Pr[\mathcal{B}_n|\mathcal{B}_{i+1}]L(h)$$

$$\overset{(e)}{=} \sum_{\mathcal{B}_n\backslash\mathcal{B}_i} \Pr[\mathcal{B}_n|\mathcal{B}_i]L(h)$$

$$= \mathbf{E}[L(h)|\mathcal{B}_i]$$

$$= X_i,$$

where (a) since $X_{i+1} = \mathbf{E}[L(h)|\mathcal{B}_{i+1}]$, (b) $\mathbf{X}_i$ is constructed based on $\mathcal{B}_i$, (c) since $\Pr[\mathbf{E}[L(h)|\mathcal{B}_{i+1}]|\mathcal{B}_i] = \Pr[\mathcal{B}_{i+1}|\mathcal{B}_i]$, (d) by expanding $\mathbf{E}[L(h)|\mathcal{B}_{i+1}]$, and (e) by using Bayes' rule.

In the above relation, the first term dominates the other two, and we thus have

$$\alpha = \Omega((l^3\chi^3/(\rho^3\tau^2))\ln(l\chi\alpha/\tau\epsilon)).  \tag{24}$$

For $\chi \gg (\tau-1)/\lambda$, we have $\rho \simeq \chi\lambda/\tau$, and (24) will be equivalent to

$$\alpha = \Omega((l^3/\lambda^3)\tau\ln(l\chi\alpha/\tau\epsilon)).^{45}  \tag{25}$$

Condition (25) is satisfied if we have

$$\alpha = \Omega((l^3/\lambda^3)\tau\ln\alpha),$$

and

$$\alpha = \Omega((l^3/\lambda^3)\tau\ln(l\chi/\tau\epsilon)).$$

The first condition is met when

$$\alpha = \Omega((l^3/\lambda^3)\tau\ln((l/\lambda)\tau)).  \tag{26}$$

Choosing $\chi$ to be a constant integer sufficiently larger than $(\tau-1)/\lambda$, the second condition can be rewritten as

$$\alpha = \Omega((l^3/\lambda^3)\tau\ln((l/\lambda\epsilon)\tau_e)).  \tag{27}$$

Thus, both conditions (26) and (27) are met when

$$\alpha = \Omega((l^3/\lambda^3)\tau\ln((l/\lambda\epsilon)\tau)).$$

## Appendix VII

### Proof of Theorem 17

Let $n = (1+\lambda)k$, and the sequence of independent random variables $h_1, ..., h_n$, be such that $h_i$ represents the same random variable as defined in the proof of Theorem 16. Let $L(h)$ be the number of bad hyperchunks given a specific $h$ (the vector of $n$ outcomes $h_i$'s). As in the proof of Theorem 16, a martingale sequence can be constructed based on $L$. However, unlike before, here, $L$ has $\chi$-Lipschitz property in that if two sets of output symbols differ in only one symbol, then the number of bad hyperchunks can differ by at most $\chi$ (each output symbol pertains to one chunk, and each chunk belongs to $\chi$ hyperchunks). The expected number of bad hyperchunks is $\epsilon q$, and applying Azuma's inequality gives

$$\Pr[L(h) \geq (1 + \gamma_a)\epsilon q] \leq e^{-(\gamma_a^2 \epsilon^2/2\alpha^2)(\tau^2/\chi)k},$$

for any $\gamma_a > 0$.

---

[45]For $\chi \gtrsim (\tau-1)/\lambda$, we have $\rho \simeq 0$, and (24) will be equivalent to $\alpha = \Omega((l^3/\lambda^3\tau\rho^3)\tau_e^3\ln(l\tau_e\alpha/\lambda\epsilon))$. Yet, for such a choice of $\chi$, since $\rho \simeq 0$, the lower bound on $\alpha$ is much larger (and not desirable) compared to that in (25).

## Appendix VIII

## Proof of Theorems 18 and 19

In the following, we study the two types of positively and negatively dependent neighborhoods. Putting the following results together, Theorems 18 and 19 are immediate for any type of dependency between hyperchunks.

*Positively Dependent Neighborhoods:* The worst case in this type occurs when for all $i \in I$, and all subsets $I_i$ of $I \setminus \{i\}$, so that $I_i \cap N_{\mathcal{G}}(i) = \emptyset$, $\Pr[G_i | \bigwedge_{j \in I_i} G_j] = \Pr[G_i]$, and for all other $I_i$'s, $\Pr[G_i | \bigwedge_{j \in I_i} G_j] = 1$.

For all $i \in I$, let $N_{\mathcal{B}}(i)$ be an ordered set (in an increasing cyclic order) of indices of the hyperchunks that overlap with the $i$th block. Each block is bad if the hyperchunks including it are all bad. Thus, for all $i \in I$, $\Pr[B_i] = \Pr[\bigwedge_{j \in N_{\mathcal{B}}(i)} G_j]$. For a given $i \in I$, consider a permutation $\{k_1, ..., k_\ell\}$ of the elements of the set $N_{\mathcal{B}}(i)$, so that $k_1 \notin N_{\mathcal{G}}(k_2)$. Such a permutation always exists as $|N_{\mathcal{B}}(i)| > |N_{\mathcal{G}}(k_j)|$, for all $1 \leq j \leq \ell$. Thus, $\Pr[G_{k_2} | G_{k_1}] = \Pr[G_{k_2}]$. It can also be verified that for such a permutation, the hyperchunks $\{\mathcal{G}_j : j \in \{k_3, ..., k_\ell\}\}$ overlap with at least one of the hyperchunks $\mathcal{G}_{k_1}$ and $\mathcal{G}_{k_2}$. The probability that a hyperchunk fails to be decodable is shown to be b.a.b. $\epsilon$, and hence, for all $k \in N_{\mathcal{B}}(i)$, $\Pr[G_k] \leq \epsilon$. By the above argument along with applying chain rule, we have $\Pr[B_i] = \Pr[\bigwedge_{j \in N_{\mathcal{B}}(i)} G_j] = \Pr[G_{k_1}] \Pr[G_{k_2} | G_{k_1}] = \Pr[G_{k_1}] \Pr[G_{k_2}] \leq \epsilon^2 := \xi$.

Each block is bad w.p. b.a.b. $\xi$, and for any choice of $\epsilon$, so long as $\xi$ goes to zero sufficiently fast with $k$, by applying a union bound, one can show that not all the blocks are recoverable w.p. b.a.b. $\xi q$. The tightness of this upper bound can be readily shown by rewriting $\Pr[\bigwedge_{j \in N_{\mathcal{B}}(i)} G_j]$ as $1 - \Pr[\bigvee_{j \in N_{\mathcal{B}}(i)} \overline{G}_j]$, and using the inclusion-exclusion principle.

Since the expected fraction of bad blocks is upper bounded by $\xi$, for larger choices of $\epsilon$ up to a constant, by constructing a martingale as before - yet this time, over the blocks, not the hyperchunks - we would be able to prove the concentration of the actual fraction of bad blocks around its expected value as follows.

*Theorem 22:* For any $\epsilon > 0$, and $\lambda > 0$, in an OCC with an aperture size $\alpha = \Omega\left((l^3/\lambda^3)\tau \ln\left((l/\lambda\epsilon)\tau\right)\right)$, and overlap parameter $\tau$, over a network of length $l$ with any one-in-one-out worst-case schedule of capacity $(1+\lambda)k$, for any type of dependency between hyperchunks with positively dependent neighborhoods, for any $\gamma_a > 0$, the fraction of blocks that are not recoverable deviates farther than $\gamma_a$ from $\xi$, w.p. b.a.b. $e^{-ck}$, for some positive constant $c = O((\gamma_a^2\xi^2/\alpha^2)(\lambda\tau))$.

*Proof:* Let $\ell$ be the number of blocks that a hyperchunk contains, i.e., $\chi + \tau - 1$. The proof is similar to that of Theorem 17, except that $L$ is $\ell$-Lipschitz in this case. The expected number of bad blocks is $\xi q$, and applying Azuma's inequality, for any $\gamma_a > 0$, we have

$$\Pr[L(h) > (1 + \gamma_a)\xi q] \le e^{-ck},$$

by choosing $c = O((\gamma_a^2\xi^2/\alpha^2)(\tau^2/\ell))$. ∎

Each block has $k/q$ message packets. Therefore, the result of Theorem 22 shows that for any type of dependency between hyperchunks with positively dependent neighborhoods, w.h.p., for any arbitrarily small constant $\gamma_a > 0$, the number of message packets that are not recoverable is upper bounded by $(1 + \gamma_a)\xi k$.

*Negatively Dependent Neighbors:* The worst case in this type occurs when for all $i \in I$, and all subsets $I_i$ of $I \setminus \{i\}$, $\Pr[G_i | \bigwedge_{j \in I_i} G_j] = \Pr[G_i]$. That is, for all $i \in I$, $G_i$'s are independent. Thus, $\Pr[B_i] = \Pr[\bigwedge_{j \in N_{\mathcal{B}}(i)} G_j] = \prod_{j \in N_{\mathcal{B}}(i)} \Pr[G_i] \le \epsilon^{|N_{\mathcal{B}}(i)|} = \epsilon^\ell := \eta$.

Now, for any choice of $\epsilon$, so long as $\eta$ goes to zero sufficiently fast with $k$, by applying a union bound, one can see that not all the blocks are recoverable w.p. b.a.b. $\eta q$. We prove the tightness of this upper bound in the following.

Suppose the problem of lower bounding the probability that a set of events $B_i$'s do not occur. For mutually independent events, $\Pr[\bigwedge_{i \in I} \overline{B_i}] = \prod_{i \in I} \Pr[\overline{B_i}]$. However, in the case that $B_i$'s are not independent, but "mostly" independent (each is only dependent on a small subset of the others), Janson's inequality provides a tight bound on $\Pr[\bigwedge_{i \in I} \overline{B_i}]$. The following is a short description.

Let $\Omega$ be a finite universal set (a set which contains all elements of interest, including itself) and let $\mathcal{A}$ be a random subset of $\Omega$ such that, for every element $r \in \Omega$, $\Pr[r \in \mathcal{A}] = p_r$, and the events $\{r \in \mathcal{A}\}_{r \in \Omega}$ are mutually independent. Let $\{A_i\}_{i \in I}$ be subsets of $\Omega$, and $I$ be a finite index set. We define the event $A_i \subseteq \mathcal{A}$ as the event $B_i$. For $i, j \in I$, we write $i \sim j$ if $i \ne j$ and $A_i \cap A_j \ne \emptyset$; otherwise, $i \not\sim j$. For $i \ne j$, and $i \not\sim j$, $B_i, B_j$ are independent events. Further, for $i \notin J \subset I$, and $i \not\sim j$, for all $j \in J$, $B_i$ is mutually independent of $\{B_j\}_{j \in J}$, and is thus independent of any Boolean function of those $B_j$'s.

Let $\Delta = \sum_{i \sim j} \Pr[B_i \wedge B_j]$, where the sum is over pairs $(i, j)$, with $(i, j)$ and $(j, i)$ are counted as one pair, and $P = \prod_{i \in I} \Pr[\overline{B_i}]$. The following is known as the Janson's inequality.

*Lemma 32:* [27, Chapter 8] Let $\{B_i\}_{i \in I}, \Delta, P$ be defined as above and assume $\Pr[B_i] \leq \delta$, for all $i \in I$. Then

$$P \leq \Pr[\bigwedge_{i \in I} \overline{B_i}] \leq P e^{\Delta/(1-\delta)}.$$

Now, let $\mathcal{A}$ be the union set of events $H_i$, for all $i \in I$, so that $H_i$ is either $G_i$, or $\overline{G_i}$. Further, let $A_i$ be the intersection set of events $\{G_j\}_{j \in N_{\mathcal{B}}(i)}$, and $B_i$ be the event $A_i \subseteq \mathcal{A}$ (i.e., $B_i$ occurs if the intersection of the events $\{G_j\}_{j \in N_{\mathcal{B}}(i)}$ occurs). Then, for $i, j \in I$ ($i \neq j$), and $N_{\mathcal{B}}(i) \cap N_{\mathcal{B}}(j) \neq \emptyset$, $i \sim j$; for $i \neq j$, and $i \not\sim j$, $B_i, B_j$ are independent due to the dependence on two disjoint subsets of $\{G_k\}_{k \in I}$. For $i, j \in I$ ($i < j$), and $i \sim j$, $N_{\mathcal{B}}(i) = \{k_1, ..., k_\ell\}$, $N_{\mathcal{B}}(j) = \{k_{j-i+1}, ..., k_\ell, ..., k_{\ell+j-i}\}$ (in cyclic order), and $N_{\mathcal{B}}(i) \cup N_{\mathcal{B}}(j) = \{k_1, ..., k_\ell, ..., k_{\ell+j-i}\}$. Thus, $|N_{\mathcal{B}}(i) \cup N_{\mathcal{B}}(j)| = \ell + j - i$, and

$$
\begin{aligned}
\Pr[B_i \wedge B_j] &= \Pr[\{\bigwedge_{k \in N_{\mathcal{B}}(i)} G_k\} \cap \{\bigwedge_{k \in N_{\mathcal{B}}(j)} G_k\}] \\
&= \Pr[\bigwedge_{k \in N_{\mathcal{B}}(i) \cup N_{\mathcal{B}}(j)} G_k] \\
&= \prod_{k \in N_{\mathcal{B}}(i) \cup N_{\mathcal{B}}(j)} \Pr[G_k] \\
&= \Pr[G_k]^{|N_{\mathcal{B}}(i) \cup N_{\mathcal{B}}(j)|} \\
&\leq \epsilon^{|N_{\mathcal{B}}(i) \cup N_{\mathcal{B}}(j)|} \\
&= \epsilon^{\ell+j-i}.
\end{aligned}
$$

Thus,

$$
\begin{aligned}
\Delta &= \sum_{i \sim j} \Pr[B_i \wedge B_j] \\
&= \sum_{i \in I} \sum_{j: i < j < i + \ell} \Pr[B_i \wedge B_j] \\
&\leq \sum_{i \in I} \sum_{j: i < j < i + \ell} \epsilon^{\ell + j - i} \\
&= \sum_{i \in I} \sum_{0 < j - i < \ell} \epsilon^{\ell + j - i} \\
&= \eta \sum_{i \in I} \sum_{0 < k < \ell} \epsilon^k \\
&= \eta \sum_{i \in I} (\epsilon - \eta) / (1 - \epsilon) \\
&= \eta q (\epsilon - \eta) / (1 - \epsilon).
\end{aligned}
$$

For sufficiently small choice of $\epsilon$, so long as $\eta q$ goes to zero, $\Delta$ goes to zero. Thus, in the limit of interest, the Janson's inequality results in $\Pr[\bigwedge_{i \in I} \overline{B_i}] = P$. Since $\Pr[B_i] \leq \eta$, $P = \prod_{i \in I} \Pr[\overline{B_i}] \geq (1 - \eta)^q \geq 1 - \eta q$.

For larger values of $\epsilon$, a block is bad w.p. b.a.b. $\eta$, and hence, an upper bound on the fraction of bad blocks is $\eta$. The actual fraction of bad blocks can be shown to be tightly concentrated around the upper bound on the expected fraction as follows.

*Theorem 23:* For any $\epsilon > 0$, and $\lambda > 0$, in an OCC with an aperture size $\alpha = \Omega\left((l^3/\lambda^3)\tau \ln\left((l/\lambda\epsilon)\tau\right)\right)$, and overlap parameter $\tau$, over a network of length $l$ with any one-in-one-out worst-case schedule of capacity $(1 + \lambda)k$, for any type of dependency between hyperchunks with negatively dependent neighborhoods, for any $\gamma_a > 0$, the fraction of blocks that are not recoverable deviates farther than $\gamma_a$ from $\eta$, w.p. b.a.b. $e^{-e^{ck}}$, for some positive constant $c = O(\lambda/\alpha)$.

Before giving the proof, we state a theorem which is useful in proving concentration results for the sum of dependent indicator random variables (see, e.g., [28]).

For a given set $I$, and a sequence of random variables $\{X_i\}_{i \in I}$, a subset $J$ of $I$ is called *independent* if $\{X_i\}_{i \in J}$ are independent. A sequence $\{I_j\}$ of subsets of $I$ is a *cover* of $I$, if $\bigcup_j I_j = I$. A sequence $\{(I_j, \omega_j)\}$ of pairs $(I_j, \omega_j)$, where $I_j \subseteq I$ and $\omega_j \in [0,1]$ is a *fractional cover* of $I$, if $\sum_{j: i \in I_j} \omega_j \geq 1$, for each $i \in I$. A cover or a fractional cover is *proper* if each set $I_j$ in it is independent. Let $M$ be the size (the number of subsets $I_j$'s) of the smallest proper cover of $I$, and $M_m$ be the smallest $\sum_j \omega_j$ over all proper fractional covers $\{(I_j, \omega_j)\}$.

*Lemma 33:* [28, Corollary 2.2] Suppose that $X = \sum_{i \in I} X_i$, and for all $i \in I$, $X_i \sim \text{Bernoulli}(p_i)$, for some $0 < p_i < 1$. Let $\mu = \mathbf{E}[X]$. For any $\gamma_a \geq 0$,

$$\Pr[X \geq (1 + \gamma_a)\mu] \leq e^{-2\gamma_a^2 \mu^2 / M_m |I|}.$$

*Proof of Theorem* 23: Let $I$ be the set of integers $[q]$. For every $i \in I$, let $X_i$ be an indicator random variable associated with the $i$th block ($\mathcal{B}_i$), so that $X_i$ is 1 if $B_i$ occurs, and it is 0, otherwise. Then, $X = \sum_{i \in I} X_i$ is the number of bad blocks. Since each block belongs to $\ell$ contiguous hyperchunks, the event of a given block being bad is a function of the events of decodability of the hyperchunks containing it. Since for all $j \in I$, $j$th block belongs to $\ell$ contiguous hyperchunks, the next block which does not belong to any of these hyperchunks is the $(j + \ell)$th block, and so forth ($\chi$ is chosen properly so that $\ell$ is a divisor of $q$). Thus, for all $j \in I$, $B_j, B_{j+\ell}, ..., B_{j+q-\ell}$ are independent. Taking $I_j = \{j, j+\ell, ..., j+q-\ell\}$, for all $j \in [\ell]$, the family $\{I_j\}$ of subsets of $I$ is the smallest proper cover of $I$, because for every $j \in [\ell]$, $I_j$ is the largest possible subset which contains all the possible indices whose associated indicator random variables $\{X_i\}_{i \in I_j}$ are independent. The size $M$ of such a proper cover of $I$ is $\ell$. To give an upper bound on $M_m$, let us expand the family $\{I_j\}$ as follows: the new family $U(\{I_j\})$ contains any non-empty subset of $I_j$, for each $j \in [\ell]$ (i.e., $U(\{I_j\})$ is the union of the powersets of subsets $I_j$'s, excluding the null set).[46] By definition, $U(\{I_j\})$ is a cover. On the other hand, there are $2^{q/\ell} - 1$ non-empty subsets of $I_j$. By symmetry, each of $q/\ell$ distinct indices is equally distributed in these subsets, and hence each index appears $(2^{q/\ell} - 1)/(q/\ell)$ times. Setting $\omega_j = (q/\ell)/(2^{q/\ell} - 1)$, the inequality $\sum_{j:i \in I_j} \omega_j \geq 1$ holds with equality. Since there are $q$ indices in $I$, $\sum_{j \in I} \omega_j = q \left[(q/\ell)/(2^{q/\ell} - 1)\right]$. Thus, $M_m \leq q \left[(q/\ell)/(2^{q/\ell} - 1)\right]$. Since $\Pr[B_i] = \eta$ (assuming the worst-case scenario), the expected number of bad blocks is $\eta q$. For any $\gamma_a > 0$, then, from Lemma 33, it follows that

$$\Pr[X \geq (1 + \gamma_a)\eta q] \leq e^{-e^{ck}},$$

by choosing $c = O(\tau/\alpha\ell) - \left(\ln(k) - O(\ln((\alpha\ell/\tau)\gamma_a^2\eta^2))\right)/k$. ∎

Therefore, for any type of dependency between hyperchunks with negatively dependent neighborhoods, w.h.p., for any arbitrarily small constant $\gamma_a > 0$, the number of unrecoverable message packets is upper bounded by $(1 + \gamma_a)\eta k$.

---

[46] The powerset of a set $S$ is the set of all subsets of $S$, including the null set and the set $S$ itself.